\documentclass[draft,jgrga]{agutex}
  \usepackage{lineno}
  \usepackage[pdftex]{graphicx}

   \setkeys{Gin}{draft=false}
\usepackage{morefloats}
\usepackage{amsmath}
\usepackage{setspace}
\usepackage{graphicx,color}
\usepackage{amssymb}
\usepackage{url}
\authorrunninghead{LIU ET AL.}

\titlerunninghead{DUST PARTICLES FROM THE GALILEAN MOONS}

\authoraddr{Corresponding author: X. Liu,
Astronomy and Space Physics, University of Oulu, Finland.
(xiaodong.liu@oulu.fi)}

\begin{document}

%% ------------------------------------------------------------------------ %%
%
%  TITLE
%
%% ------------------------------------------------------------------------ %%

\title{Dynamics and distribution of Jovian dust ejected from the Galilean satellites}
%
% e.g., \title{Terrestrial ring current:
% Origin, formation, and decay $\alpha\beta\Gamma\Delta$}
%

%% ------------------------------------------------------------------------ %%
%
%  AUTHORS AND AFFILIATIONS
%
%% ------------------------------------------------------------------------ %%

%Use \author{\altaffilmark{}} and \altaffiltext{}

% \altaffilmark will produce footnote;
% matching \altaffiltext will appear at bottom of page.

 \authors{Xiaodong Liu,\altaffilmark{1} Manuel Sachse,\altaffilmark{2}
 Frank Spahn,\altaffilmark{2} and J\"urgen Schmidt\altaffilmark{1}}

\altaffiltext{1}{Astronomy and Space Physics, University of Oulu, Finland.}

\altaffiltext{2}{Institut f{\"u}r Physik und Astronomie, Universit{\"a}t Potsdam, Germany.}

%% ------------------------------------------------------------------------ %%
%
%  KEYPOINTS
%
%% ------------------------------------------------------------------------ %%

% Key points are 1 to 3 points that the author provides,
% that are 100 characters or less, that are ultimately published
% with the article.
%% for example:
% \keypoints{\item Here is the first keypoint. what happens if it is a
% long keypoint, like this one. We want to see this wrap please.
% \item This is the second.
% \item And here is the third keypoint
% }

\keypoints{\item Circumplanetary dust dynamics \item Jupiter dust \item Galilean satellites}

%% Keypoints will print underneath the abstract.

%% ------------------------------------------------------------------------ %%
%
%  ABSTRACT
%
%% ------------------------------------------------------------------------ %%

% >> Do NOT include any \begin...\end commands within
% >> the body of the abstract.

\begin{abstract}
In this paper, the dynamical analysis of the Jovian dust originating from the four Galilean moons is presented. High accuracy orbital integrations of dust particles are used to determine their dynamical evolution. A variety of forces are taken into account, including the Lorentz force, solar radiation pressure, Poynting-Robertson drag, solar gravity, the satellites' gravity, plasma drag, and gravitational effects due to non-sphericity of Jupiter. More than 20,000 dust particles from each source moon in the size range from 0.05 micron to 1 cm are simulated over 8,000 (Earth) years until each dust grain hits a sink (moons, Jupiter, or escape from the system). Configurations of dust number density in the Jovicentric equatorial inertial frame are calculated and shown. In a Jovicentric frame rotating with the Sun the dust distributions are found to be asymmetric. For certain small particle sizes, the dust population is displaced towards the Sun, while for certain larger sizes, the dust population is displaced away from the Sun. The average lifetime as a function of particle size for ejecta from each source moon is derived, and two sharp jumps in the average lifetime are analyzed. Transport of dust between the Galilean moons and to Jupiter is investigated. Most of the orbits for dust particles from Galilean moons are prograde, while, surprisingly, a small fraction of orbits are found to become retrograde mainly due to solar radiation pressure and Lorentz force. The distribution of orbital elements is also analyzed.
\end{abstract}

%% ------------------------------------------------------------------------ %%
%
%  BEGIN ARTICLE
%
%% ------------------------------------------------------------------------ %%

% The body of the article must start with a \begin{article} command
%
% \end{article} must follow the references section, before the figures
%  and tables.

\begin{article}

%% ------------------------------------------------------------------------ %%
%
%  TEXT
%
%% ------------------------------------------------------------------------ %%

\section{Introduction}
Dust particles in the Jovian system were first discovered in the early 1970s by Pioneer 10 and 11 \citep{humes1974interplanetary}. Subsequently, from space missions and improving earth-based observations, more detailed information on the Jovian dust complex were revealed. The ring system was confirmed by Voyagers 1 and 2 \citep{smith1979jupiter, owen1979jupiter}, followed by the discoveries of Jovian dust streams \citep{grun1992ulysses, grun1996constraints}, ejecta clouds around the Galilean moons \citep{kruger1999detection}, a tenuous ring in the region of the Galilean moons \citep{krivov2002tenuous}, and dust in the region of the Jovian irregular satellites \citep{krivov2002dust}.

There are a large number of papers devoted to the study of the dynamics of the Jovian ring system including the so-called halo ring, the main ring, and two gossamer rings. In previous studies, the effect of Lorentz resonances on the Jovian ring system was investigated \citep{burns1985lorentz, schaffer1992lorentz, hamilton1994comparison}, as well as the effects of Poynting-Robertson drag \citep{burns1999formation}, plasma drag \citep{schaffer1987dynamics}, magnetospheric capture \citep{colwell1998capture, colwell1998jupiter}, cometary disruption \citep{showalter2011impact}, the role of the planetary shadow \citep{hamilton2008sculpting}, and resonant charge variations \citep{burns1989orbital, horanyi1996structure, horanyi2010plasma}. Besides, the dynamics of particles from the irregular satellites were analyzed by \citet{krivov2002dust} and \citet{bottke2013black}. The reader is also referred to the reviews by \citet{burns2001dusty}, \citet{kruger2004jovian} and \citet{burns2004jupiter}.

However, less attention has been paid to the dynamics of dust in the region of Galilean moons. Impact-generated dust clouds surrounding the Galilean moons were detected by the Galileo dust detector \citep{kruger1999detection, krueger2000, kruger2003impact, sremvcevic2005impact}. The same process was found to generate also a dust cloud around Earth's moon \citep{horanyi2015permanent}. A fraction of particles in the ejecta clouds may escape from the Galilean moons' gravity and contribute to the tenuous ring in the region of the Galilean moons. \citet{krivov2002tenuous} analyzed the dynamics of a tenuous ring formed by dust particles from Europa, Ganymede and Callisto, and \citet{zeehandelaar2007local} simulated the motions of 5 $\mu \mathrm{m}$ particles radially-launched from the Galilean moons with the satellites' escape velocity. Besides, six potential sources of dust particles in the Galilean region were analyzed using the full data set of the Galileo Dust Detection Subsystem \citep{soja2015new}. Here we present a more detailed model, including higher degree Jovian gravity, Lorentz force, solar radiation pressure, Poynting-Robertson drag, plasma drag, and gravity of the Sun and satellites. 

This paper is organized as follows. In Section \ref{section_model}, the dynamical model is described. In Section \ref{section_numerical}, the simulation procedures are explained, as well as sinks of dust and storage scheme. The details of the calculation of dust distributions in the Jovicentric equatorial inertial frame for various grain sizes are given in Section \ref{section_distri_inertial}. In Section \ref{section_lifetime}, the average lifetime for each grain size from each source moon is shown. The lifetimes of 0.05 $\mu \mathrm{m}$ (and smaller) dust particles are estimated analytically, and two sharp jumps in the average lifetime are analyzed. In Section \ref{section_asymmetry}, the azimuthal asymmetry of the dust configurations in the Jovicentric frame rotating with the Sun and the reason for that is explained. In Section \ref{section_transport}, transport of the particles between the Galilean moons and to Jupiter is demonstrated. In Section \ref{section_retrograde}, the mechanism for a fraction of orbits becoming retrograde is explained. The distributions of orbital elements are analyzed in Section \ref{section_distri_elem}.

\section{Dynamical Model} \label{section_model}
In addition to the Newtonian gravity exerted by Jupiter, the motion of dust particles in the Jovian system is influenced by a variety of forces. The most important ones are the Lorentz force, solar radiation pressure, Poynting-Robertson drag, higher degree Jovian gravity, plasma drag, and gravitational perturbations induced by the Sun and the four Galilean moons.

Several reference frames are used in our model. The integration of the equations of motion is performed in the Jupiter equatorial inertial frame (JIF) $Ox_\mathrm{i}y_\mathrm{i}z_\mathrm{i}$. Here, the $z_\mathrm{i}$ axis is defined as the spin axis of Jupiter at the J2000 epoch, the $x_\mathrm{i}$ axis is aligned with the ascending node of the Jovian orbital plane with the Jovian equator plane at the J2000 epoch, and the $y_\mathrm{i}$ axis completes an orthogonal right-handed frame. A Jupiter-centered body-fixed frame $Ox_\mathrm{bod}y_\mathrm{bod}z_\mathrm{bod}$ is also used, which is right-handed and abbreviated as JBODF. The definitions of the $x_\mathrm{bod}$ and $z_\mathrm{bod}$ axes are the same as for the left-handed Jupiter System III (1965) \citep{seidelmann1977evaluation}, i.e.~the $z_\mathrm{bod}$ axis is aligned with the spin axis of Jupiter, and the $x_\mathrm{bod}$ axis points to the prime meridian of Jupiter, but the $y_\mathrm{bod}$ axis completes an orthogonal right-handed frame. For the simulations and most part of this paper SI units are used. In some places we use Gaussian units to be consistent with the notation in previous studies.

The equations of motion of a dust particle can be expressed as
\begin{linenomath*}
\begin{equation} \label{equ_dynamic_model}
\ddot{\vec r} = \ddot{\vec r}_{\mathrm {G_J}} + \ddot{\vec r}_\mathrm L + \ddot{\vec r}_\mathrm{RP} + \ddot{\vec r}_\mathrm{PR} + \ddot{\vec r}_\mathrm{PD} + \ddot{\vec r}_{\mathrm{G_{other \ bodies}}} .
\end{equation}
\end{linenomath*}
Here, ${\vec r}$ is the instantaneous radius vector of a given grain and $\ddot{\vec r}_{\mathrm {G_J}}$ is the acceleration due to Jovian gravity including higher degrees terms,
\begin{linenomath*}
\begin{equation}
\ddot{\vec r}_{\mathrm {G_J}} = GM_\mathrm J\nabla\left\{\frac{1}{r}\left[1-\sum_{n=1}^{N_J}J_{2n}\left(\frac{R_\mathrm J}{r}\right)^{2n}P_{2n}(\cos\theta)\right]\right\} .
\end{equation}
\end{linenomath*}
The gravitational constant is denoted by $G$, $M_\mathrm J$ is the mass of Jupiter, $R_\mathrm J$ is the Jovian reference radius, $\theta$ is the colatitude in JBODF, $J_{2n}$ are even degrees zonal harmonics, and $P_{2n}$ are the normalized Legendre functions of degree $2n$. In this paper $N_J=3$, i.e.~$J_2$, $J_4$ and $J_6$ are considered, the values of which can be found in Table \ref{tab:galilean_moons}.

The term $\ddot{\vec r}_\mathrm L$ describes the Lorentz acceleration
\begin{linenomath*}
\begin{equation}
\ddot{\vec r}_\mathrm L = \frac{Q}{m_\mathrm{g}}\left(\dot{\vec r}-\vec \Omega_\mathrm{J}\times \vec r\right)\times{\vec B}
\end{equation}
\end{linenomath*}
where $m_\mathrm{g}$ is the mass of the dust grain, and $Q=4\pi\varepsilon_0r_\mathrm g\Phi$ is the grain charge with the vacuum permittivity $\varepsilon_0$. Further, $r_\mathrm g$ is the grain radius, and $\Phi$ the surface potential of the dust particles. The non-uniform rotation of Jupiter is considered, where the components of the pseudovector $\vec \Omega_\mathrm{J}$ in the inertial frame JIF $(\Omega_{\mathrm{J}x}, \ \Omega_{\mathrm{J}y}, \ \Omega_{\mathrm{J}z})$ are calculated from
\begin{linenomath*}
\begin{equation}
\begin{bmatrix}
0		& -\Omega_{\mathrm{J}z}	&  \Omega_{\mathrm{J}y} \\
 \Omega_{\mathrm{J}z}	& 0		& -\Omega_{\mathrm{J}x} \\
-\Omega_{\mathrm{J}y}	& \Omega_{\mathrm{J}x}	& 0
\end{bmatrix} = \dot A_\mathrm{ipbod}^T A_\mathrm{ipbod}.
\end{equation}
\end{linenomath*}
Here $A_\mathrm{ipbod}$ is the transformation matrix from the inertial frame JIF to the body-fixed frame JBODF. It is worth mentioning that $\Omega_{\mathrm{J}x} \ll \Omega_{\mathrm{J}z}$, $\Omega_{\mathrm{J}y} \ll \Omega_{\mathrm{J}z}$, and $\Omega_{\mathrm{J}z} \simeq \Omega_{\mathrm{J}} = |\vec \Omega_\mathrm{J}|$. To evaluate the local magnetic field $\vec B$, we use the latest Jovian magnetic field model VIPAL, up to 5th degree and 5th order \citep{hess2011model}. The scalar magnetic potential can be expressed in terms of spherical harmonics up to 5th degree and 5th order \citep{chapman1940geomagnetism}
\begin{linenomath*}
\begin{equation}
\Phi_\mathrm{mag} = R_\mathrm J\sum_{n=1}^{5}\sum_{m=0}^{n}\left(\frac{R_\mathrm J}{r}\right)^{n+1}\tilde{P}_n^m(\cos\theta)[g_n^m\cos(m\phi)+h_n^m\sin(m\phi)]
\end{equation}
\end{linenomath*}
where $\tilde{P}_n^m$ is the Schmidt seminormalized associated Legendre function of degree $n$ and order $m$, and $g_n^m$ and $h_n^m$ are the Schmidt coefficients. The angle $\phi$ is the longitude of the particle in the body-fixed frame JBODF. The magnetic field $\vec B$ is then obtained from
\begin{linenomath*}
\begin{equation}
\vec B = -\nabla\Phi_\mathrm{mag} = -R_\mathrm J\nabla\sum_{n=1}^{5}\sum_{m=0}^{n}\left(\frac{R_\mathrm J}{r}\right)^{n+1}\tilde{P}_n^m(\cos\theta)[g_n^m\cos(m\phi)+h_n^m\sin(m\phi)].
\end{equation}
\end{linenomath*}
After some algebra, the full expression of $\vec B$ reads
\begin{linenomath*}
\begin{equation}
\begin{split}
\vec B = &\sum_{n=1}^{5}\sum_{m=0}^{n}\left(\frac{R_\mathrm J}{r}\right)^{n+2}\bigg\{(n+1)\tilde{P}_n^m(\cos\theta)\left[g_n^m\cos(m\phi)+h_n^m\sin(m\phi)\right]\vec e_r \\
&+\frac{1}{\sin\theta}\left[\sqrt{(n+m)(n-m)}\tilde{P}_{n-1}^m(\cos\theta)-n\cos\theta \tilde{P}_n^m(\cos\theta)\right]\left[g_n^m\cos(m\phi)+h_n^m\sin(m\phi)\right]\vec e_{\theta} \\
&+\frac{m}{\sin\theta}\tilde{P}_n^m(\cos\theta)\left[g_n^m\sin(m\phi)-h_n^m\cos(m\phi)\right]\vec e_{\phi}\bigg\}
\end{split}
\end{equation}
\end{linenomath*}
where $\vec e_r$, $\vec e_{\theta}$ and $\vec e_{\phi}$ are the unit vectors in radial, polar and azimuthal directions in JBODF. Note that if $\theta=0$, the scalar magnetic potential reduces
\begin{linenomath*}
\begin{equation}
\Phi_\mathrm{mag} = R_\mathrm J\sum_{n=1}^{5}\left(\frac{R_\mathrm J}{r}\right)^{n+1}g_n^0
\end{equation}
\end{linenomath*}
and the magnetic field
\begin{linenomath*}
\begin{equation}
\vec B = \sum_{n=1}^{5}\left(\frac{R_\mathrm J}{r}\right)^{n+2}g_n^0(n+1)\vec e_r
\end{equation}
\end{linenomath*}
while at $\theta=\pi$ the scalar magnetic potential reads
\begin{linenomath*}
\begin{equation}
%\Phi_\mathrm{mag} = R_\mathrm J\sum_{n=1}^{5}(-1)^n\left(\frac{R_\mathrm J}{r}\right)^{n+1}g_n^0
\Phi_{\text{mag}} = R_{\mathrm{J}}\sum_{n=1}^{5}(-1)^{n}\left(\frac{R_{\mathrm{J}}}{r}\right)^{n+1}{g_{n}^{0}}
\end{equation}
\end{linenomath*}
and the magnetic field
\begin{linenomath*}
\begin{equation}
\vec B = \sum_{n=1}^{5}\left(\frac{R_\mathrm J}{r}\right)^{n+2}(-1)^ng_n^0(n+1)\vec e_r .
\end{equation}
\end{linenomath*}
The Cartesian components of $\vec B$ in the body-fixed frame JBODF can be presented as
\begin{linenomath*}
\begin{equation}
B_{x_\mathrm{bf}} = B_r\sin\theta\cos\phi+B_{\theta}\cos\theta\cos\phi-B_{\phi}\sin\phi
\end{equation}
\end{linenomath*}
\begin{linenomath*}
\begin{equation}
B_{y_\mathrm{bf}} = B_r\sin\theta\sin\phi+B_{\theta}\cos\theta\sin\phi+B_{\phi}\cos\phi
\end{equation}
\end{linenomath*}
\begin{linenomath*}
\begin{equation}
B_{z_\mathrm{bf}} = B_r\cos\theta-B_{\theta}\sin\theta
\end{equation}
\end{linenomath*}
where $B_r=\vec B \cdot \vec e_r$, $B_{\theta}=\vec B \cdot \vec e_{\theta}$, and $B_{\phi}=\vec B \cdot \vec e_{\phi}$. Finally, the components of $\vec B$ in the inertial frame JIF are obtained from
\begin{linenomath*}
\begin{equation}
\begin{bmatrix}
B_x \\
B_y \\
B_z
\end{bmatrix} = A_\mathrm{bodpi}
\begin{bmatrix}
B_{x_\mathrm{bf}} \\
B_{y_\mathrm{bf}} \\
B_{z_\mathrm{bf}}
\end{bmatrix}.
\end{equation}
\end{linenomath*}
Here $A_\mathrm{bodip}$ is the transformation matrix from JBODF to JIF.

The Lorentz force can be considered to be composed of two parts, the electric part $F_\mathrm{e}$ and the magnetic part $F_\mathrm{m}$ \citep[e.g.][]{northrop1982adiabatic, hamilton1993motion}
\begin{linenomath*}
\begin{equation}
{\vec F_\mathrm{e}} = -Q(\vec \Omega_\mathrm{J}\times \vec r)\times{\vec B} = -Q\begin{bmatrix}
										\Omega_{\mathrm{J}y}z - \Omega_{\mathrm{J}z}y \\
										\Omega_{\mathrm{J}z}x - \Omega_{\mathrm{J}x}z \\
										\Omega_{\mathrm{J}x}y - \Omega_{\mathrm{J}y}x
										\end{bmatrix} \times{\vec B}
\end{equation}
\end{linenomath*}
and
\begin{linenomath*}
\begin{equation}
{\vec F_\mathrm{m}} = Q(\dot{\vec r} \times{\vec B} ).
\end{equation}
\end{linenomath*}
Both $\vec F_\mathrm{e}$ and $\vec F_\mathrm{m}$ have radial, azimuthal and polar components.

\begin{figure}
\centering 
\noindent\includegraphics[width=0.55\textwidth, angle=90]{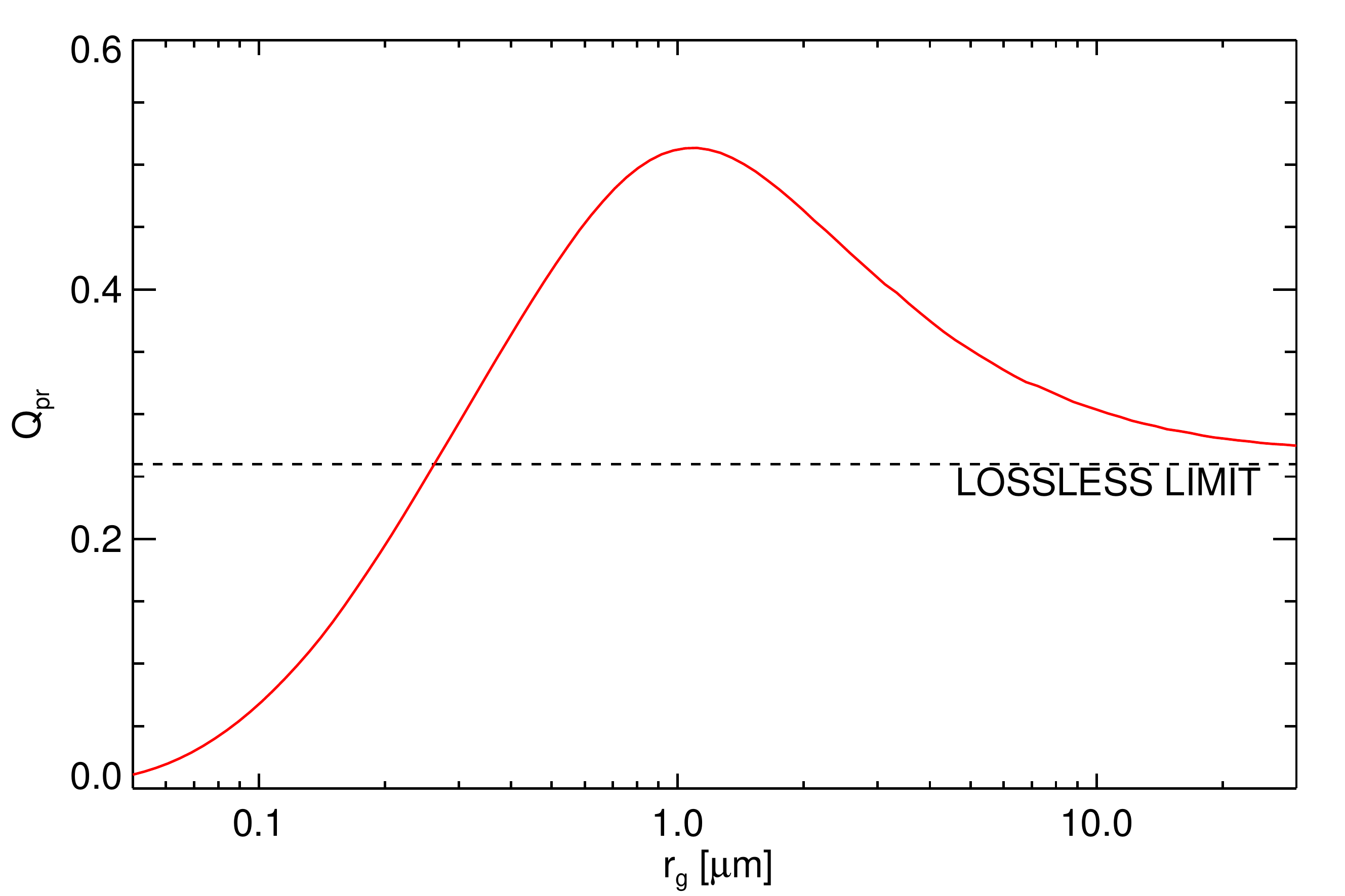}
\caption{\label{fig_MakeQprWarren} Size-dependent radiation pressure efficiency $Q_\mathrm{pr}$. The values are computed from Mie theory \citep{mishchenko2002scattering, mishchenko1999bidirectional}, averaging over the solar spectrum and using an experimentally determined, wavelength dependent refractive index for water ice \citep{warren1984optical}. The lossless limit for spherical grains with a refractive index of 1.333 is shown as a dashed line \citep{van1981light}. For a non-zero imaginary part of the refractive index the value of $Q_\mathrm{pr}$ will rise again for larger particle sizes (not shown here), for which, however, the effect of radiation pressure is negligibly small.}
\end{figure}

The acceleration due to solar radiation pressure is given by \citep{BURNS:1979wg}
\begin{linenomath*}
\begin{equation}
\ddot {\vec r}_\mathrm{RP} = \frac{3Q_\mathrm SQ_\mathrm {pr}\mathrm{AU}^2}{4(\vec r-{\vec r}_\mathrm S)^2\rho_\mathrm gr_\mathrm gc}\left[1-\frac{(\dot{\vec r}-\dot{\vec r}_\mathrm S)\cdot\hat{\vec r}_\mathrm {Sd}}{c}\right]\hat{\vec r}_\mathrm {Sd}
\end{equation}
\end{linenomath*}
where $Q_\mathrm S$ is the solar radiation energy flux at one AU (astronomical unit), $c$ is the speed of light, ${\vec r}_\mathrm S$ is the position vector from Jupiter to Sun, and $\hat{\vec r}_\mathrm {Sd}$ is the unit vector in the direction of the incident photon beam \citep{BURNS:1979wg}
\begin{linenomath*}
\begin{equation}
\hat{\vec r}_\mathrm {Sd} = \frac{\vec r-{\vec r}_\mathrm S}{|\vec r-{\vec r}_\mathrm S|}.
\end{equation}
\end{linenomath*}
The variable $Q_\mathrm{pr}$ is the solar radiation pressure efficiency factor. The dependence of $Q_\mathrm{pr}$ on grain size for icy particles, calculated from Mie theory for spherical grains, is shown in Fig.~\ref{fig_MakeQprWarren}. We find that $Q_\mathrm{pr}$ attains a maximum of 0.51 at about 1 $\mathrm{\mu m}$. This curve for $Q_\mathrm{pr}$ is practically identical to the result shown in Fig.~7b by \citet{BURNS:1979wg}.

The term $\ddot{\vec r}_\mathrm{PR}$ gives the acceleration due to the Poynting-Robertson drag \citep{BURNS:1979wg}
\begin{linenomath*}
\begin{equation}
\ddot {\vec r}_\mathrm{PR} = -\frac{3Q_\mathrm SQ_\mathrm{pr}\mathrm{AU}^2}{4(\vec r-{\vec r}_\mathrm S)^2\rho_\mathrm gr_\mathrm gc^2}(\dot{\vec r}-\dot{\vec r}_\mathrm S).
\end{equation}
\end{linenomath*}
If a grain is in the planetary shadow, the effect of solar radiation pressure and Poynting-Robertson drag will disappear. Note that we do not include the variation of photoelectric charging due to planetary shadow, treating the grain charge as a constant throughout (see also discussion at the end of this Section). The components of $\vec r_\mathrm{S}$ are denoted by $(S_x, \ S_y, \ S_z)$. When the particle is in the planetary shadow, the following two conditions are satisfied: (1) The grain and the Sun are on different sides of Jupiter, i.e., $\vec r \cdot \vec r_\mathrm{S} < 0 $, and (2) the distance between grain and the shadow center line, $d_\mathrm{act}$ must be less than the distance between the boundary of the shadow and the shadow center line $d_\mathrm{crit}$, i.e., $ d_\mathrm{act} < d_\mathrm{crit} $. After some algebra, the planetary shadow conditions can be written as
\begin{linenomath*}
\begin{equation}
xS_x+yS_y+zS_z<0, \ \ \ r^2-\frac{(xS_x+yS_y+zS_z)^2}{|\vec r_\mathrm{S}|^2}<R_\mathrm{J}^2.
\end{equation}
\end{linenomath*}
The acceleration due to plasma drag is denoted by $\ddot{\vec r}_\mathrm{PD}$. Far from the synchronous radius in the Jovian system, the direct drag due to heavy ions dominates. If the relative motion is subsonic or comparable to the thermal speed, the Coulomb drag may become important \citep{morfill1980dust, northrop1989gyrophase}. For our supersonic case, i.e., when the relative velocity of the grain is generally much larger than the local sound speed in the plasma, it is sufficient to consider the supersonic approximation of the direct drag due to heavy ions \citep{morfill1980dust, dikarev1999dynamics}
\begin{linenomath*}
\begin{equation} \label{equ_plasma_drag}
%\ddot{\vec r}_\mathrm{PD} = -\sum_{\mathrm H=1}^{N_\mathrm H}\frac{3n_\mathrm Hm_\mathrm H|\vec v_\mathrm{rel}|}{4\rho_\mathrm gr_\mathrm g}\vec v_\mathrm{rel}.
\ddot{\vec r}_{\text{PD}} = -\sum_{\text{H=1}}^{N_{\mathrm{H}}}\frac{3n_{\mathrm{H}}m_{\mathrm{H}}|\vec v_{\text{rel}}|}{4\rho_{\mathrm{g}}r_{\mathrm{g}}}\vec v_{\text{rel}}.
\end{equation}
\end{linenomath*}
Here $N_\mathrm H$ is the number of different species of heavy ions (including $\mathrm{O}^+$, $\mathrm{O}^{++}$, $\mathrm{S}^+$, $\mathrm{S}^{++}$, $\mathrm{S}^{+++}$ and $\mathrm{Na}^+$) in the plasma, $n_\mathrm H$ is the respective heavy ion number density, and $m_\mathrm H$ the ion mass. $\vec v_\mathrm{rel}$ is the velocity of the dust relative to the plasma. The values of $n_\mathrm H$ of different species are taken from the DG83 model \citep{divine1983charged}. Note that the plasma distributions of this model are defined with respect to the left-handed frame Jupiter System III (1965) \citep{seidelmann1977evaluation}.

The term $\ddot{\vec r}_{\mathrm{G_{other \ bodies}}}$ describes the acceleration due to gravity perturbations from other bodies. In the inertial frame, the acceleration due to the gravitational perturbations from the Sun and the four Galilean moons reads \citep{murray1999solar}
\begin{linenomath*}
\begin{equation} \label{equ_third_body}
\ddot{\vec r}_{G_\mathrm{other \ bodies}} = Gm_\odot\left(\frac{\vec r_\mathrm{dS}}{r_\mathrm{dS}^3}-\frac{\vec r_\mathrm S}{r_\mathrm S^3}\right) + \sum_{\mathrm i=1}^{4}Gm_{\mathrm m_i}\left(\frac{\vec r_{\mathrm {dm}_i}}{r_{\mathrm {dm}_i}^3}-\frac{\vec r_{\mathrm m_i}}{r_{\mathrm m_i}^3}\right)
\end{equation}
\end{linenomath*}
where $m_\odot$ is the mass of the Sun, $\mathrm m_i$ the mass of the $i$th Galilean moon, $\vec r_\mathrm{dS} = - \vec r_\mathrm{Sd}$, and $\vec r_{\mathrm {dm}_i}$ is the vector from the particle to the respective moon, and $\vec r_{\mathrm m_i}$ is the vector from Jupiter to the moon.

The physical parameters and ephemerides for Jupiter, the Sun and four Galilean moons are obtained from high accuracy data provided by the NAIF SPICE toolkit, where the kernels ``naif0011.tls", ``pck00010.tpc", ``gm\_de431.tpc" and ``jup310.bsp" are used. The satellite ephemeris kernel file ``jup310.bsp" is significantly improved over previous versions \citep{jup310}. Beyond the time coverage of the SPICE kernels, the orbits of celestial bodies are assumed to be Keplerian. We use a Gragg-Bulirsch-Stoer integrator with adaptive stepsize \citep{wanner1993solving, bulirsch1980introduction} for the numerical computation of dust trajectories, which provides a good compromise between speed and accuracy. For the convenience of the reader, physical and orbital properties of the Galilean moons and Jupiter are shown in Table \ref{tab:galilean_moons}.

Finally, we like to discuss the limitations of our model. A constant equilibrium surface potential of +5 V is adopted for simplicity and to be consistent with previous work on the dynamics of particles in the region of the Galilean moons \citep{krivov2002tenuous, zeehandelaar2007local}. With this assumption the effect of variations of grain charge induced by the traversals of the planetary shadow is neglected. The significance of this effect on the grain dynamics in the gossamer ring was discussed by \citet{hamilton2008sculpting}, but will be less important at the larger radial distances considered in this paper, because the time spent by a particle in the shadow, if any, is then much smaller. Typical charging times for tiny Jovian stream particles were calculated by \citet{graps2001io} (see also \citet{kruger2004jovian}). For the stream particles the assumption of a fixed charge can lead to the underestimation of the grain speed and the sensitivity to the expansion order of the magnetic field. In our simulations with larger grain sizes and thus shorter charging time, the difference between results for a fixed charge and a spatially variable charge established as an equilibrium of the relevant currents will be much smaller. In our model, the magnetic field and the ambient plasma are assumed to co-rotate rigidly with Jupiter. The corotation breaks down in the vicinity of Io due to mass loading \citep{thomas2004io}, which can affect the dynamics of grains in the Io region. However, the assumption of rigid co-rotation should still be a valid approximation for the investigation of the global dynamics of particles in the region of the Galilean moons. Further, the effect of sputtering on the grain size is neglected. To date the sputtering lifetime is not very well constrained from previous studies. Nevertheless, rough numbers for sputtering lifetime are given in the literature. According to Table II in \citet{burns2001dusty}, a one micron particle at 1.8 $R_\mathrm{J}$ has a sputtering lifetime of $10^{3\pm1}$ years. This estimate is larger than the dynamical (sink) lifetime (less than 100 years) in our simulations (see Section \ref{section_lifetime}). Therefore, we believe that the neglect of sputtering in our study gives us meaningful estimates for the dust configuration. These limitations will be investigated in future in more detail.

\section{Numerical Simulation Scheme} \label{section_numerical}
All the dust particles we use in the analysis are assumed to be spherical icy grains of density $\rho = 1\mathrm{g/cm^3}$, consistent with previous modelling approaches \citep{1997GeoRL..24.2175H, horanyi1998dust,kruger2003impact, bottke2013black}. Overall, the numerical procedure is divided into three steps. First we determine the ejecta distribution at the Hill sphere of each source moon. To this aim, we start a large number of particles from a moon's surface and integrate their trajectories until they either re-impact their parent moon or leave its gravitational sphere of influence (Hill sphere). In the latter case we store their phase space coordinates from a large number of such integrations, and calculate an intermediate ejecta distribution at the Hill sphere (see Section \ref{sec:step_one} for details). Second we start a reduced number of particles with uniformly distributed starting conditions from the Hill sphere and continue integration over 8,000 Earth years until the particle hits a sink (Section \ref{sec:step_two}). In the last step we weight the trajectories from the long-term integrations with the ejecta distribution at the Hill sphere obtained by the simulations inside the Hill Sphere of the source moon.

The main advantage of this procedure is the possibility to apply changes to the starting conditions from the surface of the source moon without the necessity to recalculate the full trajectories. Only the distributions at the Hill sphere need to be re-calculated. The second step, which is most CPU expensive (in the order of 100,000 CPU hours), needs only be done once, because the individual trajectories from this step can be reused for different surface starting conditions. The dynamics in the vicinity of the source moon can be calculated to a good approximation using a reduced set of forces. We simulate a much larger number of particles (more than 100 million per moon) in the first step with lower computational cost, in order to get ejecta distributions at the source moon's Hill sphere with good statistical significance; and use a smaller number of particles (26,754 particles per moon) in the second step reducing the computational effort significantly.

\subsection{Evaluation of the Distribution of Ejecta at the Hill Sphere of a Source Moon} \label{sec:step_one}
The primary process of dust production on the Galilean moons is impact-ejection, the generation of dust particles due to the bombardment by fast micrometeoroids. The dominant family of projectiles are hypervelocity interplanetary dust particles \citep{kruger1999detection,krueger2000,kruger2003impact, sremvcevic2005impact}. Of the four Galilean satellites, Io carries a tenuous atmosphere of dominantly $\mathrm{SO_2}$, which is likely patchy and partly driven by volcanism \citep{mcgrath2004satellite}. The column densities of $\mathrm{SO_2}$ found from different measurements are around $10^{16} \ \mathrm{cm^{-2}}$. Interplanetary particles at Jupiter can hit Io with typical velocities around 26 km/s, boosted by gravitational focusing by the planet \citep{sremvcevic2003impact}, so that the thin atmosphere might be expected to have some effect on the impact-ejecta process. Employing an appropriate formula for atmospheric drag \citep{bertotti2003physics} we can estimate the acceleration of a grain of given size. Dynamically, for micron sized grains the acceleration due to atmospheric drag is comparable to the gravitational acceleration of Io. But atmospheric drag acceleration becomes negligible relative to Io gravity for particles larger than about 10 microns. In either case, both accelerations induce only a minor change in velocity of the projectiles, compared to the 26 km/s, before they impact the surface. Assuming an atmospheric height of 300 km (which corresponds to the height of the Pele plume) we estimate that the ram pressure on the projectile in this atmosphere lies several orders of magnitude below the tensile strength of ice (and much more below the one for silicates), so that break up can be safely neglected. Thermally, the atmospheric drag may lead to melting of micron sized projectiles formed of silicates but not to their evaporation. Grains larger than a few microns will not melt. The ejecta that escape Io have starting velocities around two to three km/s. For these velocities the influence of atmospheric drag is entirely negligible. In conclusion we neglect the effect of Io's atmosphere for our modelling. For Europa and Ganymede $\mathrm{O_2}$ atmospheres are reported of column densities around $10^{15} \ \mathrm{cm^{-2}}$ \citep{mcgrath2004satellite}. This implies densities that are by a factor of 20 lower than the atmospheric density of Io, so that also for these moons the neglect of the effect of atmospheric drag is justified.

\begin{figure}
 \centering
 \noindent\includegraphics[width=0.5\textwidth, angle=270]{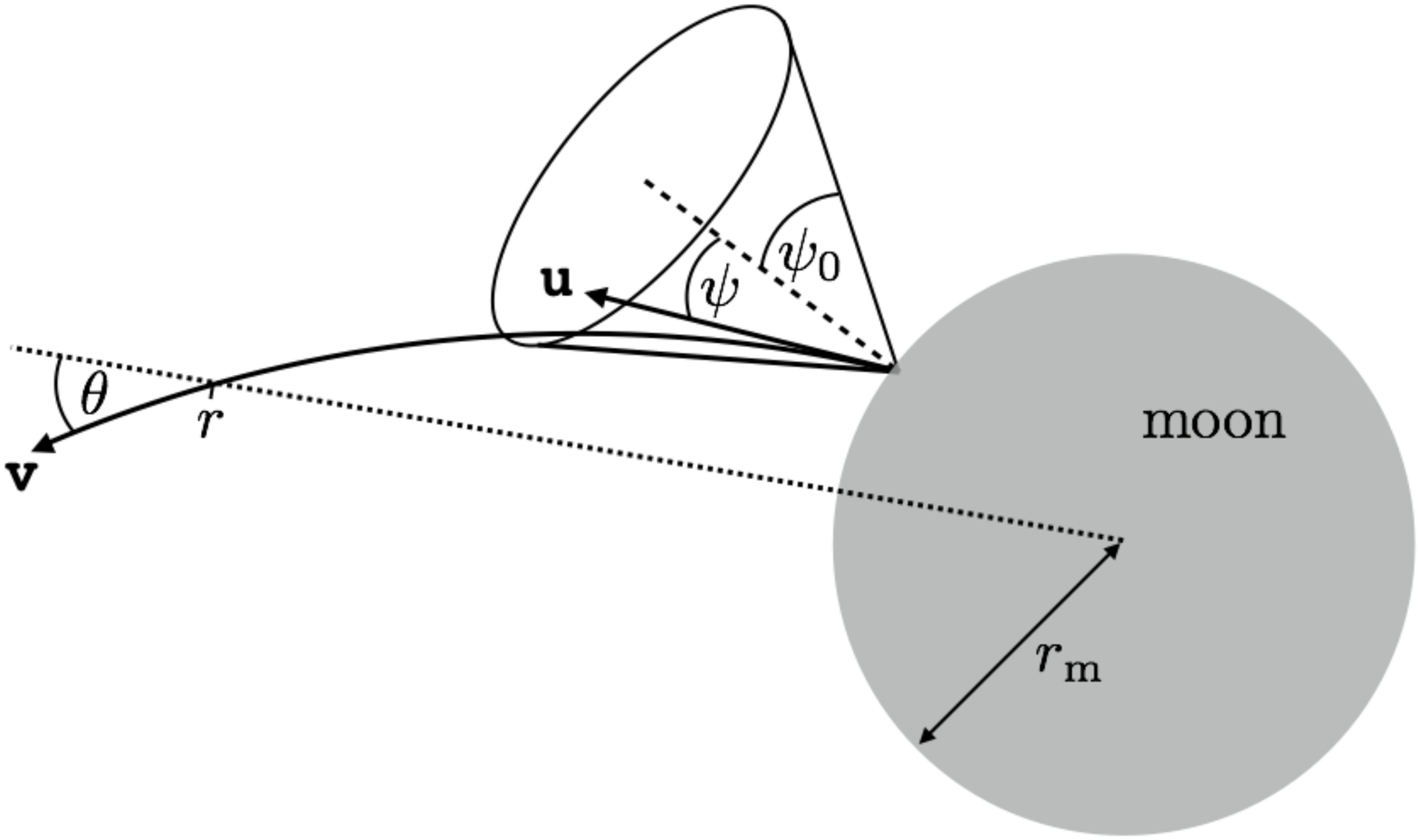}
 \caption{Sketch of the impact-ejecta process. A fast projectile hits a moon and releases material from its surface within a cone of half-opening angle $\psi_0\in\left[0^\circ,90^\circ\right]$. Individual particles are ejected at a distance $r_\mathrm{m}$ from the centre of the moon with velocity $\vec{u}$ and direction $\psi$ (distribution $f_\psi\left(\psi\right)$) measured from the normal to the surface at the ejection point. Some ejecta can escape the moon's gravity and contribute to the planetary dust environment, while others fall back to the surface.}
 \label{fig:sketch}
\end{figure}

For the initial conditions we assume a simple model, where fragment size and speed at the moment of ejection are uncorrelated. In this case all ejecta are launched with the same speed distribution regardless of their size. The variables used in the following are depicted in Fig.~\ref{fig:sketch}. Impact experiments and scaling laws \citep{housen2011} show that the differential speed distribution can be approximated by a power law with exponent $\gamma$
\begin{linenomath*}
\begin{equation}
 f\left(u\right)=\frac{\gamma-1}{u_0}\left(\frac{u}{u_0}\right)^{-\gamma}\Theta_{\mathrm{H}}\left(u-u_0\right) \label{eq:speed_dist}
\end{equation}
\end{linenomath*}
where $\Theta_{\mathrm{H}}\left(x\right)$ is the Heaviside unit step function, which is unity for $x\geq0$ and zero otherwise. The index $\gamma$ depends on properties of the target material ranging from $\gamma=2$ for (highly) porous to $\gamma=3$ for nonporous materials \citep{krivov2003impact}. Considering that the surfaces of the Galilean moons (except for Io) primarily consist of water ice, we use a value of $\gamma=3$. The minimal ejection speed $u_0$ is chosen so that the kinetic energy of the ejecta is a few tens of percent of the impact energy depending on impact velocity and surface properties of the target \citep{1985JGR....9012445A,hartmann1985}. Hard surfaces (e.g.~ice) are generally less dissipative than soft surfaces (e.g.~snow, regolith).

It is further assumed that the fragments are ejected uniformly distributed (in terms of solid angle) in a cone of half-opening angle $\psi_0\in\left[0^\circ,90^\circ\right]$
\begin{linenomath*}
\begin{equation}
 f_\psi\left(\psi\right)=\frac{\sin\left(\psi\right)}{1-\cos\left(\psi_0\right)}\Theta_{\mathrm{H}}\left(\psi_0-\psi\right). \label{eq:angle_dist}
\end{equation}
\end{linenomath*}
The particles are started one by one from ejecta cones uniformly distributed across the surface of the source moon. The ejection speeds and angles of individual particles are randomly chosen from the distributions given by Eqs.~(\ref{eq:speed_dist}) and (\ref{eq:angle_dist}).

The integration domain is the Hill sphere of the source moon given by
\begin{linenomath*}
\begin{equation}
 r_\mathrm{h}=a_\mathrm{m}\left[\frac{m_\mathrm{m}}{3\left(M_\mathrm{J}+m_\mathrm{m}\right)}\right]^\frac13. \label{eq:hill_radius}
\end{equation}
\end{linenomath*}
where $a_\mathrm{m}$ and $m_\mathrm{m}$ are the semi-major axis and the mass of the source moon, respectively. The dynamics of dust particles in the vicinity of their parent body is primarily governed by the satellite's gravity. Compared to the satellite's gravity, other forces only constitute small perturbations and can be neglected within the moon's Hill sphere \citep{krivov2003impact, sremvcevic2003impact}. 

\begin{figure}
\centering
 \noindent\includegraphics[width=1.\textwidth]{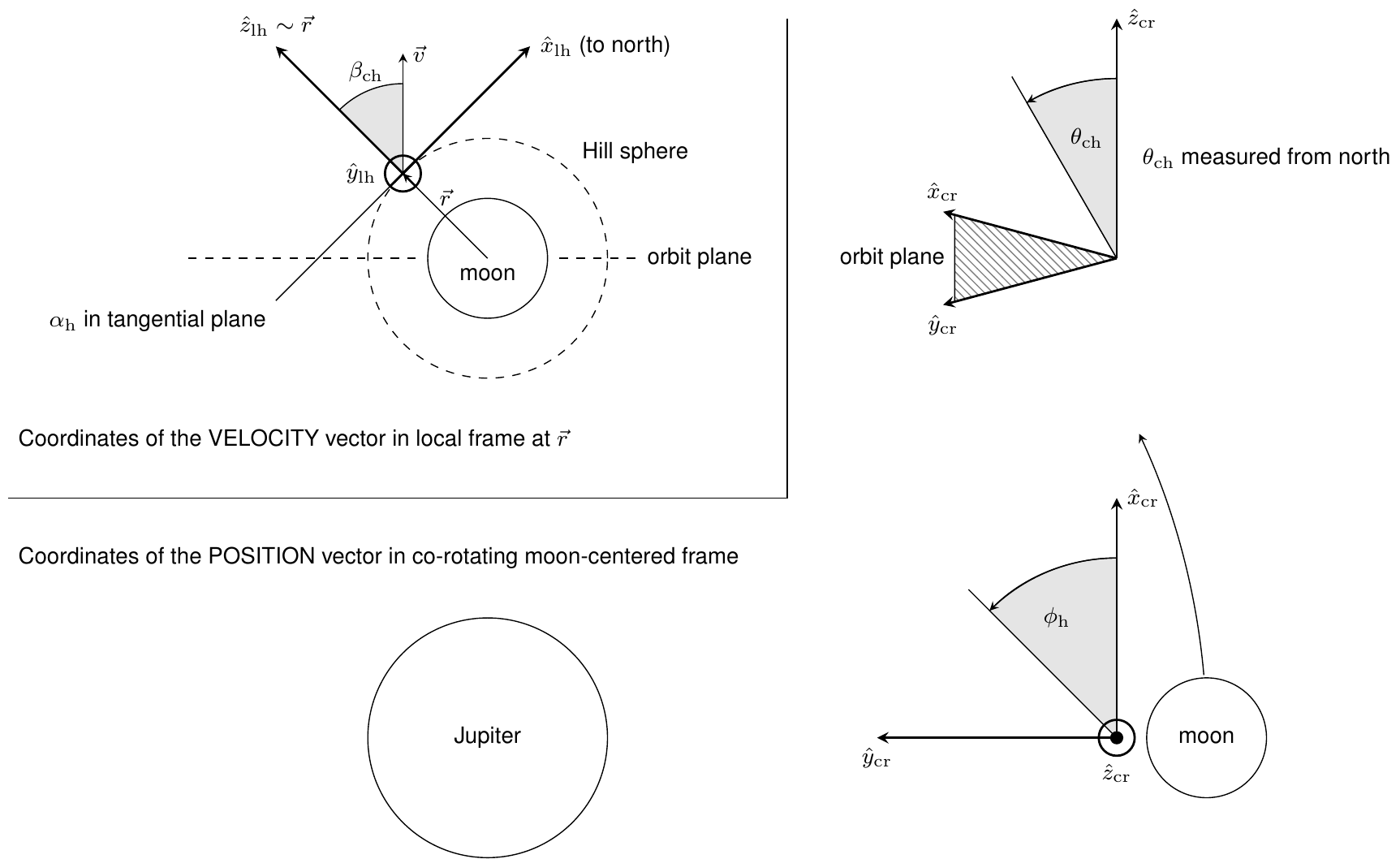}
 \caption{Definition of the coordinate systems used for the ejecta positions and velocities at the moon's Hill sphere. The position of the particle is stored in co-rotating moon-centered frame MCRF ($Ox_\mathrm{cr}y_\mathrm{cr}z_\mathrm{cr}$). The $z_\mathrm{cr}$-axis is defined as the normal to the moon's orbital plane around Jupiter. The $y_\mathrm{cr}$-axis always points from the moon to Jupiter, and the $x_\mathrm{cr}$-axis completes the orthogonal right-handed frame. For the velocity we use a local frame MLHF ($Ox_\mathrm{lh}y_\mathrm{lh}z_\mathrm{lh}$) attached to the position of the particle at the Hill sphere. The $z_\mathrm{lh}$-axis points from the moon's center to the particle. The $y_\mathrm{lh}$-axis is perpendicular to both the $z_\mathrm{lh}$-axis and the $z_\mathrm{cr}$-axis, and points to the local western direction from the given point on the Hill sphere. The $x_\mathrm{lh}$ axis, pointing to the local north, completes the orthogonal right-handed frame.}
 \label{fig:ejecta_angles}
\end{figure}

Such simulations are performed for all four Galilean moons. At the moment when a particle crosses the Hill sphere its phase coordinates are stored. For the position we use a \underline{m}oon-centered \underline{c}o-\underline{r}otating \underline{f}rame (MCRF) $Ox_\mathrm{cr}y_\mathrm{cr}z_\mathrm{cr}$ (Fig.~\ref{fig:ejecta_angles}). The $z_\mathrm{cr}$-axis is defined as the normal to the orbital plane of the moon around Jupiter. The $y_\mathrm{cr}$-axis always points from the moon to Jupiter, and the $x_\mathrm{cr}$-axis completes the orthogonal right-handed reference frame.

For the velocity we define a local frame (MLHF) $Ox_\mathrm{lh}y_\mathrm{lh}z_\mathrm{lh}$, attached to the point where the particle crosses the moon's Hill sphere (Fig.~\ref{fig:ejecta_angles}). The $z_\mathrm{lh}$-axis points from the satellite center to the particle, and the $y_\mathrm{lh}$-axis is perpendicular to both the $z_\mathrm{lh}$-axis and the $z_\mathrm{cr}$-axis, i.e.
\begin{linenomath*}
\begin{equation}
\hat{\vec y}_\mathrm{lh} = \frac{\hat{\vec z}_\mathrm{lh}\times\hat{\vec z}_\mathrm{cr}}{|\hat{\vec z}_\mathrm{lh}\times\hat{\vec z}_\mathrm{cr}|}.
\end{equation}
\end{linenomath*}
The so defined $\hat{\vec y}_\mathrm{lh}$ points to the local western direction from the given point on the Hill sphere. The $x_\mathrm{lh}$ axis, pointing to the local north, completes the orthogonal right-handed reference frame. The standard basis of the MLHF frame, $\hat{\vec x}_\mathrm{lh}$, $\hat{\vec y}_\mathrm{lh}$, and $\hat{\vec z}_\mathrm{lh}$, can be expressed in terms of the coordinates in the MCRF frame as
\begin{linenomath*}
\begin{equation}
\hat{\vec x}_\mathrm{lh} = \left[-\frac{x_\mathrm{cr}z_\mathrm{cr}}{r_\mathrm{h}r_{\mathrm{h},xy}}, \ -\frac{y_\mathrm{cr}z_\mathrm{cr}}{r_\mathrm{h}r_{\mathrm{h},xy}}, \ \frac{r_{\mathrm{h},xy}}{r_\mathrm{h}}\right]^\mathrm{T}
\end{equation}
\end{linenomath*}
\begin{linenomath*}
\begin{equation}
\hat{\vec y}_\mathrm{lh} = \left[\frac{y_\mathrm{cr}}{r_{\mathrm{h},xy}}, \ -\frac{x_\mathrm{cr}}{r_{\mathrm{h},xy}}, \ 0\right]^\mathrm{T}
\end{equation}
\end{linenomath*}
\begin{linenomath*}
\begin{equation}
\hat{\vec z}_\mathrm{lh} = \left[\frac{x_\mathrm{cr}}{r_\mathrm{h}}, \ \frac{y_\mathrm{cr}}{r_\mathrm{h}}, \ \frac{z_\mathrm{cr}}{r_\mathrm{h}}\right]^\mathrm{T}
\end{equation}
\end{linenomath*}
where $r_{\mathrm{h},xy} = \sqrt{x_\mathrm{cr}^2+y_\mathrm{cr}^2}$ is the projection of the grain's position in the $x_\mathrm{cr}$-$y_\mathrm{cr}$-plane. 

The six Cartesian phase space coordinates on the Hill sphere are transformed into spherical coordinates, which are a more appropriate representation of the ejecta distributions. For a fixed distance ($r_\mathrm{h}$) from the center of the source moon the remaining five coordinates are: position colatitude (polar angle) $\theta_\mathrm{ch}$, position longitude (azimuthal angle) $\phi_\mathrm{h}$, speed relative to the source moon $v_\mathrm{h}$, velocity colatitude (polar angle) $\beta_\mathrm{ch}$, and velocity longitude (azimuthal angle) $\alpha_\mathrm{h}$. The definitions of the ejecta angles are depicted in Fig.~\ref{fig:ejecta_angles}. Note the angle $\alpha_\mathrm{h}$ is in the tangential plane, and is measured from local north in anti-clockwise direction.

\begin{figure}
 \centering
 \noindent\includegraphics[width=1.\textwidth]{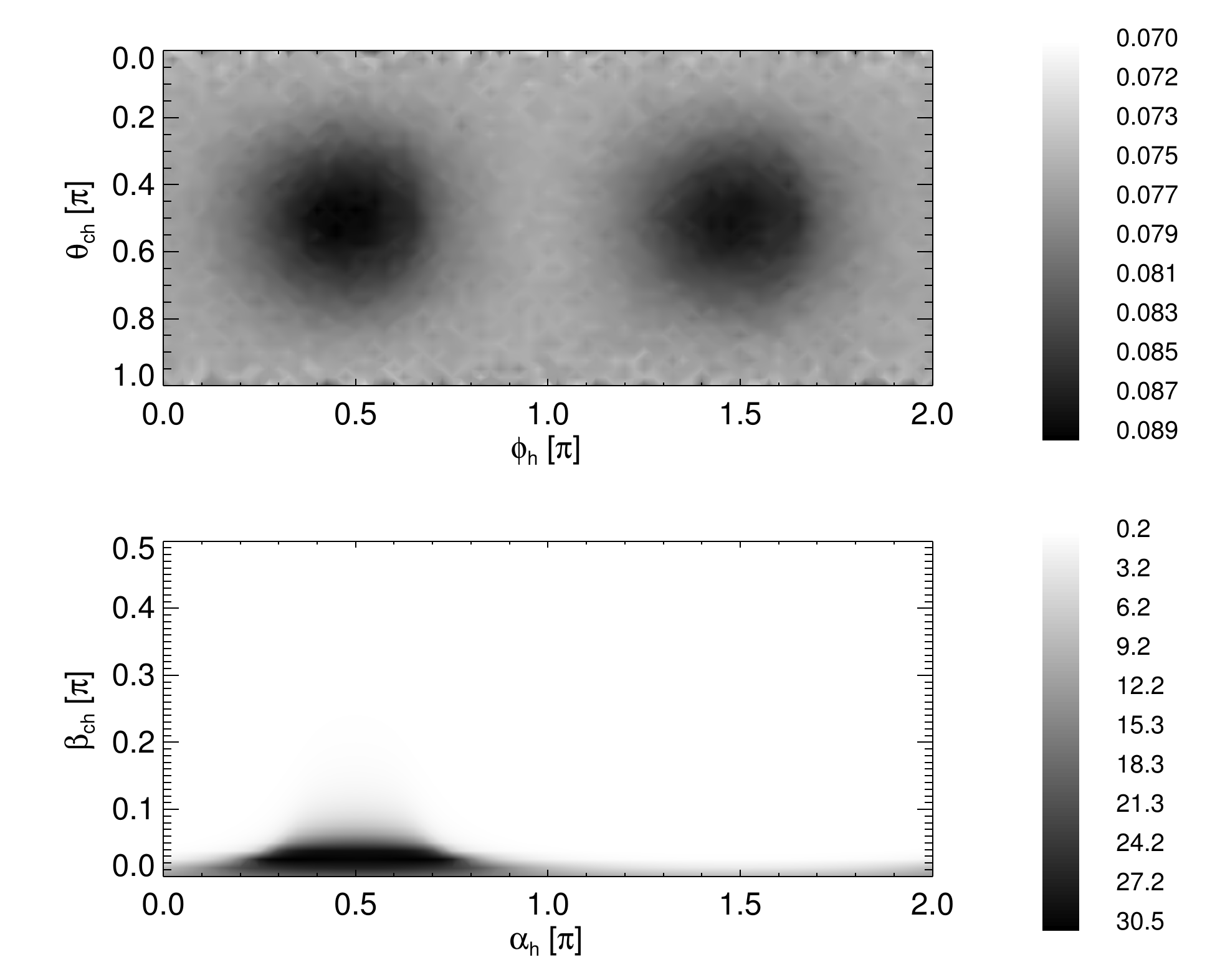}
 \caption{Distribution of ejecta at the Hill sphere of Ganymede. Top panel: Angular distribution of the position vector, showing a larger probability to escape at the Lagrangian points $\mathcal{L}_1$ (at $\phi_\mathrm{h}=\pi/2$) and $\mathcal{L}_2$ (at $\phi_\mathrm{h}=3\pi/2$). Bottom panel: Distribution of angles of the velocity vector. The distribution of $\alpha_\mathrm{h}$ is peaked around $\pi/2$, i.e.~particles are ejected preferentially (at $\mathcal{L}_1$ and $\mathcal{L}_2$) in western direction, which is the local direction of Keplerian motion in the field of the planet. The deviation of the velocity polar angle $\beta_\mathrm{ch}$ from the normal to the Hill sphere, however, is small throughout and $\beta_\mathrm{ch}$ is sharply peaked about this small value. The normalization is to unit solid angle, e.g. $\int\mathrm{d}\phi_\mathrm{h}\int\mathrm{d}\theta_\mathrm{ch}\sin\theta_\mathrm{ch}\,p_1\left(\phi_\mathrm{h},\theta_\mathrm{ch}\right)$ for the angle of the position vector.}
 \label{fig:ejecta_ganymede_1}
\end{figure}

\begin{figure}
\centering
\noindent\includegraphics[width=0.65\textwidth, angle=90]{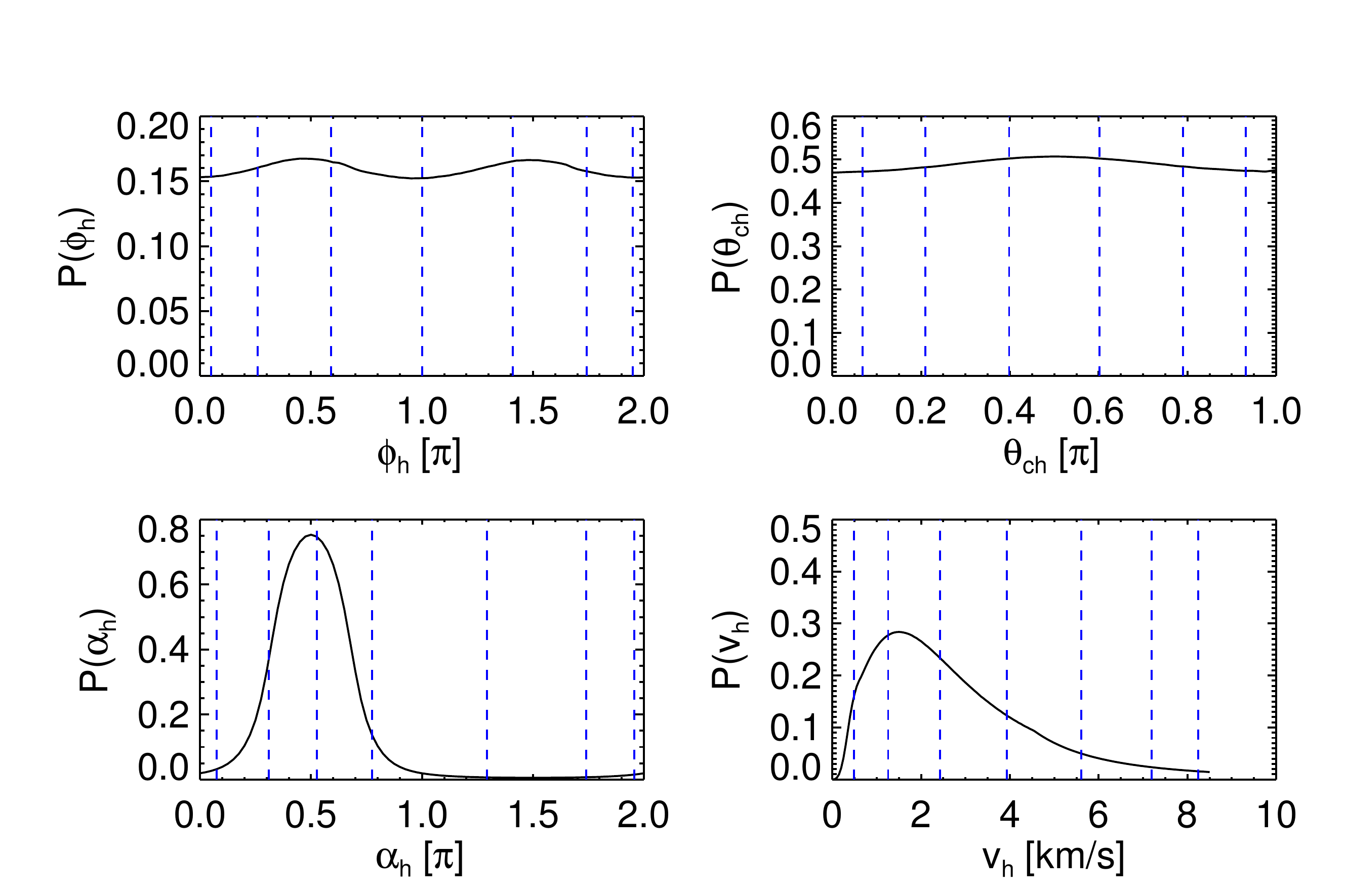}
\caption{\label{fig_ganymede_weight} Relative distributions of starting conditions at the Hill sphere of Ganymede.}
\end{figure}

We have integrated the trajectories of more than $100$ million particles per satellite. Note that the initial conditions and gravitational forces are grain size-independent. Even in this case the number of particles is too small to properly resolve a five-variable density function with not too coarse binning. Instead we assume that the angular coordinates of the position and velocity as well as the speed separate. Then the ejecta distribution can be approximated by
\begin{linenomath*}
\begin{equation}
p\left(\phi_\mathrm{h},\theta_\mathrm{ch},\alpha_\mathrm{h},\beta_\mathrm{ch},v_\mathrm{h}\right)\simeq p_1\left(\phi_\mathrm{h},\theta_\mathrm{ch}\right)p_2\left(\alpha_\mathrm{h},\beta_\mathrm{ch}\right)p_3\left(v_\mathrm{h}\right)
\end{equation}
\end{linenomath*}
where the reduced distributions $p_1$, $p_2$, and $p_3$ have been obtained by integration of the full distribution $p$ over the remaining coordinates. The reduced distributions will be used to calculate the weights for the long-term integrations. An example is shown in Fig.~\ref{fig:ejecta_ganymede_1} for the moon Ganymede. The distributions of locations on the Hill sphere show an increased density at the Lagrangian points $\mathcal{L}_1$ and $\mathcal{L}_2$, meaning that particles are more likely to escape at these points than at any other point. In the end, we need to weight the results from the long-term integrations with the distribution at the Hill sphere. For simplicity, we assume that the distribution factorizes completely as
\begin{linenomath*}
\begin{equation} \label{equ_fac_distri}
p\left(\phi_\mathrm{h},\theta_\mathrm{ch},\alpha_\mathrm{h},\beta_\mathrm{ch},v_\mathrm{h}\right)  \simeq   p_\phi\left(\phi_\mathrm{h}\right) p_\theta\left(\theta_\mathrm{ch}\right) p_\alpha\left(\alpha_\mathrm{h}\right) \delta\left(\beta_\mathrm{ch}\right) p_v\left(v_\mathrm{h}\right)
\end{equation}
\end{linenomath*}
where we identify again the individual distributions with the proper projections of the original, correlated distribution. The choice of a Dirac delta function for the distribution of the cone angle $\beta_{ch}$ (Fig.~\ref{fig:ejecta_ganymede_1}) is justified, because this quantity varies only in a very narrow range. We will later perform the numerical integration over these distributions in terms of Gaussian quadrature \citep[e.g.][]{press1992numerical}. The reduced distributions for Ganymede are shown in Fig.~\ref{fig_ganymede_weight} as an example. The abscissas for Gauss quadrature determined for these distributions are shown as dashed vertical lines and we store the weights for the quadrature formula for the integration at a later step. We will use these abscissa values of angles and speed as the starting conditions for the long-term integrations performed in the second step of the modelling. The orders of the quadrature formula (equaling the number of abscissa values) are equal to the number of the dashed lines in the plot. For parameters $\theta_\mathrm{ch}$, $\phi_\mathrm{h}$, $\alpha_\mathrm{h}$ and $v_\mathrm{h}$, the orders of the related polynomials, that are chosen for the Gaussian quadratures, are 6, 7, 7 and 7, respectively. Thus, the number of integrated dust particles for each grain size from each source moon is 2,058 to cover all possible combinations of the discrete values of $\theta_\mathrm{ch}$, $\phi_\mathrm{h}$, $\alpha_\mathrm{h}$, $\beta_\mathrm{ch}$ and $v_\mathrm{h}$.

\subsection{Long-term Integration of Individual Trajectories of Particles Starting From the Hill Sphere} \label{sec:step_two}
The second step is the integration of individual trajectories from a source moon's Hill sphere until the grain hits a sink. Dust grains from sub-microns to centimeters are included. Specifically, we carry out integrations for the 13 different sizes 0.05 $\mathrm{\mu m}$, 0.1 $\mathrm{\mu m}$, 0.3 $\mathrm{\mu m}$, 0.6 $\mathrm{\mu m}$, 1 $\mathrm{\mu m}$, 2 $\mathrm{\mu m}$, 5 $\mathrm{\mu m}$, 10 $\mathrm{\mu m}$, 30 $\mathrm{\mu m}$, 100 $\mathrm{\mu m}$, 300 $\mathrm{\mu m}$, 1 $\mathrm{mm}$, and 1 $\mathrm{cm}$. For each grain size 2,058 particles are simulated (Section \ref{sec:step_one}). Therefore, in total, from each moon 26,754 particles of 13 different sizes are simulated for a maximum of 8,000 (Earth) years until each grain hits a sink. 

The starting positions and velocities in the MCRF frame at the Hill sphere of the source moon can be expressed in terms of the Gaussian abscissas of $\theta_\mathrm{ch}$, $\phi_\mathrm{h}$, $\alpha_\mathrm{h}$, $\beta_\mathrm{ch}$ and $v_\mathrm{h}$
\begin{linenomath*}
\begin{equation}
x_\mathrm{cr} = r_\mathrm{h}\sin\theta_\mathrm{ch}\cos\phi_\mathrm{h}
\end{equation}
\end{linenomath*}
\begin{linenomath*}
\begin{equation}
y_\mathrm{cr} = r_\mathrm{h}\sin\theta_\mathrm{ch}\sin\phi_\mathrm{h}
\end{equation}
\end{linenomath*}
\begin{linenomath*}
\begin{equation}
z_\mathrm{cr} = r_\mathrm{h}\cos\theta_\mathrm{ch}.
\end{equation}
\end{linenomath*}
The initial velocity vector of dust expressed in the standard basis of the MLHF frame can be obtained as
\begin{linenomath*}
\begin{equation}
\vec v_\mathrm{cr} = v_\mathrm{h}\sin\beta_\mathrm{ch}\cos\alpha_\mathrm{h}\,\hat{\vec x}_\mathrm{lh} + v_\mathrm{h}\sin\beta_\mathrm{ch}\sin\alpha_\mathrm{h}\,\hat{\vec y}_\mathrm{lh} + v_\mathrm{h}\cos\beta_\mathrm{ch}\,\hat{\vec z}_\mathrm{lh}.
\end{equation}
\end{linenomath*}
Using the fact that all Galilean moons are tidally locked to Jupiter, the grain's initial states ($\vec r_\mathrm{cr}$, $\vec v_\mathrm{cr}$) can be transformed into the Jupiter equatorial inertial frame JIF where the long-term integrations are performed using the dynamical model described in Section \ref{section_model}. The cubic Hermite interpolation is used to detect collisions with Jupiter and its satellites (see details in Section \ref{subsection_sinks}). The dust trajectories are stored in the form of osculating orbital elements (see details in Section \ref{subsection_storage}). As mentioned previously, the computational burden is this step is very large, so the large computer cluster located at the CSC -- IT Center for Science Ltd.~is used for this long-term integration.

\subsection{Detecting Collisions with Planet and Satellites} \label{subsection_sinks}
\begin{figure}
\centering 
\noindent\includegraphics[width=0.32\textwidth]{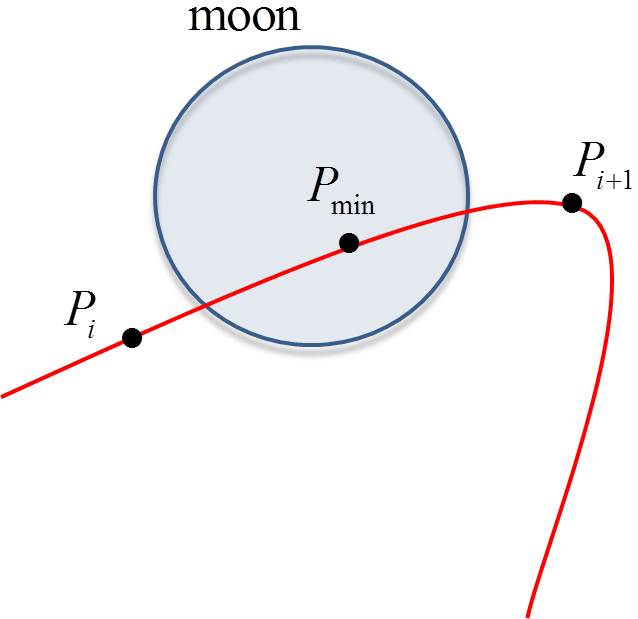}
\caption{\label{fig_sinks} Schematic illustration how cubic interpolation is used to detect a collision of a particle with a moon that takes place between two consecutive time steps of the integrator, $t_i$ and $t_{i+1}$.}
\end{figure}

As sinks for the particles we consider impact on the planet and the Galilean moons, as well as escape from the Jovian system. In the latter case, a simulation is terminated when the distance of the particle to Jupiter is larger than Jupiter's Hill radius $R_\mathrm{H}$ ($\approx$ 743 $R_\mathrm{J}$). The case of collisions with moons and planet is complicated by the fact that numerical integrators always generate positions of dust particles in discrete time steps, so that impacts in between may be easily overlooked. Here, we consider the collision of a dust grain and a moon as an example. We denote the position of the particle in the $i$th time step $t_i$ as $P_i$ and the distance between $P_i$ and the moon as $R_i$. An impact occurred, if at any time the distance to the center of the moon is less than $R_\mathrm{crit}$. For a spherical target moon, $R_\mathrm{crit}$ is identical to the radius of the moon. Sometimes this condition is fulfilled between two consecutive steps $t_i$ and $t_{i+1}$, even if both $R_i>R_\mathrm{crit}$ and $R_{i+1}>R_\mathrm{crit}$ (Fig.~\ref{fig_sinks}). If we only evaluate the condition at the two discrete steps, we will miss this collision.

To solve this problem we use a cubic Hermite interpolation, which was used by \citet{chambers1999hybrid} in N-body simulations. To this end, we assume that the separation between the dust particle and the center of the moon can be expressed as a cubic polynomial
\begin{linenomath*}
\begin{equation}
g(t)=at^3+bt^2+ct+d.
\end{equation}
\end{linenomath*}
Using the boundary conditions
\begin{linenomath*}
\begin{equation}
g(t_i)=R_i,  \ \ \  \dot g(t_i)=\dot R_i
\end{equation}
\end{linenomath*}
\begin{linenomath*}
\begin{equation}
g(t_{i+1})=R_{i+1},  \ \ \  \dot g(t_{i+1})=\dot R_{i+1}
\end{equation}
\end{linenomath*}
the values of the parameters $a, \ b, \ c$ and $d$ can be determined. From this cubic polynomial method, we find an approximation to the minimum separation, and compare it with the critical distance $R_\mathrm{crit}$. If the minimum separation is smaller than the critical distance, we assume that a collision occurred between $t_i$ and $t_{i+1}$.

\subsection{Storage Scheme for Dust Trajectories} \label{subsection_storage}
Practically, it is not feasible to store the output from the integration of the dust trajectories in the form of fast changing phase space coordinates (positions and velocities) because it would require huge storage space which cannot be handled. Alternatively, the slowly changing orbital elements semi-major axis $a$, eccentricity $e$, inclination $i$, argument of pericenter $\omega$, and longitude of ascending node $\Omega$ are stored in dust trajectory library files. Generally, we store one set of such osculating orbital elements per orbit. For the storage at time $t_j$, the actual set of orbital elements stored is obtained by averaging six sets from roughly equal time intervals $T_j/5$ in the same orbital cycle. Here $T_j = 2\pi\sqrt{\frac{a_{j,0}^3}{GM_\mathrm{J}}}$ is an estimate of the orbital period that is determined from the semi-major axis $a_{j,0}$ at the beginning of the cycle. We write $\sigma=$ $(a, \ e, \ i, \ \Omega, \ \omega)$ for the set of orbital elements. The averaged $a$ and $e$ at time $t_j$ can be expressed as
\begin{linenomath*}
\begin{equation}\label{equ_elem_storage1}
\overline{\sigma_j} = \frac{1}{N+1}\sum_{l=0}^{N}\sigma_{j,l}
\end{equation}
\end{linenomath*}
where $N=5$. For angles $i$, $\Omega$ and $\omega$, the averaged $\overline{\sigma_j}$ can be calculated from the following relations
\begin{linenomath*}
\begin{equation}\label{equ_elem_storage2}
\sin\overline{\sigma_j} = \frac{1}{N+1}\sum_{l=0}^{N}\sin\sigma_{j,l}
\end{equation}
\end{linenomath*}
and
\begin{linenomath*}
\begin{equation}\label{equ_elem_storage3}
\cos\overline{\sigma_j} = \frac{1}{N+1}\sum_{l=0}^{N}\cos\sigma_{j,l}.
\end{equation}
\end{linenomath*}
After these calculations, the set ($t_j$, $\overline\sigma_j$, $f_j$) is stored at time $t_j$, where $f_j$ is true anomaly. When the step size of the integrator $h$ becomes very large and $h>T_j$, which usually occurs for hyperbolic and parabolic orbits, we just use directly the output of the integrator. In some cases, the time interval between two stored points $\Delta T$ is much less than $T_j$ (for example, if a grain will hit a sink within the next orbit). The averaged orbital elements can then still be calculated from Eqs.~(\ref{equ_elem_storage1})-(\ref{equ_elem_storage3}), but here $N = \mathrm{round}(5\Delta T/T_j)$, and $N<5$. Binary format is used for all trajectory library files for efficient storage. In this way, the disc space required to store all trajectories of all 13 grain sizes considered for each source moon is on the order of 10 GB.

\section{Dust Distributions in the Jovicentric Equatorial Inertial Frame} \label{section_distri_inertial}
Using the trajectory library files for dust particles, described in Section \ref{subsection_storage}, we can calculate for each moon the (non-normalized) number density of particles (of given size) in the Jovicentric equatorial inertial frame as follows. The data stored for two consecutive sets of orbital elements, ($t_j$, $\overline\sigma_j$, $f_j$) and ($t_{j+1}$, $\overline\sigma_{j+1}$, $f_{j+1}$), are used as an example here.

The true anomalies $f_j$ and $f_{j+1}$ are first transformed into mean anomalies $M_j$ and $M_{j+1}$, respectively. The mean anomaly has the advantage that equidistant steps $\Delta M$ correspond to equidistant steps in time. Using the time step $\Delta t$ (chosen uniquely for all grain sizes and all source moons), the arc between $M_j$ and $M_{j+1}$ can be divided into a series of $m$ sets ($t_{j,k}$, $\overline\sigma_j$, $M_{j,k}$), where $t_{j,k}$ = $t_j+k\Delta t$, $M_{j,k}$ = $M_j+k\Delta M$, $\Delta M$ = $\Delta t\sqrt{GM_\mathrm{J}/a_{j,0}^3}$, and $k$ = 0, 1, $\ldots$, $m-1$, $m$ = round$\left[(t_{j+1}-t_j)/\Delta t\right]$. The value of $\Delta t$ is taken as eightieth of Io's orbital period, i.e.~$\Delta t = \frac{1}{80}T_\mathrm{Io}$, to compromise between accuracy and efficiency. Also note that $t_j$ = $t_{j,0}$ and $M_j$ = $M_{j,0}$. One set corresponds to one particle in space. Each set ($t_{j,k}$, $\overline\sigma_j$, $M_{j,k}$) in the form of orbital elements is then transformed into position and velocity, evaluated at equidistant times along the trajectory.

In order to express dust distributions in space, we define a cylindrical grid $(\rho_\mathrm{cl}$, $\phi_\mathrm{cl}$, $z_\mathrm{cl})$ in the JIF frame, such that $\rho_\mathrm{cl}$ = $\sqrt{x^2+y^2}$, $\phi_\mathrm{cl}$ = $\mathrm{atan2}(y,x)$ and $z_\mathrm{cl}$ = $z$. For each set ($t_{j,k}$, $\overline\sigma_j$, $M_{j,k}$), we determine the index $(i_\mathrm{bin}, j_\mathrm{bin}, k_\mathrm{bin})$ of the grid cell where the point is located. For a trajectory of a particle of given size $r_\mathrm{g}$, started from a certain moon, we obtain the number of occurrences in the grid cell ($i_\mathrm{bin}$, $j_\mathrm{bin}$, $k_\mathrm{bin}$)
\begin{linenomath*}
\begin{equation}
N_\mathrm{moon}(\xi; i_\mathrm{bin}, j_\mathrm{bin}, k_\mathrm{bin}; r_\mathrm{g})
\end{equation}
\end{linenomath*}
where $\xi$ = ($\phi_\mathrm{h}$, $\theta_\mathrm{ch}$, $\alpha_\mathrm{h}$, $\beta_\mathrm{ch}$, $v_\mathrm{h}$) 
labels one of the 2,058 starting conditions from the Hill sphere of the source moon. Integrating over the distribution of starting conditions $P_\mathrm{moon}(\xi)$, obtained in the first step of the modeling (Section \ref{sec:step_one}), we get the particle number density as
\begin{linenomath*}
\begin{equation} \label{equ_moon_density}
\tilde n_\mathrm{moon} (i_\mathrm{bin}, j_\mathrm{bin}, k_\mathrm{bin}; r_\mathrm{g})= \int \frac{N_\mathrm{moon}(\xi; i_\mathrm{bin}, j_\mathrm{bin}, k_\mathrm{bin};r_\mathrm{g})}{V_\mathrm{bin}(i_\mathrm{bin},j_\mathrm{bin},k_\mathrm{bin})} P_\mathrm{moon}(\xi) \mathrm{d}\xi
\end{equation}
\end{linenomath*}
where $V_\mathrm{bin}(i_\mathrm{bin},j_\mathrm{bin},k_\mathrm{bin})$ is the volume of the grid cell $(i_\mathrm{bin}, j_\mathrm{bin}, k_\mathrm{bin})$. In practise, we carry out the integration over $\xi$ numerically in terms of the factorized distribution (Eq.~\ref{equ_fac_distri}) with the Gaussian weights determined from the starting distributions (Section \ref{sec:step_one}), using the previously determined Gaussian abscissa values $\xi$. This step requires computation time of the order of 1,000 CPU hours, also carried out on the large cluster of CSC. 

\begin{figure}
\centering
\noindent\includegraphics[width=1.0\textwidth]{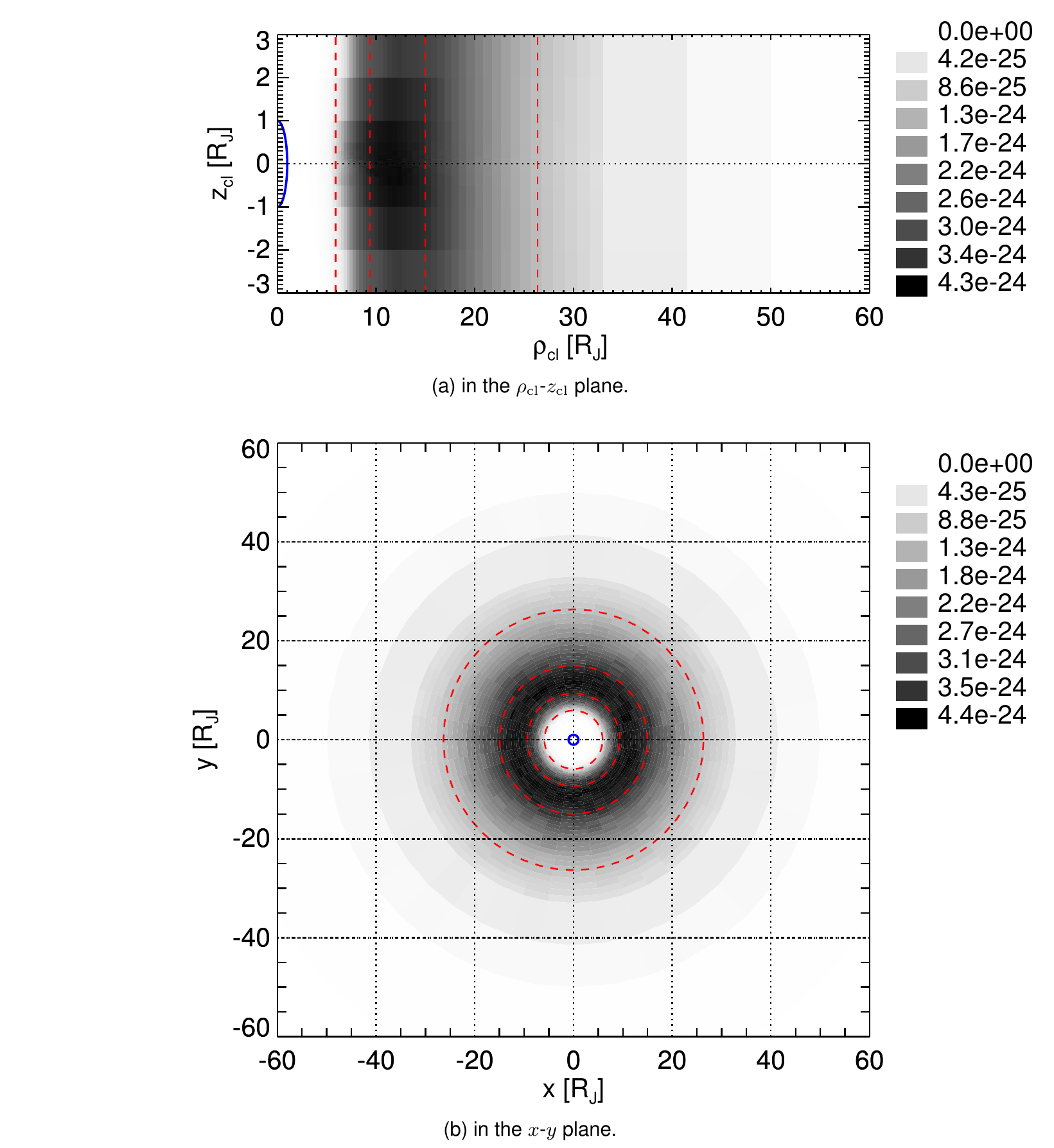}
\vspace{-7 mm}
\caption{Relative number density of 0.3 $\mathrm{\mu m}$ particles from Europa. The blue solid line denotes Jupiter, and the red dashed lines correspond to the orbital distance of Io, Europa, Ganymede and Callisto. The relative number density is in arbitrary units.}
\label{fig_p3_density}
\end{figure}

\begin{figure}
\centering
\noindent\includegraphics[width=1.0\textwidth]{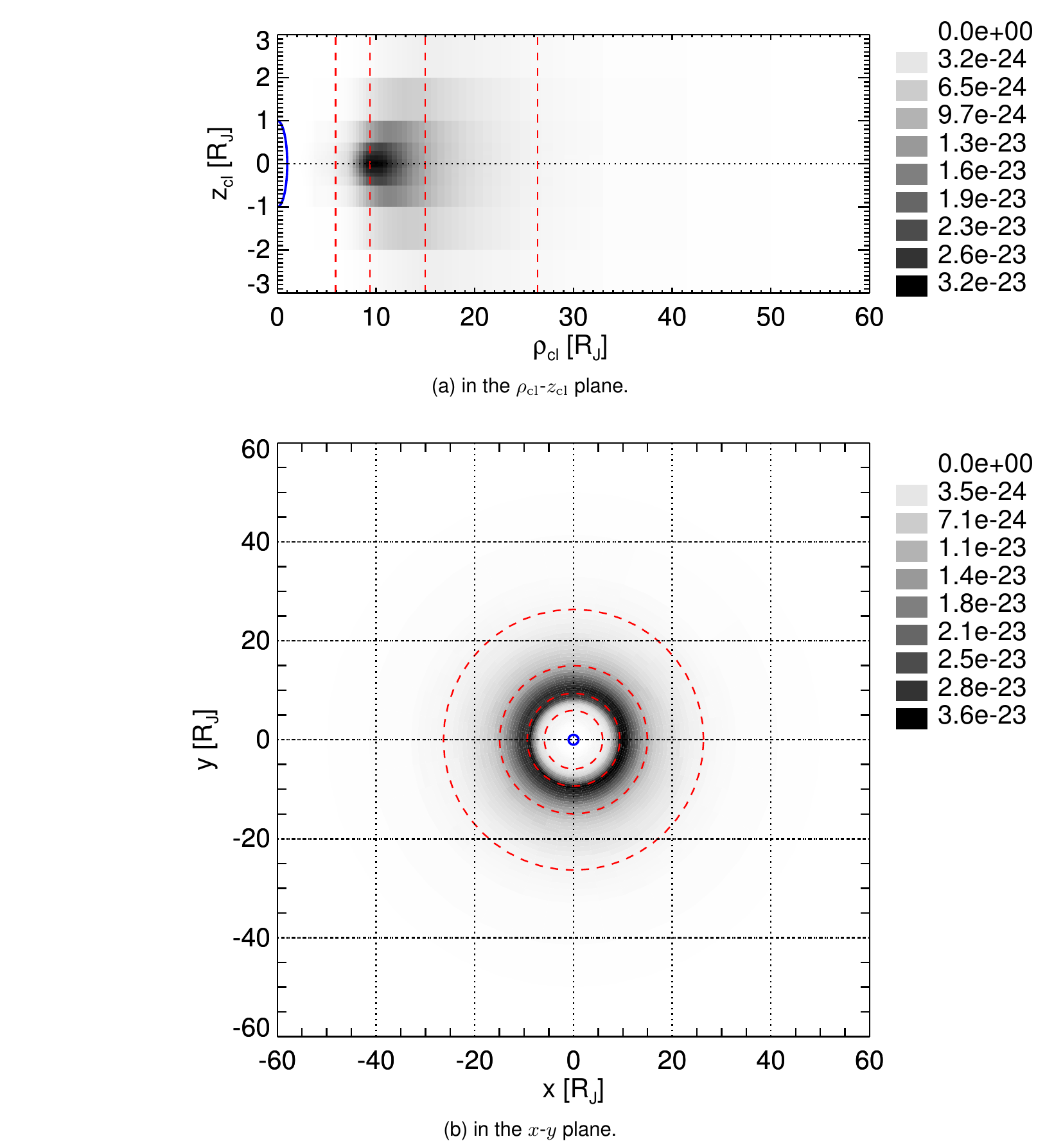}
\vspace{-7 mm}
\caption{Relative number density of 0.6 $\mathrm{\mu m}$ particles from Europa. The blue solid line denotes Jupiter, and the red dashed lines correspond to the orbital distance of Io, Europa, Ganymede and Callisto. The relative number density is in arbitrary units.}
\label{fig_p6_density}
\end{figure}

Examples are shown in Figs.~\ref{fig_p3_density}--\ref{fig_p6_density} for 0.3 $\mathrm{\mu m}$ and 0.6 $\mathrm{\mu m}$ particles from Europa, displaying their (relative) number density in the $\rho_\mathrm{cl}$-$z_\mathrm{cl}$ and $x$-$y$ planes. The number densities are azimuthally averaged (Figs.~\ref{fig_p3_density}(a) and \ref{fig_p6_density}(a)) or vertically averaged (Figs.~\ref{fig_p3_density}(b) and \ref{fig_p6_density}(b)) to make the distribution plot smoother. These figures show that in the inertial frame JIF, the dust distribution is approximately axisymmetric and symmetric with respect to the equatorial plane. It can be seen that the dust dust particles are preferentially transported outwards due to the effects of plasma drag and the outward radial component of the Lorentz force.

\section{The Average Lifetime of Dust Particles as a Function of Size} \label{section_lifetime}
The average lifetime of dust particles can be easily calculated from the results of numerical integrations. Here, the average is taken over the distribution of starting conditions (Eq.~\ref{equ_fac_distri}) on the Hill sphere, for each moon and each grain size separately. The result is shown in Fig.~\ref{fig_lifetime}. The lifetime of 0.05 $\mathrm{\mu m}$ grains (and smaller) can also be estimated analytically (Section \ref{subsection_smallest}). Generally, we find for this size group very short lifetimes (less than 0.1 year as seen from Fig.~\ref{fig_lifetime}). Particles in the size range of one micron exhibit lifetimes on the order of 10 to 100 years. Particles larger than a few microns have typical lifetimes on the order of 100 to 1000 years. The nature of these two sharp jumps in the average lifetime will be addressed (Section \ref{subsection_lifetime_jump}). 

\subsection{Lifetime of the Smallest Dust Particles} \label{subsection_smallest}
Almost all of the smallest particles in the simulations, i.e., 0.05 $\mathrm{\mu m}$, are found to escape from the Jovian system very quickly (Jupiter's Hill radius $R_\mathrm{H}$ defines the distance at which we consider the particle having escaped the system (see Section \ref{subsection_sinks})). For this particle size, the Lorentz force dominates the central gravity. Considering the Lorentz force alone, the change of energy is equal to the work done by the Lorentz force. For the magnetic field, only the dominating dipole component $g_{1,0}$ is considered. Because only the electric field part of Lorentz force $F_\mathrm{e}$ can do work, we have
\begin{linenomath*}
\begin{equation} \label{equ_energy_equ}
\int_{r_{\mathrm{m}_i}}^{R_\mathrm{H}} F_\mathrm{e} \mathrm{d}r = \int_{r_{\mathrm{m}_i}}^{R_\mathrm{H}} \frac{GM_\mathrm{J}L}{r^2} \mathrm{d}r = E_\mathrm{H} - E_{\mathrm{m}_i}
\end{equation}
\end{linenomath*}
giving the energy difference between the radial distance of the source moon at the moment of particle ejection, $r_{\mathrm{m}_i}$, and Jupiter's Hill radius $R_\mathrm{H}$. Here $r_{\mathrm{m}_i}$ is 
taken to be equal to the semi-major axis of the source moon $a_{\mathrm{m}_i}$, and $E_{\mathrm{m}_i}$ and $E_\mathrm{H}$ are the energies of the dust grain at $r_{\mathrm{m}_i}$ and $R_\mathrm{H}$, respectively. The parameter $L$ is the ratio of electric part of the Lorentz force relative to the planet's gravity \citep{hamilton1993motion}
\begin{linenomath*}
\begin{equation}
L = \frac{Qg_{1,0}R_\mathrm{J}^3\Omega_\mathrm{J}}{c m_\mathrm{g}GM_{\mathrm{J}}}
\end{equation}
\end{linenomath*}
where $g_{1,0}$ is the dipole component of the magnetic field. 

With this, Eq.~(\ref{equ_energy_equ}) gives
\begin{linenomath*}
\begin{equation}
-\frac{GM_\mathrm{J}L}{R_\mathrm{H}} + \frac{GM_\mathrm{J}L}{a_{\mathrm{m}_i}} = -\frac{GM_\mathrm{J}}{2a_\mathrm{H}} + \frac{GM_\mathrm{J}}{2a_{\mathrm{m}_i}}
\end{equation}
\end{linenomath*}
where $a_\mathrm{H}$ is the semi-major axis of the particle at the moment when it traverses Jupiter's Hill radius. After some algebra, the semi-major axis of the dust, when it arrives at Jupiter's Hill sphere, reads
\begin{linenomath*}
\begin{equation} \label{equ_a_final}
a_\mathrm{H} = \frac{a_{\mathrm{m}_i}}{1 - 2L + 2L\frac{a_{\mathrm{m}_i}}{R_\mathrm{H}} } \simeq \frac{a_{\mathrm{m}_i}}{1-2L}.
\end{equation}
\end{linenomath*}
The latter relation follows because $a_{\mathrm{m}_i}$ $\ll$ $R_\mathrm{H}$ for all the four Galilean moons. Thus, the value of the semi-major axis of 0.05 $\mathrm{\mu m}$ particles at Jupiter's Hill sphere is nearly proportional to the semi-major axis of their source moon. For $r_\mathrm{g}$ = 0.05 $\mathrm{\mu m}$, we have $L$ = 11.3, so that $a_\mathrm{H}$ $<$ 0, and the orbits of all 0.05 $\mathrm{\mu m}$ dust particles become hyperbolic long before they arrive at Jupiter's Hill sphere. Based on Eq.~(\ref{equ_a_final}), the approximate values of semi-major axis at $R_\mathrm{H}$ for particles from Io, Europa, Ganymede and Callisto can be derived analytically as --0.273 $R_\mathrm{J}$, --0.435 $R_\mathrm{J}$, --0.693 $R_\mathrm{J}$ and --1.22 $R_\mathrm{J}$, respectively, which match closely the values obtained from the full numerical simulations (Fig.~\ref{fig_lifetime_a_p05}).

Because the electric force has the same radial dependence as the (point mass) gravity of the planet, the effect can be expressed in terms of an effective mass \citep{krivov2002dust}
\begin{linenomath*}
\begin{equation} \label{equ_effective_mass}
M'_\mathrm{J} = M_\mathrm{J}(1-L)
\end{equation}
\end{linenomath*}
leading to the conserved effective energy integral $E'$ \citep{krivov2002dust}
\begin{linenomath*}
\begin{equation}
E' = \frac{\dot {\vec r}^2}{2} - \frac{GM'_\mathrm{J}}{r}.
\end{equation}
\end{linenomath*}
Because the effective energy integral at the location of the source moon is equal to the energy at any other time, we have
\begin{linenomath*}
\begin{equation}
\frac{\big|\dot {\vec r}\,\big|^2}{2} - \frac{GM'_\mathrm{J}}{r} = \frac{\big|\dot {\vec r}_{\mathrm{m}_i}\big|^2}{2} - \frac{GM'_\mathrm{J}}{a_{\mathrm{m}_i}}
\end{equation}
\end{linenomath*}
where $\big|\dot {\vec r}_{\mathrm{m}_i}\big|^2$ $\simeq$ $\frac{GM_\mathrm{J}}{a_{\mathrm{m}_i}}$. Note that when $L$ $>$ 1 (corresponding to $r_\mathrm{g}$ $<$ 0.168 $\mathrm{\mu m}$), $M'_\mathrm{J}$ $<$ 0. As just mentioned, $L$ = 11.3 for $r_\mathrm{g}$ = 0.05 $\mathrm{\mu m}$, which means that the effective central gravity force $\frac{GM'_\mathrm{J}}{r^2} \hat{\vec r}$ is strongly radially repelling. Hence, the approximation $\big|\dot {\vec r}\,\big|$ $\simeq$ $\dot r$ holds for most of the time. As a result we obtain an estimate for the particle lifetime as
\begin{linenomath*}
\begin{equation}
\int_{a_{\mathrm{m}_i}}^{R_\mathrm{H}} \frac{\mathrm{d}r}{\sqrt{\left(1-\frac{M_\mathrm{J}}{2M'_\mathrm{J}}\right) / a_{\mathrm{m}_i} - \frac{1}{r} }} = t_\mathrm{lf} \sqrt{2G|M'_\mathrm{J}|}.
\end{equation}
\end{linenomath*}
Neglecting small terms, we obtain
\begin{linenomath*}
\begin{equation} \label{equ_lifetime_final}
t_\mathrm{lf} \simeq \frac{R_\mathrm{H}}{2} \sqrt{\frac{a_{\mathrm{m}_i}(2L-1)}{GM_\mathrm{J}(L-1)^2}}.
\end{equation}
\end{linenomath*}
Thus, the lifetime of the 0.05 $\mathrm{\mu m}$ particle is approximately proportional to the square root of the source moon's semi-major axis, i.e., $t_\mathrm{lf}$ $\propto$ $\sqrt{a_{\mathrm{m}_i}}$. From Eq.~(\ref{equ_lifetime_final}), the approximate lifetimes for 0.05 $\mathrm{\mu m}$ dust particles from Io, Europa, Ganymede and Callisto are 2.19$\times10^{-2}$, 2.76$\times10^{-2}$, 3.49$\times10^{-2}$ and 4.63$\times10^{-2}$ years, respectively. These values match closely the average lifetime for this particle size shown in Fig.~\ref{fig_lifetime}, and also the $\sqrt{a_{\mathrm{m}_i}}$ dependence is well confirmed by the full numerical integrations.

Combining Eqs.~(\ref{equ_a_final}) and (\ref{equ_lifetime_final}), it can be seen that $a_\mathrm{H}$ and $t_\mathrm{lf}$ follow a parabolic relationship
\begin{linenomath*}
\begin{equation} \label{equ_aH_tlf}
a_\mathrm{H} \simeq -\frac{4GM_\mathrm{J}(L-1)^2}{R_\mathrm{H}^2(2L-1)^2} t_\mathrm{lf}^2
\end{equation}
\end{linenomath*}
in good agreement with the numerical results (Fig.~\ref{fig_lifetime_a_p05}). The above analysis is also applicable to particles smaller than 0.05 $\mathrm{\mu m}$.

\subsection{Two Sharp Jumps in the Average Lifetime} \label{subsection_lifetime_jump}
The first jump occurs between 0.1 $\mathrm{\mu m}$ and 0.3 $\mathrm{\mu m}$ for particles from all four moons. It occurs as a result of the Lorentz force dominating the dynamics of very small particles. Because of the radially aligned corotational electric field, the primary effect is a radial acceleration, which is directed outward for positively charged grains. Associated with the effective mass $M'_\mathrm{J}$ (Eq.~\ref{equ_effective_mass}), the effective semi-major axis $a'$ and eccentricity $e'$ are defined for a circular orbit as \citep{krivov2002dust}
\begin{linenomath*}
\begin{equation}
a' = a_0\frac{1-L}{1-2L}, \ \ e' = \frac{L}{1-L}.
\end{equation}
\end{linenomath*}
Here, 
$a_0$ is the semi-major axis associated with the mass of Jupiter $M_\mathrm{J}$.

\begin{figure}
\centering
\noindent\includegraphics[width=0.65\textwidth, angle=90]{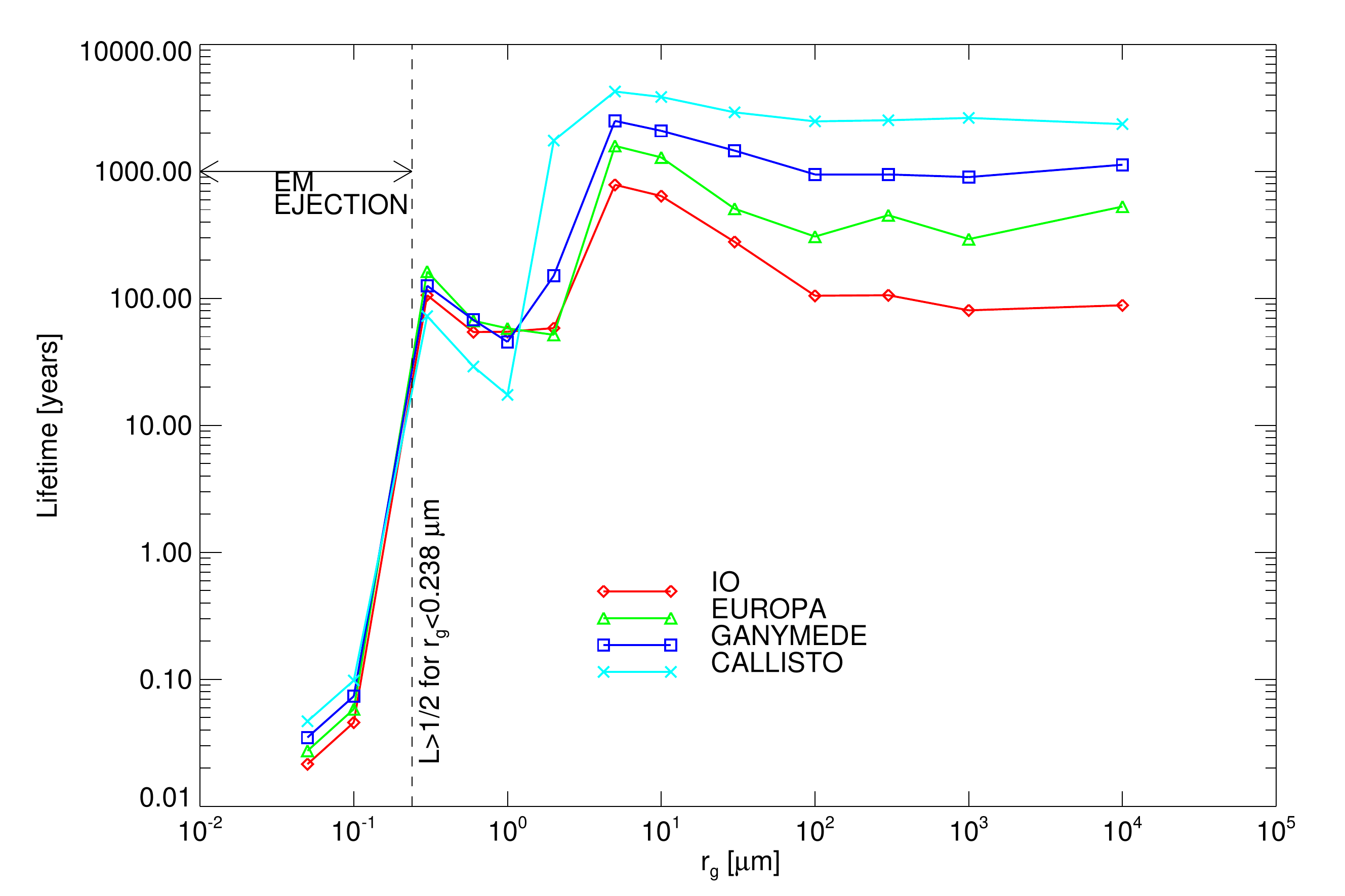}
\caption{\label{fig_lifetime} The average lifetime as a function of grain size from each source moon in (Earth) years.}
\end{figure}

\begin{figure}
\centering
\noindent\includegraphics[width=1\textwidth]{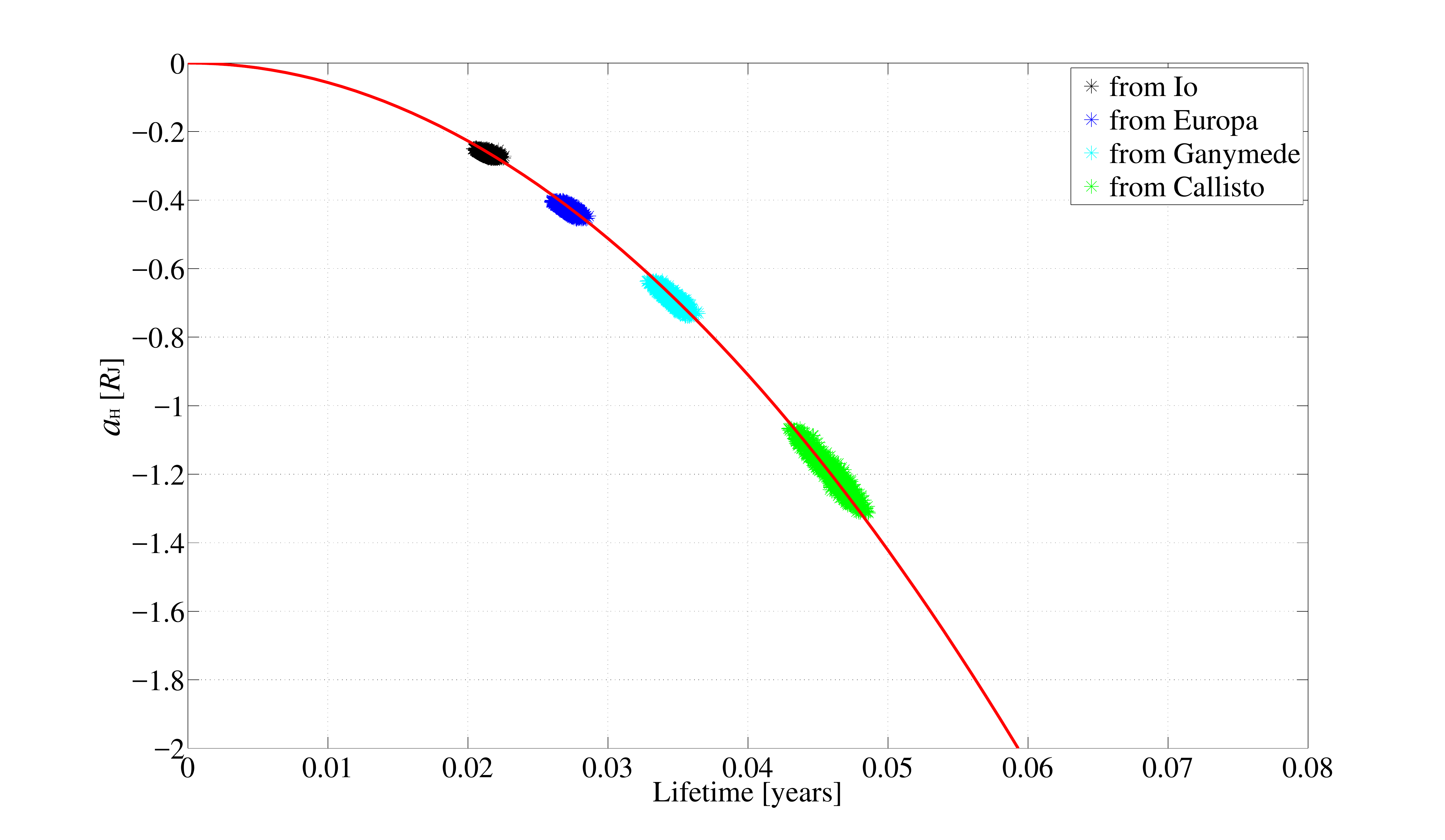}
\caption{\label{fig_lifetime_a_p05} Semi-major axis (at the moment of crossing Jupiter's Hill sphere) as a function of lifetime for electromagnetically dominated 0.05 $\mathrm{\mu m}$ particles. The points marked with asterisks correspond to the results obtained by numerical integration (Section \ref{section_numerical}). The red line presents the parabolic relationship from Eq.~(\ref{equ_aH_tlf}).}
\end{figure}

If $L>1$, the combined effect of the central gravity and electric field is repelling; if $1/2<L<1$, then $a'<0$ and $e'>1$, which means that the effective osculating orbit is hyperbolic. Thus, when $L > 1/2$, the dust will be expelled very quickly from the Jupiter system by the Lorentz force \citep{hamilton1993ejection}. For $\phi = +5$ V, this limit corresponds to the grain size $r_\mathrm{g}$ $<$ 0.238 $\mathrm{\mu m}$ \citep{krivov2002tenuous}. This is the reason for the very short lifetimes of grain sizes $r_\mathrm{g}$ $<$ 0.238 $\mathrm{\mu m}$ (Fig.~\ref{fig_lifetime}), explaining the first sharp jump in the particle lifetime.

\begin{figure}
\centering
\noindent\includegraphics[width=0.66\textwidth]{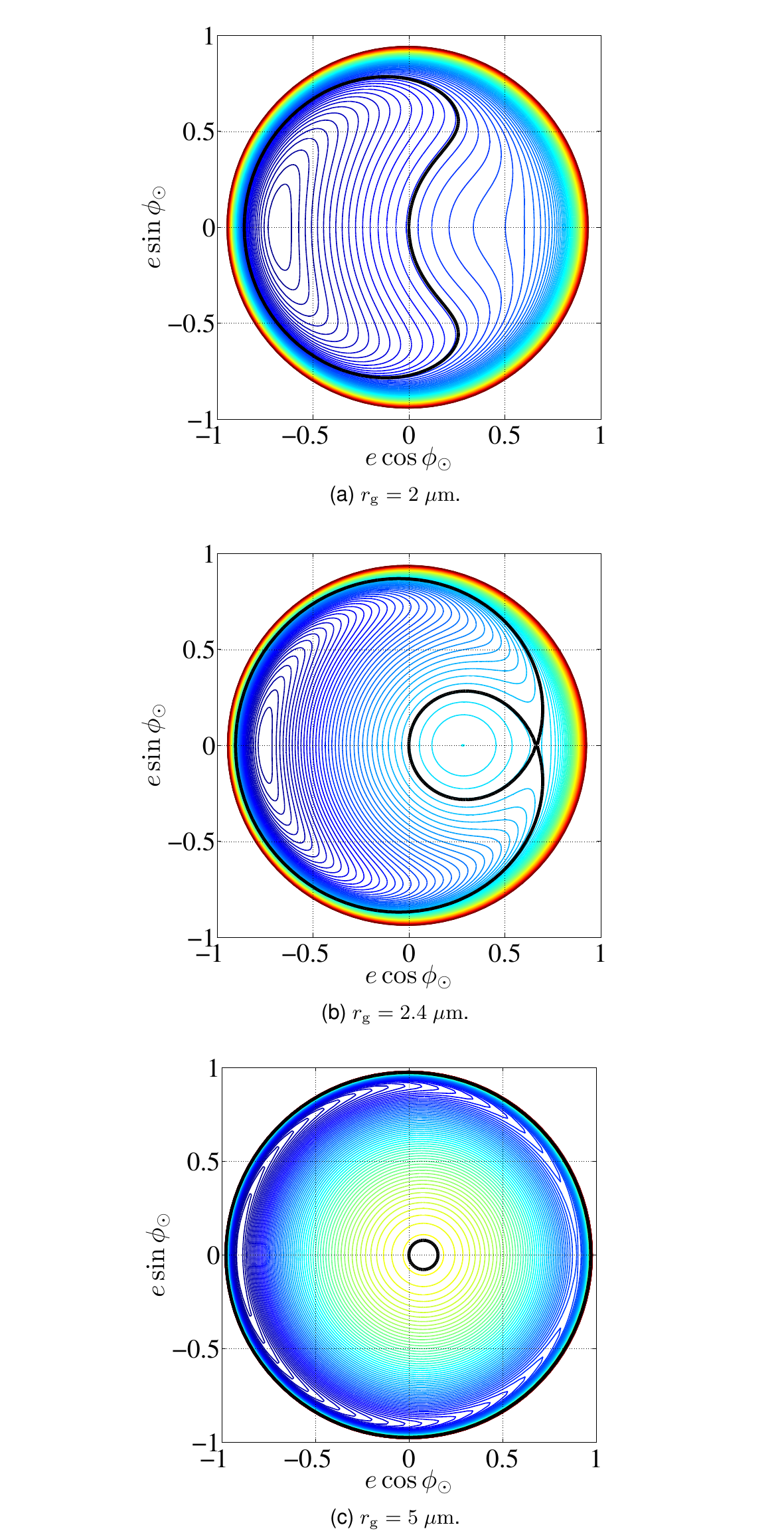}
\vspace{-5 mm}
\caption{Phase portraits for dust particles from Ganymede.}
\label{fig_ganymede_bifurcation}
\end{figure}

\begin{figure}
\centering
\noindent\includegraphics[width=0.66\textwidth]{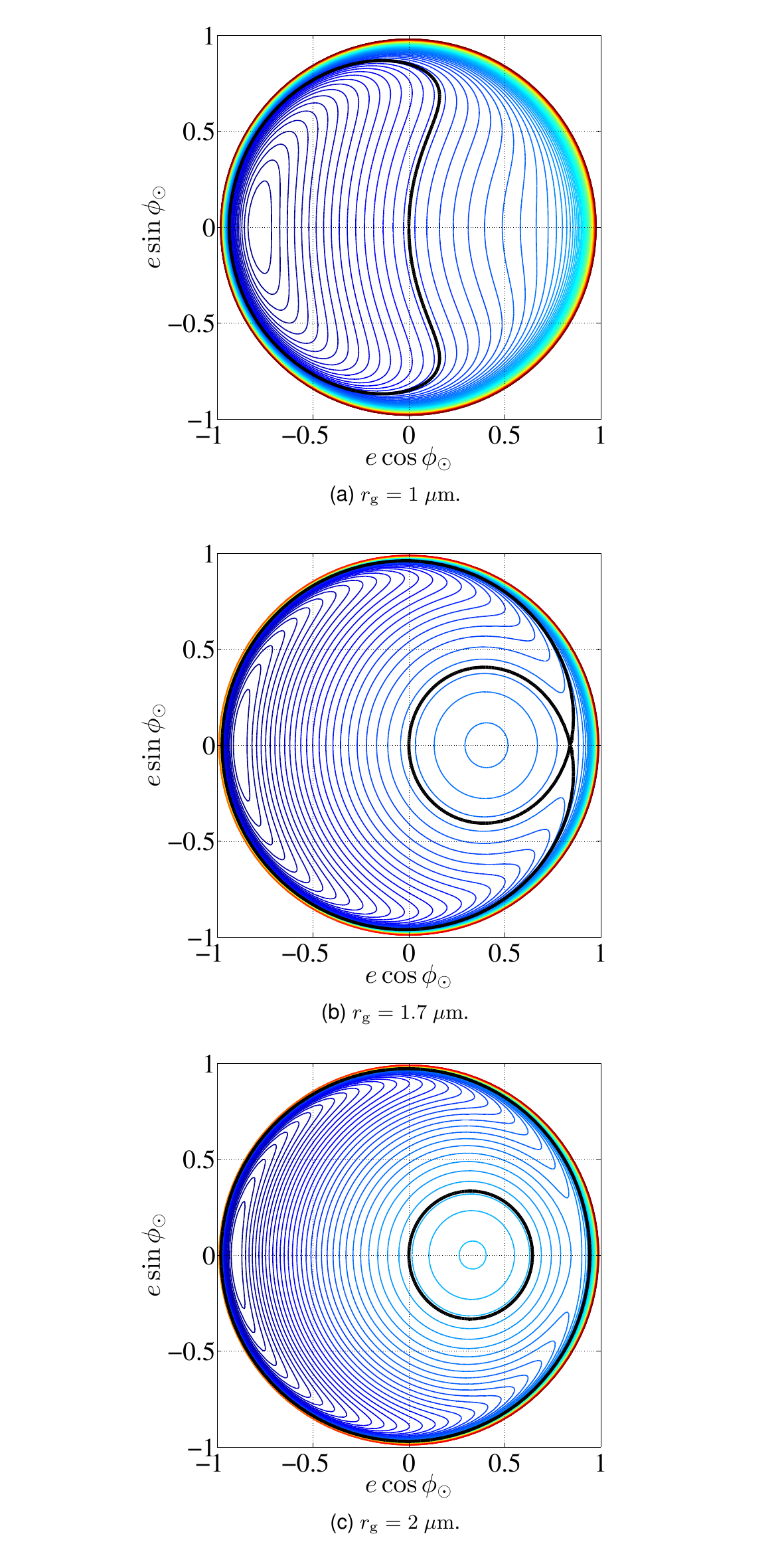}
\vspace{-5 mm}
\caption{Phase portraits for dust particles from Callisto.}
\label{fig_callisto_bifurcation}
\end{figure}

The second jump occurs between 2 $\mathrm{\mu m}$ and 5 $\mathrm{\mu m}$ for Io, Europa and Ganymede, while it occurs between 1 $\mathrm{\mu m}$ and 2 $\mathrm{\mu m}$ for Callisto. It can be explained by a bifurcation in the integral of motion that leads to a sharp drop in orbital eccentricity. Dust particles with higher eccentricity will collide with Jupiter rather quickly, so they usually have a shorter lifetime. The sharp drop in eccentricity was noted by \citet{1996Icar..123..503H} and \citet{krivov2002tenuous}. \citet{1996Icar..123..503H} defined a conserved ``Hamiltonian" if only the dominant perturbations of solar radiation pressure and Lorentz force are in the analysis
\begin{linenomath*}
\begin{equation} \label{equ_hamilton}
\mathcal H = \sqrt{1-e^2} + Ce\cos\phi_\odot + \frac{\tilde L}{2(1-e^2)}.
\end{equation}
\end{linenomath*}
Here $\phi_\odot$ is the solar angle, defined as the angle between the direction towards the Sun and the direction to the grain's pericenter, i.e.~$\phi_\odot = \Omega+\omega-\lambda_\odot$, where $\lambda_\odot$ is the solar longitude. Here $C$ and $\tilde L$ are parameters labelling the relative strengths of radiation pressure and Lorentz force, both dependent on grain size (see \citet{1996Icar..123..503H} for definitions). The bifurcation occurs for a certain combination of grain size and eccentricity \citep{1996Icar..123..503H}
\begin{linenomath*}
\begin{equation} \label{equ_bifurcation1}
\sqrt{1-e^2} + Ce + \frac{\tilde L}{2(1-e^2)} - 1 - \frac{\tilde L}{2} = 0
\end{equation}
\end{linenomath*}
\begin{linenomath*}
\begin{equation} \label{equ_bifurcation2}
\frac{e}{\sqrt{1-e^2}}\left[\frac{\tilde L}{(1-e^2)^{3/2}}-1\right] + C = 0.
\end{equation}
\end{linenomath*}
For Io, Europa and Ganymede, the second sharp jump in the average lifetime occurs between 2 $\mathrm{\mu m}$ and 5 $\mathrm{\mu m}$, while for Callisto, it occurs between 1 $\mathrm{\mu m}$ and 2 $\mathrm{\mu m}$ (Fig.~\ref{fig_lifetime}). This can be understood taking Ganymede and Callisto as examples. Figs.~\ref{fig_ganymede_bifurcation}--\ref{fig_callisto_bifurcation} show the phase portraits for different grain sizes ejected from Ganymede and Callisto. Trajectories with $e$ close to 1 are not shown for clarity. At the time of ejection, the dust particle has nearly zero eccentricity, which corresponds to the origins ($e=0$) of these figures \citep{1996Icar..123..503H}. Thus, the phase contours passing through the origin correspond to the initial trajectories of particles ejected from the moons, which are shown in these figures with thick black lines. The colors of other phase contours correspond to different values of the ``Hamiltonian" $\mathcal H$, calculated from Eq.~(\ref{equ_hamilton}) for different combinations of $e$ and $\phi_\odot$.

For the dust particles launched at Ganymede the bifurcation size is about $r_\mathrm{g} = 2.4 \ \mathrm{\mu m}$ (from Eqs.~(\ref{equ_bifurcation1}) and (\ref{equ_bifurcation2})). The phase portrait of this case is shown in Fig.~\ref{fig_ganymede_bifurcation}(b). When $r_\mathrm{g} = 2 \ \mathrm{\mu m}$ $< 2.4 \ \mathrm{\mu m}$, the grain can reach very large eccentricities, as shown in Fig.~\ref{fig_ganymede_bifurcation}(a), i.e. the grain has a high probability to hit Jupiter or escape from the system. For this reason these particles have a shorter average lifetime. When $r_\mathrm{g} = 5 \ \mathrm{\mu m}$ $> 2.4 \ \mathrm{\mu m}$, the maximum eccentricity of the grain is reduced significantly (Fig.~\ref{fig_ganymede_bifurcation}(c)). This stabilizes the trajectories, explaining the second sharp jump in the average lifetime between sizes $2 \ \mathrm{\mu m} <$ $r_\mathrm{g}$ $<5 \ \mathrm{\mu m}$ for Ganymede. 

In contrast, for dust particles from Callisto, the bifurcation size is smaller, about $r_\mathrm{g} = 1.7 \ \mathrm{\mu m}$ based on Eqs.~(\ref{equ_bifurcation1}) and (\ref{equ_bifurcation2}). These phase portraits are shown in Fig.~\ref{fig_callisto_bifurcation}(b). When $r_\mathrm{g} = 1 \ \mathrm{\mu m}$ $< 1.7 \ \mathrm{\mu m}$ the maximum eccentricity is close to 1 (Fig.~\ref{fig_callisto_bifurcation}(a)), so that these particles tend to collide with Jupiter and thus have short lifetimes. When $r_\mathrm{g} = 2 \ \mathrm{\mu m}$ $> 1.7 \ \mathrm{\mu m}$, the maximum eccentricity is much smaller (Fig.~\ref{fig_callisto_bifurcation}(c)), which will significantly reduce the probability of an impact with Jupiter or escape from the system. The phase portraits shown in Figs.~\ref{fig_callisto_bifurcation}(a)--(c) are consistent with the sharp jump in the average lifetime between sizes $1 \ \mathrm{\mu m}<$ $r_\mathrm{g}$ $<2 \ \mathrm{\mu m}$ for Callisto (Fig.~\ref{fig_lifetime}).

Note that a small portion of large particles ($\geqslant$ 5 $\mathrm{\mu m}$) is still in orbit around Jupiter even after 8,000 years of integration, for which we take their lifetime as 8,000 years approximately. Thus, the lifetimes of these large particles are undervalued. However, this approximation does not affect our analysis since the sharp jumps occur at smaller sizes ($<$ 5 $\mathrm{\mu m}$).

\section{Appearance of the Dust Population in a Frame Rotating with the Sun} \label{section_asymmetry}
Generally, dust rings in the region of the Galilean moons tend to maintain a fixed orientation with respect to the Sun. For instance, \citet{krivov2002tenuous} showed that rings formed by 0.6 $\mathrm{\mu m}$ particles from Europa and Ganymede are displaced towards the Sun by analyzing scatterplots for one Jovian season in the Jovicentric equatorial inertial frame. In this section, this azimuthal asymmetry of dust in the region of the Galilean moons is discussed in a more detailed and general way, also taking into account trajectories over much longer integration time (up to 8,000 Earth years). 

The asymmetry becomes apparent when the dust configuration obtained from our model is plotted in a Jovicentric frame that co-rotates with the Sun. Thus, the Jovicentric rotating frame $Ox_\mathrm{sun}y_\mathrm{sun}z_\mathrm{sun}$ is defined such that the $x_\mathrm{sun}$ axis points towards the Sun, the $z_\mathrm{sun}$ axis is aligned with the orbital normal of the Sun around Jupiter, and the $y_\mathrm{sun}$ axis completes an orthogonal right-handed reference frame. From our numerical simulations, we find that for smaller particles the dust distribution is shifted towards the Sun, while, for larger sizes, the dust distribution is shifted away from the Sun. Specifically, for Io, Europa and Ganymede, all dust rings formed by 0.6 $\mathrm{\mu m}$, 1 $\mathrm{\mu m}$ and 2 $\mathrm{\mu m}$ are shifted towards the Sun, and rings formed by 5 $\mathrm{\mu m}$ and 10 $\mathrm{\mu m}$ are shifted away from the Sun. For Callisto, the behavior is similar, only for a bit different grain size, such that rings formed by 0.6 $\mathrm{\mu m}$ and 1 $\mathrm{\mu m}$ grains are displaced towards the Sun, and rings formed by 2 $\mathrm{\mu m}$, 5 $\mathrm{\mu m}$ and 10 $\mathrm{\mu m}$ grains are displaced away from the Sun.

Note that the inversion of the preferred orientation for Callisto occurs between 1 $\mathrm{\mu m}$ and 2 $\mathrm{\mu m}$, i.e.~at a smaller grain size than for the other moons for which the inversion occurs between 2 $\mathrm{\mu m}$ and 5 $\mathrm{\mu m}$. This is analogous to the second jump in lifetime (Fig.~\ref{fig_lifetime}), occurring for Callisto also at a smaller grain size than for the other three moons. The inversion of orientation arises as a consequence of the same bifurcation point of the ``Hamiltonian" $\mathcal H$ as discussed in Section \ref{subsection_lifetime_jump}. In what follows, we again use the ``Hamiltonian" phase portraits to analyze also the asymmetry. Two cases with pronounced asymmetry are selected and presented in the following.

\begin{figure}
\centering
\noindent\includegraphics[width=0.65\textwidth, angle=90]{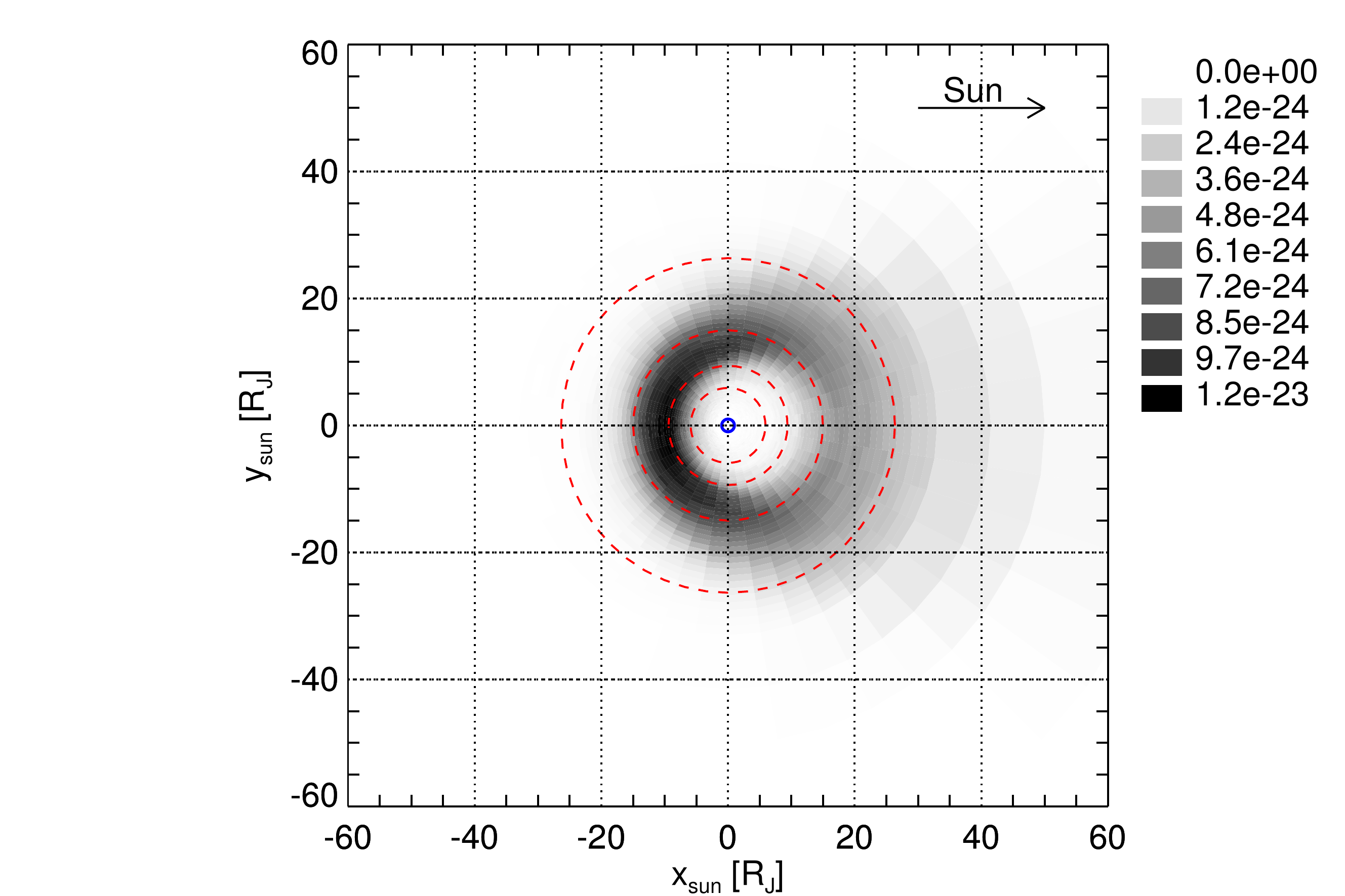}
\caption{\label{fig_ganymede_p6_asym} Relative number density (in arbitrary units) in a Jovicentric frame rotating with the Sun for $r_\mathrm{g} = 0.6 \ \mathrm{\mu m}$ particles from Ganymede.}
\end{figure}

\begin{figure}
\centering
\noindent\includegraphics[width=0.7\textwidth]{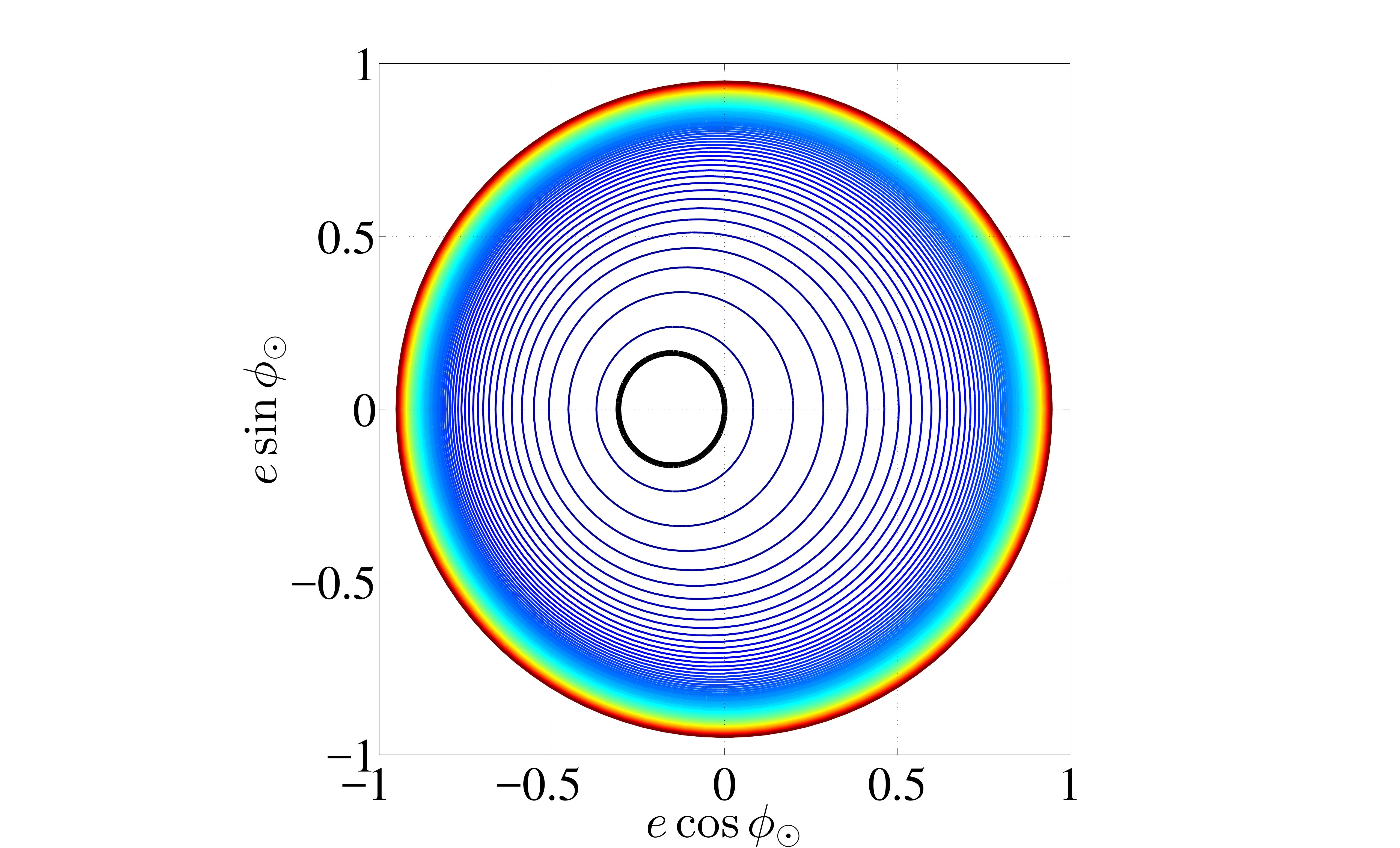}
\caption{\label{fig_ganymede_p6_phase} Phase portraits for $r_\mathrm{g} = 0.6 \ \mathrm{\mu m}$ dust from Ganymede.}
\end{figure}

\begin{figure}
\centering
\noindent\includegraphics[width=0.65\textwidth, angle=90]{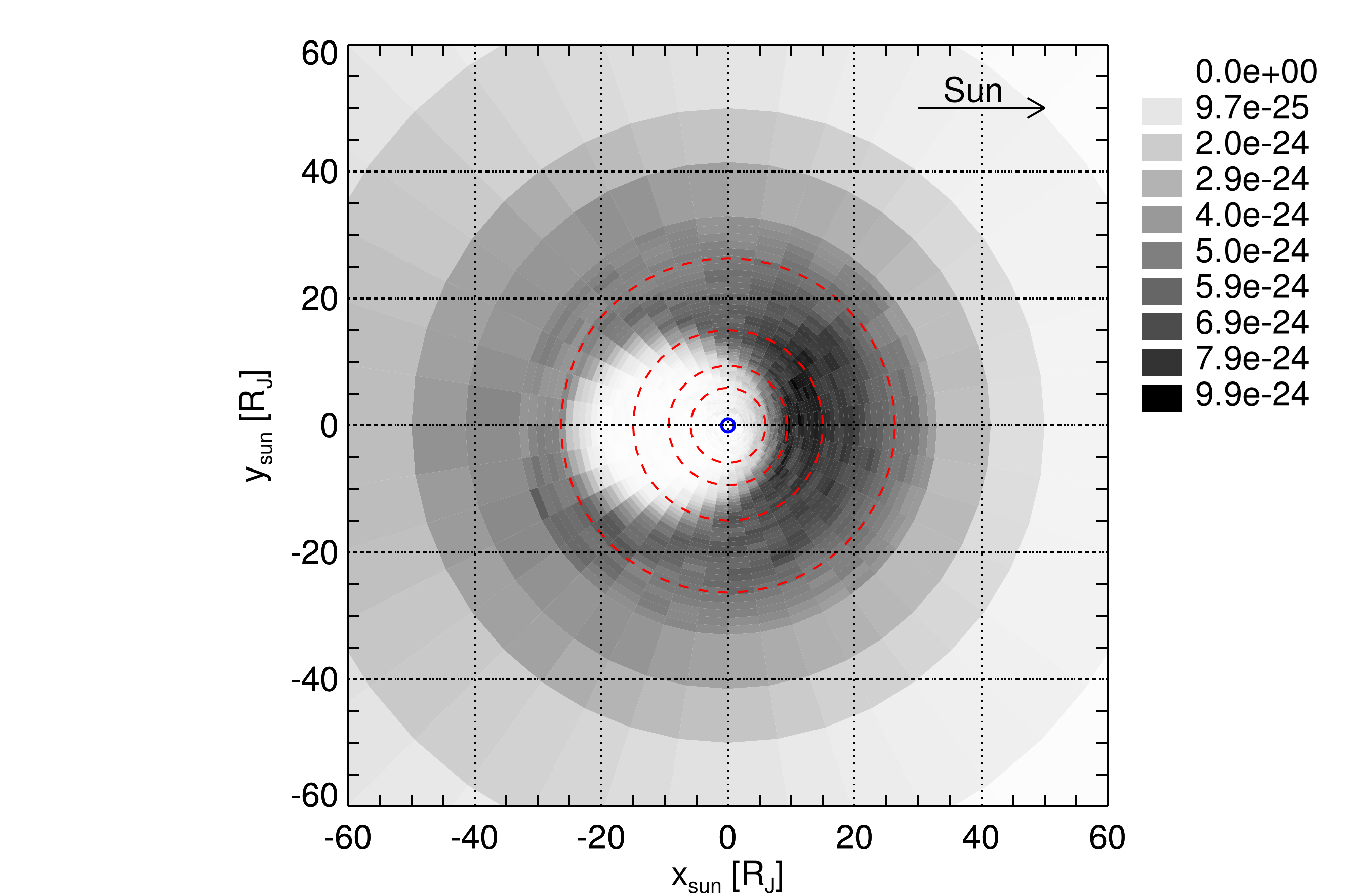}
\caption{\label{fig_callisto_2_asym} Relative number density (in arbitrary units) in a Jovicentric frame rotating with the Sun for $r_\mathrm{g} = 2 \ \mathrm{\mu m}$ particles from Callisto.}
\end{figure}

To evaluate our results, a new set of grid bins $(\rho_\mathrm{cl}$, $\phi_\mathrm{cl}$, $z_\mathrm{cl})$ in the Jovicentric frame rotating with the Sun has to be employed. We perform similar operations as described in Section \ref{section_distri_inertial} to determine the dust configurations in this frame. Note that in the following figures the number density is vertically averaged in order to make the plotted distributions smoother. Fig.~\ref{fig_ganymede_p6_asym} shows the relative number density distribution in the $x_\mathrm{sun}$-$y_\mathrm{sun}$ plane for 0.6 $\mathrm{\mu m}$ particles from Ganymede. There is an enhanced dust density in the $-x_\mathrm{sun}$ direction that corresponds to a concentration of the grains' pericenters. As a result, the dust ring is offset towards the Sun. The reason for this asymmetry can again be understood with the help of the phase portraits for this dust population (Fig.~\ref{fig_ganymede_p6_phase}). The maximum eccentricity is attained when $\phi_\odot$ = $180^\circ$, which implies that this dust population is offset towards the Sun because $\phi_\odot$ is the angle between the Sun and the orbital pericenter as seen from Jupiter. The opposite orientation is seen for 2 $\mathrm{\mu m}$ dust from Callisto (Fig.~\ref{fig_callisto_2_asym}). For this case, the maximum eccentricity is attained when $\phi_\odot$ = $0^\circ$ (Fig.~\ref{fig_callisto_bifurcation}(c)), which means that the dust population is offset away from the Sun (Fig.~\ref{fig_callisto_2_asym}). As a result, the azimuthal asymmetry is consistent with phase portraits of the ``Hamiltonian" $\mathcal H$ taking into account the effect of Lorentz force combined with solar radiation pressure. Due to different relative strength of these two forces, which depend on grain size, the dust ring can be offset towards the Sun or away from Sun. This behavior has been reported for dust around Mars \citep{hamilton1996asymmetric}, Saturn's E ring \citep{Horanyi:1992ww, hedman2012three}, Saturn's Phoebe ring \citep{verbiscer2009saturn}, Saturn's Charming Ringlet \citep{hedman2010shape}, and dust in the outer part of the Jupiter system \citep{krivov2002dust}. Besides, a planetary shadow resonance can also produce asymmetry of Jupiter's gossamer rings \citep{hamilton2008sculpting}.

\section{Transport of Dust Between the Galilean Moons and to Jupiter} \label{section_transport}
The fate of dust particles from the four Galilean moons after 8,000 Earth years is summarized in Tables.~\ref{tab_Io_sink}--\ref{tab_Callisto_sink}. In general, small particles, including 0.05 $\mathrm{\mu m}$, 0.1 $\mathrm{\mu m}$ and 0.3 $\mathrm{\mu m}$ are much more easily expelled from the Jupiter system, as a consequence of a positive grain surface potential producing a strong, outward directed electric field part of the Lorentz force (Section \ref{section_lifetime}). Especially, all the 0.05 $\mathrm{\mu m}$ and 0.1 $\mathrm{\mu m}$ dust particles escape the Jupiter system to interplanetary space. For particles of intermediate size, including 0.6 $\mathrm{\mu m}$, 1 $\mathrm{\mu m}$ and 2 $\mathrm{\mu m}$, the fraction of particles that impact Jupiter becomes large because for these sizes, solar radiation pressure becomes more important. Solar radiation pressure pumps the orbital eccentricity, until the grain collides with Jupiter. For particles of 1 $\mathrm{\mu m}$ and 2 $\mathrm{\mu m}$, the transport efficiencies to Jupiter range from $46.4\%$ to $81.7\%$ for the four source moons. For still larger grain sizes, the sharp drop in orbital eccentricity discussed earlier (Section \ref{subsection_lifetime_jump}), reduces the number of particles that impact Jupiter. As a result, 5 $\mathrm{\mu m}$ and 10 $\mathrm{\mu m}$ dust particles have the largest fraction remaining in orbit around Jupiter after 8,000 years compared to other grain sizes. For particles of sizes 30 $\mathrm{\mu m}$, 100 $\mathrm{\mu m}$, 300 $\mathrm{\mu m}$, $10^3$ $\mathrm{\mu m}$ (1 $\mathrm{mm}$), and $10^4$ $\mathrm{\mu m}$ (1 $\mathrm{cm}$), the gravitational forces become important, especially the gravity of the parent moon. As a result, the fraction of the particles that collide with the parent moon becomes high. For $\geqslant$ 30 $\mathrm{\mu m}$ particles from Io, Europa, Ganymede, and Callisto, this fraction is about $69.9\%$ to $87.1\%$, $45.6\%$ to $52.5\%$, $56.3\%$ to $58.7\%$, and $34.6\%$ to $43.5\%$, respectively. In most cases, the number of grains hitting the parent moon is larger than those hitting other moons because initially the semi-major axis remains close to the one of the parent moon. Also note that the transport efficiency from the adjacent moons to Ganymede (the Galilean moon with largest mass and size) is high. This is obvious for $r_\mathrm{g}\geqslant$ 5 $\mathrm{\mu m}$ particles from Europa where the transport efficiency to Ganymede is about $21.8\%$ to $30.3\%$, and also for $r_\mathrm{g}\geqslant$ 30 $\mathrm{\mu m}$ particles from Callisto where the transport efficiency to Ganymede is about $20.3\%$ to $30.3\%$. Moreover, the transport efficiency of dust particles (from other Galilean moons) to 
Europa is of special interest to future space missions (Europa Multiple-Flyby Mission and JUpiter ICy moons Explorer). For both $r_\mathrm{g}\geqslant$ 1 $\mathrm{\mu m}$ particles from Io, and $r_\mathrm{g}\geqslant$ 5 $\mathrm{\mu m}$ particles from Ganymede, the transport efficiency to Europa is on the order of $10\%$, which will contaminate the surface of Europa.

\section{Dust on Retrograde Orbits} \label{section_retrograde}
In previous studies, it was found that there are several sources of retrograde dust in the Jupiter system: magnetospherically captured interplanetary and interstellar dust \citep{colwell1998capture, colwell1998jupiter}, particles from the Shoemaker-Levy 9 comet \citep{horanyi1994new}, and grains from irregular satellites of Jupiter \citep{krivov2002dust, bottke2013black}. In this paper, we find that the Galilean moons can also be a source of retrograde dust. Namely, a small fraction of particles from the Galilean moons may become retrograde in course of the orbital evolution under the action of the perturbing forces, although all four moons are prograde. From Table \ref{tab_retro_frac}, it can be concluded that for grains from Io, Europa and Ganymede in the size range 0.3-2 $\mathrm{\mu m}$, the fraction of retrograde particles is on the order of 1\% to 5\%; for grains from Callisto, in the size range 0.3-1 $\mathrm{\mu m}$, this fraction is on the order of 1\% to 10\%. This implies that there is a sharp drop in the fraction of retrograde particles between 2 $\mathrm{\mu m}$ and 5 $\mathrm{\mu m}$ for Io, Europa and Ganymede, while for Callisto, the sharp drop occurs earlier between 1 $\mathrm{\mu m}$ and 2 $\mathrm{\mu m}$. The abrupt change vs.~grain size appears again analogous to the changes of the average lifetime (Section \ref{subsection_lifetime_jump}) and orientation of the dust configuration with respect to the Sun (Section \ref{section_asymmetry}), i.e.~these properties are dynamically correlated.

\begin{figure}
\centering
\noindent\includegraphics[width=0.95\textwidth]{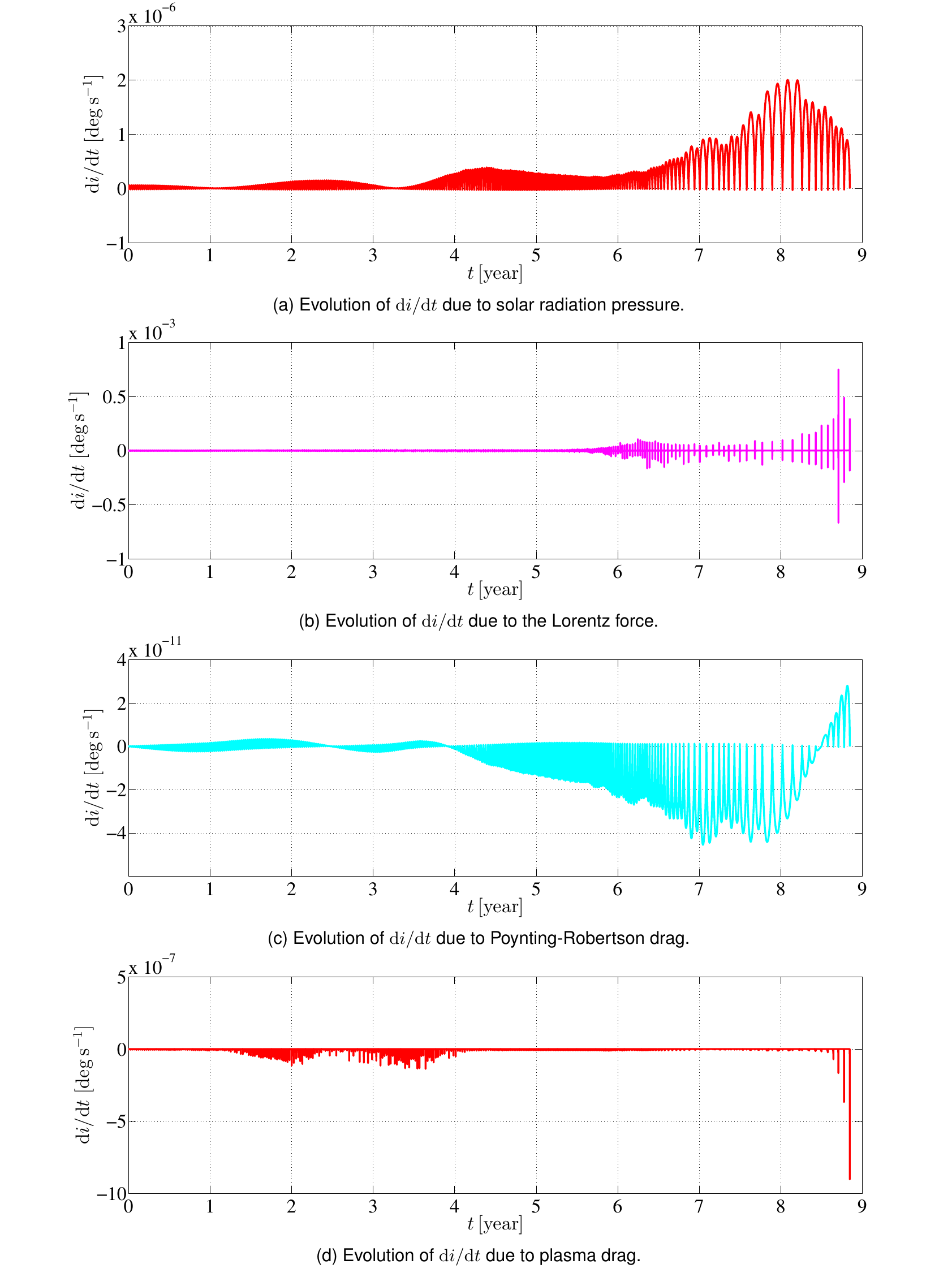}
\caption{Evolution of $\mathrm di/\mathrm dt$ due to non-gravitational perturbation forces for a 2 $\mu \mathrm{m}$ particle from Europa.}
\label{fig_didt_nongravity}
\end{figure}

\begin{figure}
\centering
\noindent\includegraphics[width=0.95\textwidth]{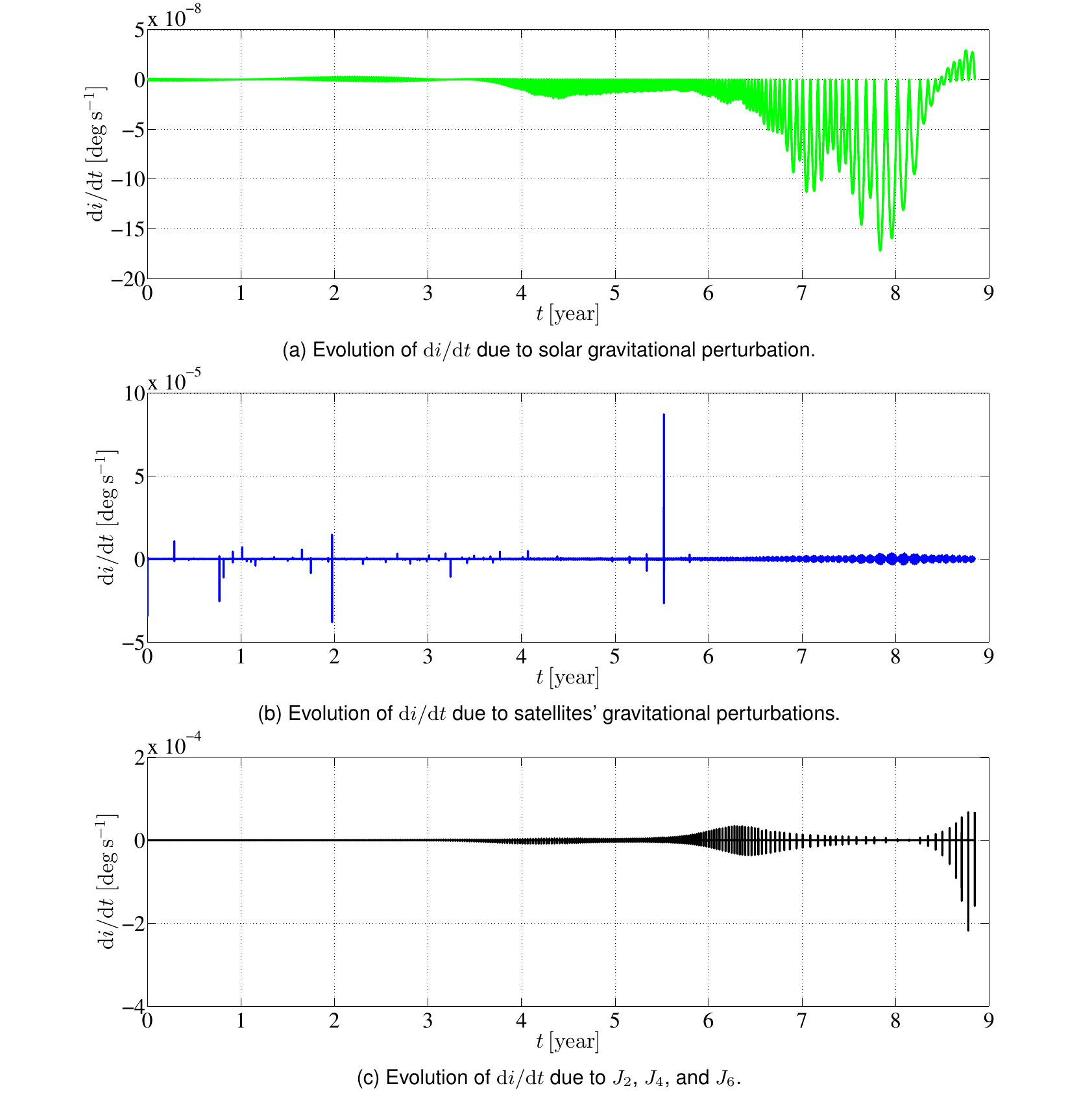}
\caption{Evolution of $\mathrm di/\mathrm dt$ due to gravitational perturbation forces for a 2 $\mu \mathrm{m}$ particle from Europa.}
\label{fig_didt_gravity}
\end{figure}

\begin{figure}
\centering
\noindent\includegraphics[width=1\textwidth]{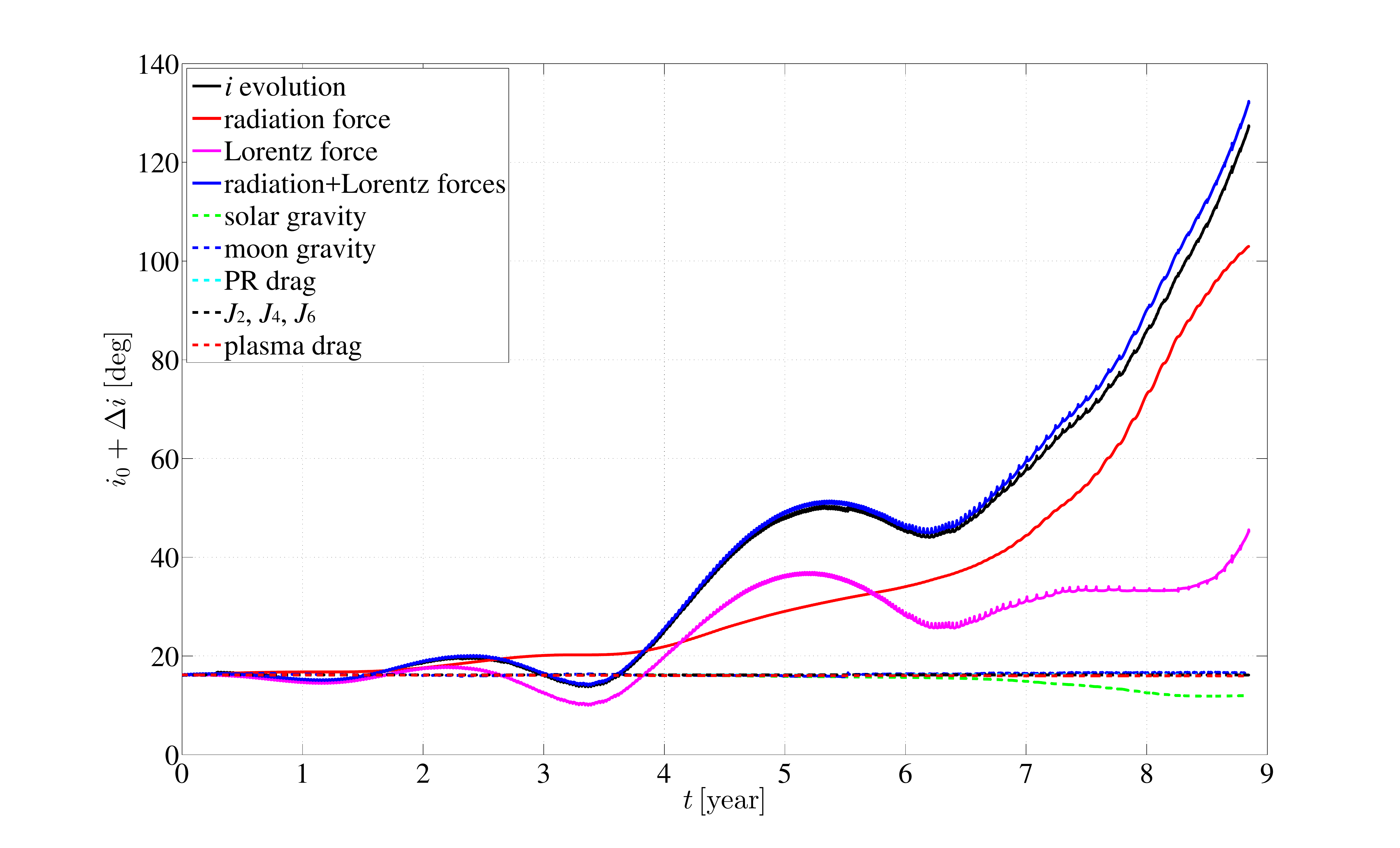}
\caption{\label{fig_i_evolution} $i_0 + \Delta i$, where $\Delta i$ is the integral of the contribution to $\mathrm di/\mathrm dt$ due to respective perturbation force for a 2 $\mu \mathrm{m}$ particle from Europa.}
\end{figure}

Here we take a 2 $\mathrm{\mu m}$ particle from Europa as an example, which shows a typical evolution from a prograde to a retrograde orbit. Its initial inclination is about $16.17^\circ$, and its maximum inclination is $127.08^\circ$. From the perturbation equations of motion \citep{murray1999solar}, the contribution of a given perturbation force on $\mathrm di/\mathrm dt$ can be calculated as
\begin{linenomath*}
\begin{equation} \label{equ_i_contri}
\frac{\mathrm{d}i}{\mathrm{d}t} \Big|_\mathrm{force}  = \frac{r\cos(\omega+f)}{na^2\sqrt{1-e^2}}N |_\mathrm{force}
\end{equation}
\end{linenomath*}
where $N |_\mathrm{force}$ is the normal component (with respect to the orbital plane) of the acceleration due to this force. We denote with $\hat {\vec h} = (\sin i\sin\Omega, -\sin i\cos\Omega, \cos i)^\mathrm{T}$ the unit vector of the dust particle's angular momentum, so that
\begin{linenomath*}
\begin{equation} \label{equ_normal_force}
N |_\mathrm{force} = \hat {\vec h} \cdot \ddot{\vec r}_\mathrm{force}.
\end{equation}
\end{linenomath*}
The expressions $\ddot{\vec r}_\mathrm{force}$ for different perturbation forces are given in Section \ref{section_model}. The separate contributions of non-gravitational and gravitational perturbation forces on $\mathrm di/\mathrm dt$ are shown in Figs.~\ref{fig_didt_nongravity} and \ref{fig_didt_gravity}, respectively. From Fig.~\ref{fig_didt_nongravity}(a), it is seen that solar radiation pressure increases the inclination nearly continuously (see also Fig.~\ref{fig_i_evolution}).

In contrast, the Lorentz force increases and decreases the inclination in an alternating manner, the largest positive and negative peaks corresponding to encounters with Jupiter (Fig.~\ref{fig_didt_nongravity}(b)). The cumulative effect of the Lorentz force on the inclination (Fig.~\ref{fig_i_evolution}) is not monotonic, yet it is important. Fig.~\ref{fig_i_evolution} demonstrates that the sum of the effects of solar radiation pressure and Lorentz force comes very close to the results of numerical simulations with full dynamics, indicating that they are the dominant perturbation forces. The solar gravitational perturbation makes a less important, yet visible contribution to the variation of the inclination (Figs.~\ref{fig_didt_gravity}(a) and \ref{fig_i_evolution}). This is because for highly eccentric orbits the distance between dust and Jupiter is large around apocenter. Moreover, the effects of solar gravitational perturbation and Poynting-Robertson drag (PR drag) on $\mathrm di/\mathrm dt$ (Figs.~\ref{fig_didt_gravity}(a) and \ref{fig_didt_nongravity}(c)) are opposing the one of solar radiation pressure for most of the time, but with much smaller amplitude. 

In general, the amplitudes of $\mathrm di/\mathrm dt$ due to satellites' gravitational perturbations and $J_2, J_4, J_6$ are quite large (Figs.~\ref{fig_didt_gravity}(b) and \ref{fig_didt_gravity}(c)). Sharp peaks occur at encounters with the Galilean moons or Jupiter. For example, the highest positive peak in Fig.~\ref{fig_didt_gravity}(b) corresponds to an encounter with Callisto, while the largest negative peak corresponds to an encounter with Io. The largest negative peak in Fig.~\ref{fig_didt_gravity}(c) corresponds to an encounter with Jupiter. However, the long-term average effect of these two perturbations is close to zero. For the average effect of $J_2$ on $i$ see \citet{hamilton1993motion}. Thus, the cumulative effects of the satellites' gravitational perturbations, and $J_2, J_4, J_6$ on the evolution of $i$ remain small (Fig.~\ref{fig_i_evolution}). Also the effect of plasma drag on $\mathrm di/\mathrm dt$ is about an order of magnitude smaller than the one of solar radiation pressure for most of the time (Fig.~\ref{fig_didt_nongravity}(d)), and the cumulative effect due to plasma drag on inclination is small (Fig.~\ref{fig_i_evolution}). In conclusion, solar radiation pressure combined with the action of the Lorentz force can significantly change the inclination over long times, in some cases leading to the retrograde orbits we observe.

It is worth noting that the plasma drag will always decrease the orbital inclination with respect to the planetary equatorial frame. In our simulations, the value of $\mathrm di/\mathrm dt$ due to the direct drag is always negative and often close to zero (Fig.~\ref{fig_didt_nongravity}(d)). This can be explained by Eq.~(\ref{equ_i_contri})
\begin{linenomath*}
\begin{equation}
\frac{\mathrm{d}i}{\mathrm{d}t} \Big|_\mathrm{PD}  = \frac{r\cos(\omega+f)}{na^2\sqrt{1-e^2}}N |_\mathrm{PD}
\end{equation}
\end{linenomath*}
where $N|_\mathrm{PD}$ is the component of the acceleration due to plasma drag in the direction of the orbital normal. Combining Eqs.~(\ref{equ_plasma_drag}) and (\ref{equ_normal_force}), we get
\begin{linenomath*}
\begin{equation}
N |_\mathrm{PD} = \hat {\vec h} \cdot \ddot{\vec r}_\mathrm{PD} = -\hat {\vec h} \cdot \sum_{\mathrm H=1}^{N_\mathrm H}\frac{3n_\mathrm Hm_\mathrm H|\vec v_\mathrm{rel}|}{4\rho_\mathrm gr_\mathrm g}\vec v_\mathrm{rel}
\end{equation}
\end{linenomath*}
where 
\begin{linenomath*}
\begin{equation}
\vec v_\mathrm{rel} = \dot{\vec r}-\vec \Omega_\mathrm{J}\times \vec r.
\end{equation}
\end{linenomath*}
In the planetary equatorial frame, we have $\vec \Omega_\mathrm{J} \simeq (0,0,\Omega_\mathrm{J})^\mathrm{T}$, thus, it becomes
\begin{linenomath*}
\begin{equation}
\vec v_\mathrm{rel} \simeq \dot{\vec r} - (0,0,\Omega_\mathrm{J})^\mathrm{T}\times \vec r
\end{equation}
\end{linenomath*}
where
\begin{linenomath*}
\begin{equation}
\begin{bmatrix}
0 \\
0 \\
\Omega_\mathrm{J}
\end{bmatrix} \times \vec r =r\Omega_\mathrm{J}
\begin{bmatrix}
-\sin\Omega\cos(\omega+f)-\cos\Omega\sin(\omega+f)\cos i \\
\cos\Omega\cos(\omega+f)-\sin\Omega\sin(\omega+f)\cos i \\
0
\end{bmatrix}.
\end{equation}
\end{linenomath*}
With all this, we finally obtain
\begin{linenomath*}
\begin{equation}
\frac{\mathrm{d}i}{\mathrm{d}t} \Big|_\mathrm{PD} = - \left(\frac{r^2\Omega_\mathrm{J}}{na^2\sqrt{1-e^2}} \right) \left(\sum_{\mathrm H=1}^{N_\mathrm H}\frac{3n_\mathrm Hm_\mathrm H|\vec v_\mathrm{rel}|}{4\rho_\mathrm gr_\mathrm g}\right) \sin i \cos^2(\omega+f) \leqslant 0
\end{equation}
\end{linenomath*}
demonstrating that the direct plasma drag always decreases the inclination, no matter whether the particle is inside or outside the corotational radius.

\section{The Distribution of Orbital Elements} \label{section_distri_elem}
\begin{figure}
\centering
\noindent\includegraphics[width=1\textwidth]{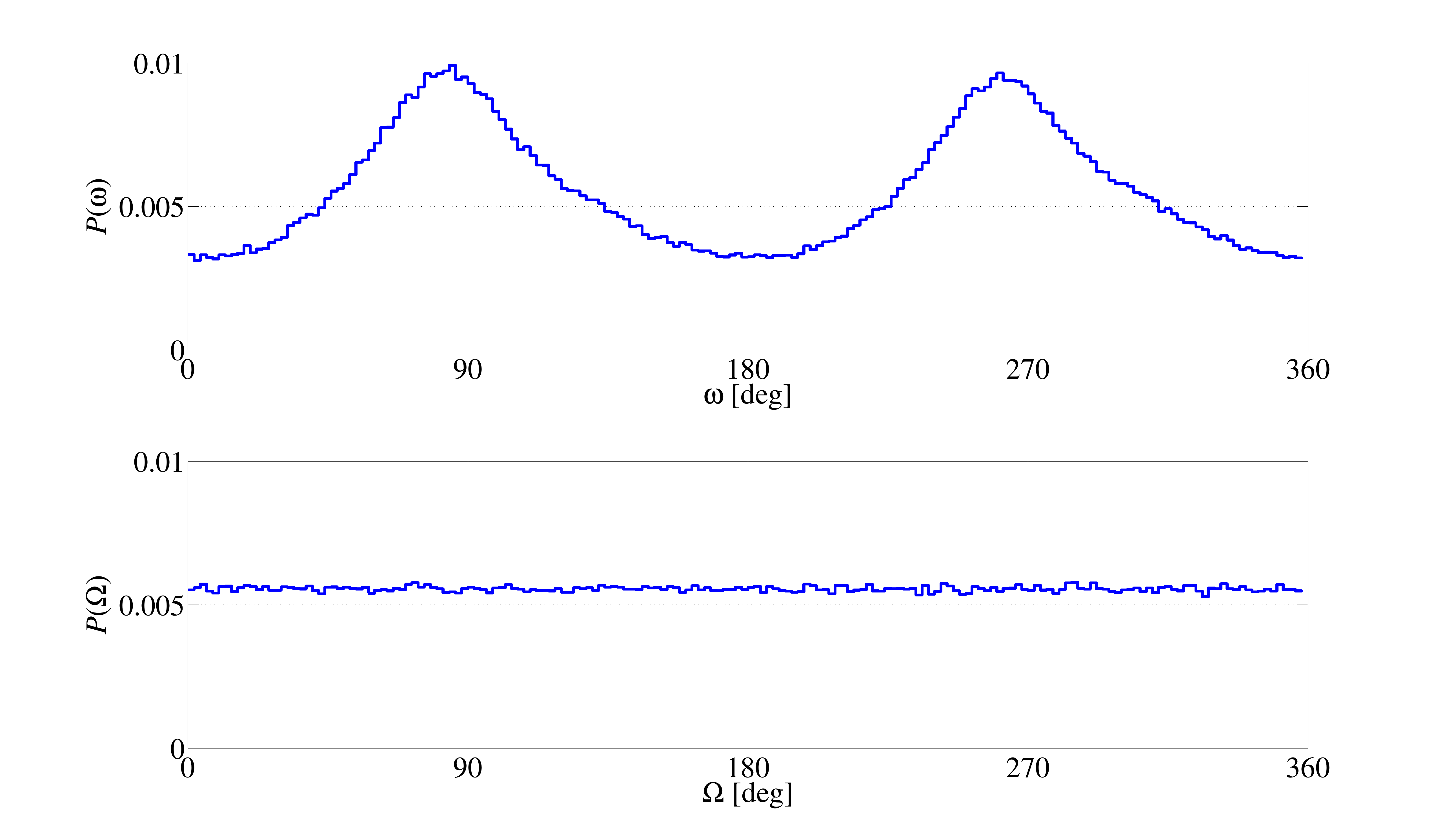}
\caption{\label{fig_PwO_p3} Relative distributions of $\omega$ and $\Omega$ for 0.3 $\mathrm{\mu m}$ particles from Europa.}
\end{figure}

\begin{figure}
\centering
\noindent\includegraphics[width=1\textwidth]{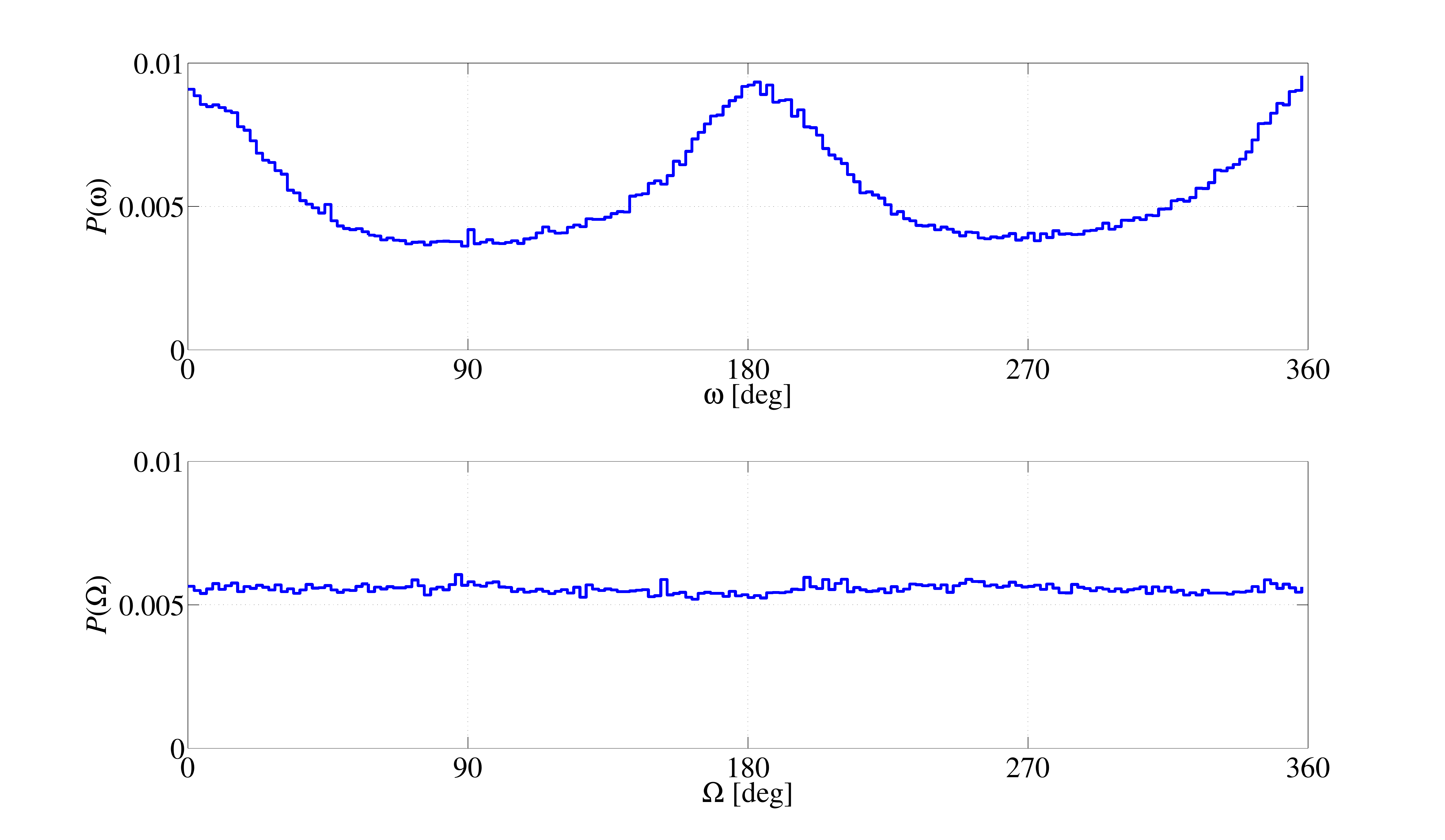}
\caption{\label{fig_PwO_2} Relative distributions of $\omega$ and $\Omega$ for 2 $\mathrm{\mu m}$ particles from Europa.}
\end{figure}

From the numerical integrations, it is found that for 0.3 $\mathrm{\mu m}$ particles from all four Galilean moons the distribution of the argument of pericenter $\omega$ peaks around $90^\circ$ and $270^\circ$, and has troughs around $0^\circ$ and $180^\circ$ (see Fig.~\ref{fig_PwO_p3} for an example, showing 0.3 $\mathrm{\mu m}$ particles from Europa). In contrast, for larger particles, the distribution of the argument of pericenter $\omega$ peaks around $0^\circ$ and $180^\circ$, and has troughs around $90^\circ$ and $270^\circ$, respectively (see Fig.~\ref{fig_PwO_2} for an example, showing 2 $\mathrm{\mu m}$ particles from Europa). We use the orbit averaged Gaussian equation to describe the long-term evolution orbital elements and analyze their distributions.

For dust particles in the size range 0.3-2 $\mathrm{\mu m}$, Lorentz force and solar radiation pressure are the most important perturbation forces for small particles. For the following semi-analytical analysis, it is assumed that the solar orbit is circular and equatorial, i.e.~$e_\odot$ $\approx$ 0 and $i_\odot$ $\approx$ 0. The average variation of orbital elements due to solar radiation pressure is given by Eq.~(10) of \citet{hamilton1993motion}. \citet{hamilton1993motion} also provided the average variation of orbital elements due to Lorentz force with the strongest dipole term $g_{1,0}$ (see his Eq.~(23)) by neglecting high order terms in $e$ and $i$. Here we generalize the results for all eccentricities and inclinations because high order terms in $e$ and $i$ are important for our case
\begin{linenomath*}
\begin{equation}
\left\langle\frac{\mathrm{d}a}{\mathrm{d}t}\right\rangle_{g_{1,0}} = 0
\end{equation}
\end{linenomath*}
\begin{linenomath*}
\begin{equation}
\left\langle\frac{\mathrm{d}e}{\mathrm{d}t}\right\rangle_{g_{1,0}} = -nL\frac{e\sqrt{1-e^2}}{(1+\sqrt{1-e^2})^2}\sin^2i\sin(2\omega)
\end{equation}
\end{linenomath*}
\begin{linenomath*}
\begin{equation}
\left\langle\frac{\mathrm{d}i}{\mathrm{d}t}\right\rangle_{g_{1,0}} = \frac{nL}{\sqrt{1-e^2}}\frac{e^2}{(1+\sqrt{1-e^2})^2}\sin i\cos i\sin(2\omega)
\end{equation}
\end{linenomath*}
\begin{linenomath*}
\begin{equation}
\left\langle\frac{\mathrm{d}\Omega}{\mathrm{d}t}\right\rangle_{g_{1,0}} = \frac{nL}{\sqrt{1-e^2}}\left\{\cos i\left[1-\frac{e^2\cos(2\omega)}{(1+\sqrt{1-e^2})^2}\right]-\frac{1}{1-e^2}\left(\frac{n}{\Omega_\mathrm{J}}\right)\right\}
\end{equation}
\end{linenomath*}
\begin{linenomath*}
\begin{equation}
\begin{split}
\left\langle\frac{\mathrm{d}\omega}{\mathrm{d}t}\right\rangle_{g_{1,0}} = \frac{nL}{\sqrt{1-e^2}} \bigg\{ & -\cos^2i\left[1-\frac{e^2\cos(2\omega)}{(1+\sqrt{1-e^2})^2}\right]+\frac{3\cos i}{1-e^2}\left(\frac{n}{\Omega_\mathrm{J}}\right) \\
& -\frac{\sin^2i\cos(2\omega)\sqrt{1-e^2}}{(1+\sqrt{1-e^2})^2}\bigg\}
\end{split}
\end{equation}
\end{linenomath*}
\begin{linenomath*}
\begin{equation}
\left\langle\frac{\mathrm{d}M}{\mathrm{d}t}-n\right\rangle_{g_{1,0}} = -nL\left[2-\sin^2i+\sin^2i\cos(2\omega)\frac{e^2-\sqrt{1-e^2}}{(1+\sqrt{1-e^2})^2}\right]
\end{equation}
\end{linenomath*}
where $M$ is the mean anomaly of the particle. Note that here Gaussian units are used in the quantity $L$ in order to be consistent with the formula given by \citet{hamilton1993motion}.

The average changing rates of argument $\omega$ and $\Omega$ can be written as
̈́\begin{linenomath*}
\begin{equation}
\left\langle\frac{\mathrm{d}\omega}{\mathrm{d}t}\right\rangle = \left\langle\frac{\mathrm{d}\omega}{\mathrm{d}t}\right\rangle_\mathrm{g_{1,0}} + \left\langle\frac{\mathrm{d}\omega}{\mathrm{d}t}\right\rangle_\mathrm{radiation}
\end{equation}
\end{linenomath*}
̈́\begin{linenomath*}
\begin{equation}
\left\langle\frac{\mathrm{d}\Omega}{\mathrm{d}t}\right\rangle = \left\langle\frac{\mathrm{d}\Omega}{\mathrm{d}t}\right\rangle_\mathrm{g_{1,0}} + \left\langle\frac{\mathrm{d}\Omega}{\mathrm{d}t}\right\rangle_\mathrm{radiation}
\end{equation}
\end{linenomath*}
where the expressions of $\left\langle\frac{\mathrm{d}\omega}{\mathrm{d}t}\right\rangle_\mathrm{radiation}$ and $\left\langle\frac{\mathrm{d}\Omega}{\mathrm{d}t}\right\rangle_\mathrm{radiation}$ are given by \citet{hamilton1993motion}.

For dust particles in the size range 0.3-2 $\mathrm{\mu m}$, the distribution of $\Omega$ is found to be practically uniform (see Figs.~\ref{fig_PwO_p3} and \ref{fig_PwO_2}, for example), i.e., $\mathrm P(\Omega) \simeq \mathrm{const}$. Under this condition the distribution of $\omega$ is independent of the solar longitude. Therefore, we may choose any solar longitude for the average method. The locations of peaks and troughs in Figs.~\ref{fig_PwO_p3} and \ref{fig_PwO_2} can be understood by analyzing the fixed points or extrema of $\left|\left\langle\frac{\mathrm{d}\omega}{\mathrm{d}t}\right\rangle\right|$. Fixed points of $\left|\left\langle\frac{\mathrm{d}\omega}{\mathrm{d}t}\right\rangle\right|$ will emerge in pairs, in a saddle-node bifurcation, where one fixed point is stable and one is unstable. At the stable fixed point, if any, we expect to see a peak in the distribution of $\omega$. But even if no fixed points exist, a small value of $\left|\left\langle\frac{\mathrm{d}\omega}{\mathrm{d}t}\right\rangle\right|$ can imply that the system stays longer at a value of $\omega$ where the minimum of $\left|\left\langle\frac{\mathrm{d}\omega}{\mathrm{d}t}\right\rangle\right|$ occurs. This behavior is called the ``ghost" of a saddle-node bifurcation \citep{strogatz2014nonlinear}. Therefore, also in this case we may expect to see a peak in the distribution of $\omega$, where the minimum of $\left|\left\langle\frac{\mathrm{d}\omega}{\mathrm{d}t}\right\rangle\right|$ occurs. In contrast, the system stays only a very short time close to the value of $\omega$ where the maximum of $\left|\left\langle\frac{\mathrm{d}\omega}{\mathrm{d}t}\right\rangle\right|$ occurs. Thus, we may expect to see a minimum in the distribution of $\omega$, at the value where the maximum of $\left|\left\langle\frac{\mathrm{d}\omega}{\mathrm{d}t}\right\rangle\right|$ occurs.

\begin{figure}
\centering
\noindent\includegraphics[width=1.05\textwidth]{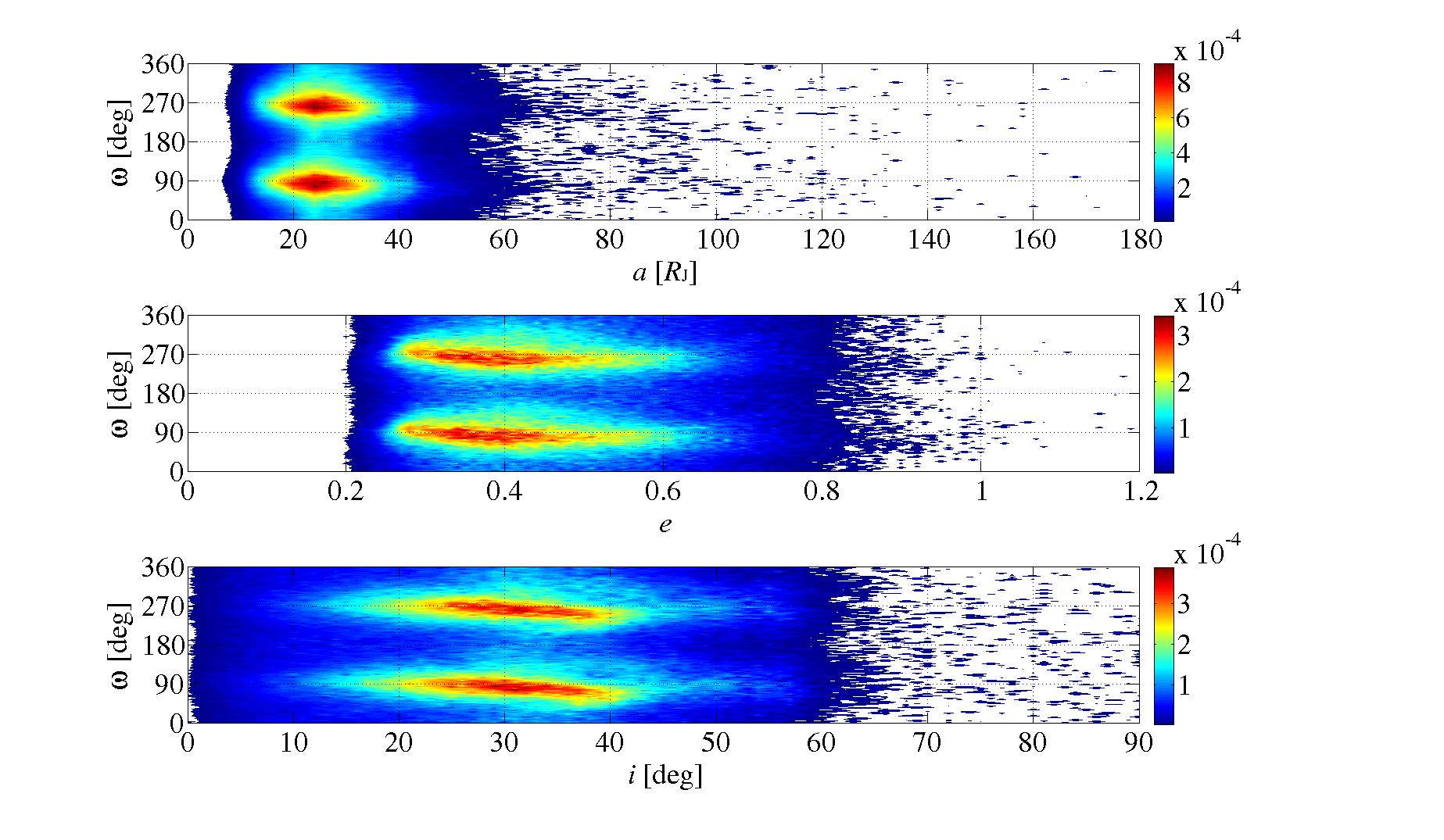}
\caption{\label{fig_aeiw_p3} Relative distributions of dust in the $a-\omega$, $e-\omega$, and $i-\omega$ planes for 0.3 $\mathrm{\mu m}$ particles from Europa.}
\end{figure}

\begin{figure}
\centering
\noindent\includegraphics[width=0.65\textwidth,angle=90]{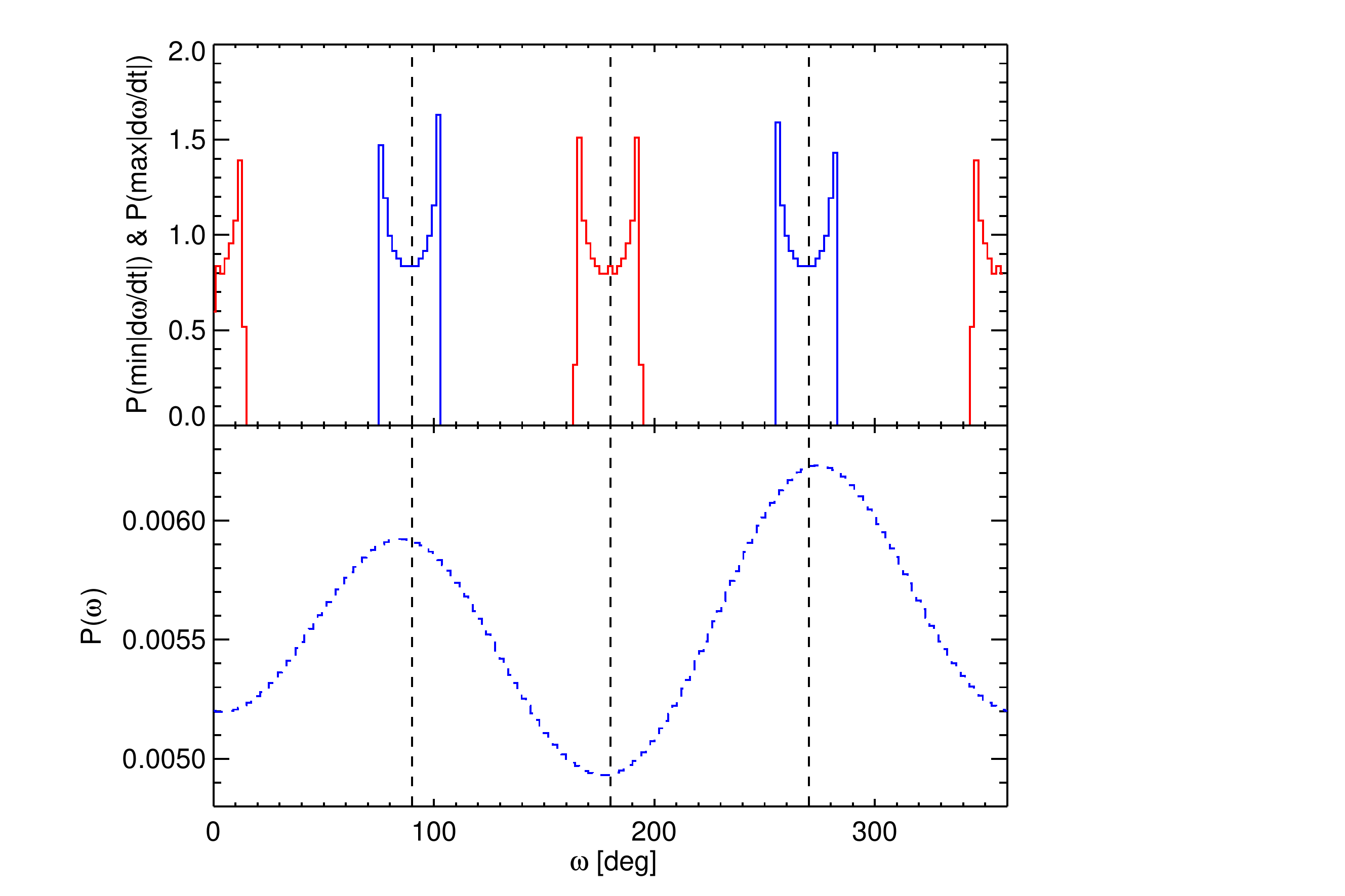}
\caption{\label{fig_Pw_recon_p3} Top panel: (relative) $P\left(\mathrm{min}\left|\frac{\mathrm{d}\omega}{\mathrm{d}t}\right|\right)$ (blue) and $P\left(\mathrm{max}\left|\frac{\mathrm{d}\omega}{\mathrm{d}t}\right|\right)$ (red) for 0.3 $\mathrm{\mu m}$ particles from Europa. Bottom panel: constructed (relative) $P (\omega)$.}
\end{figure}

\begin{figure}
\centering
\noindent\includegraphics[width=1.05\textwidth]{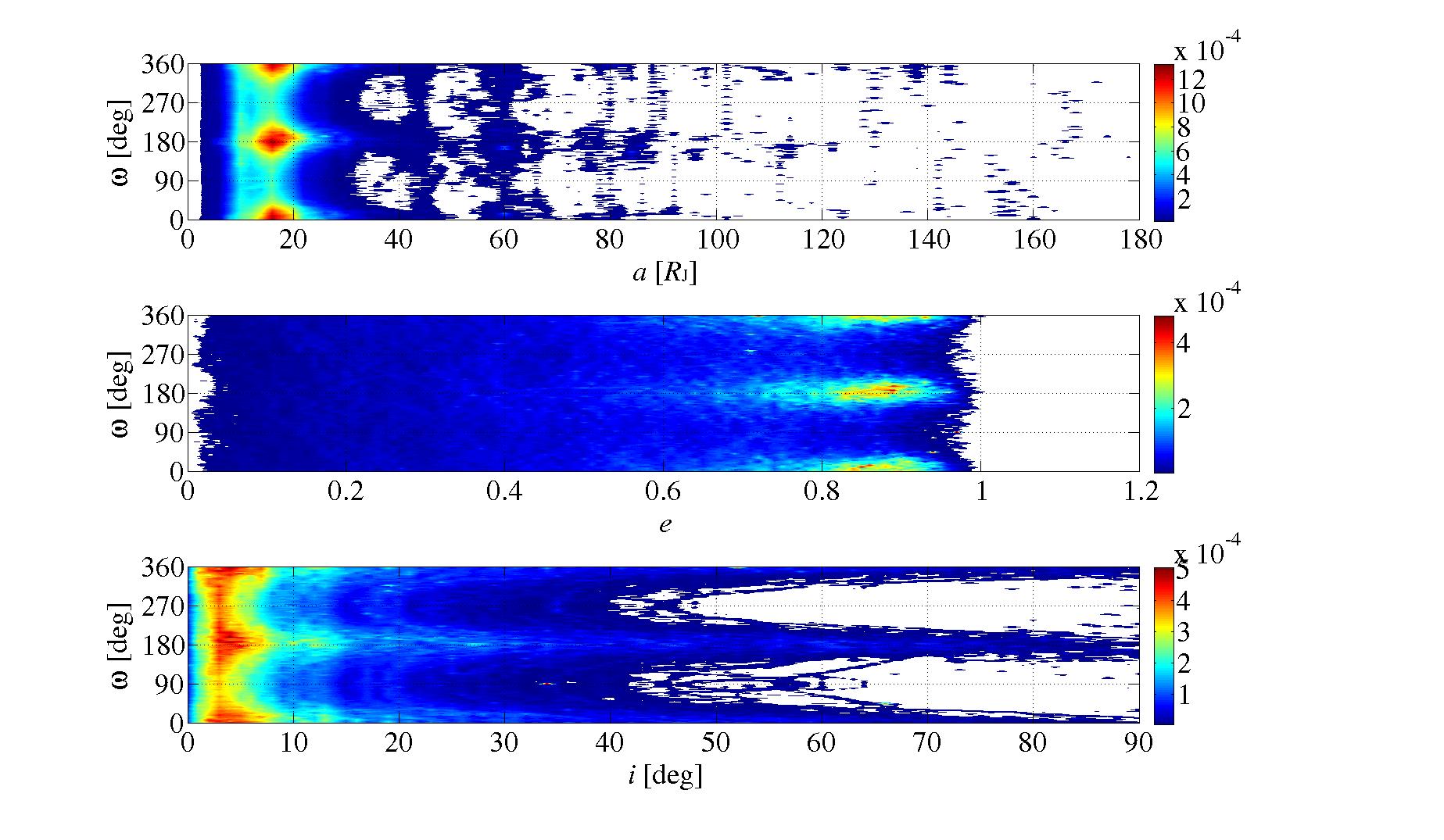}
\caption{\label{fig_aeiw_2} Relative distributions of dust in the $a-\omega$, $e-\omega$, and $i-\omega$ planes for 2 $\mathrm{\mu m}$ particles from Europa.}
\end{figure}

\begin{figure}
\centering
\noindent\includegraphics[width=0.65\textwidth,angle=90]{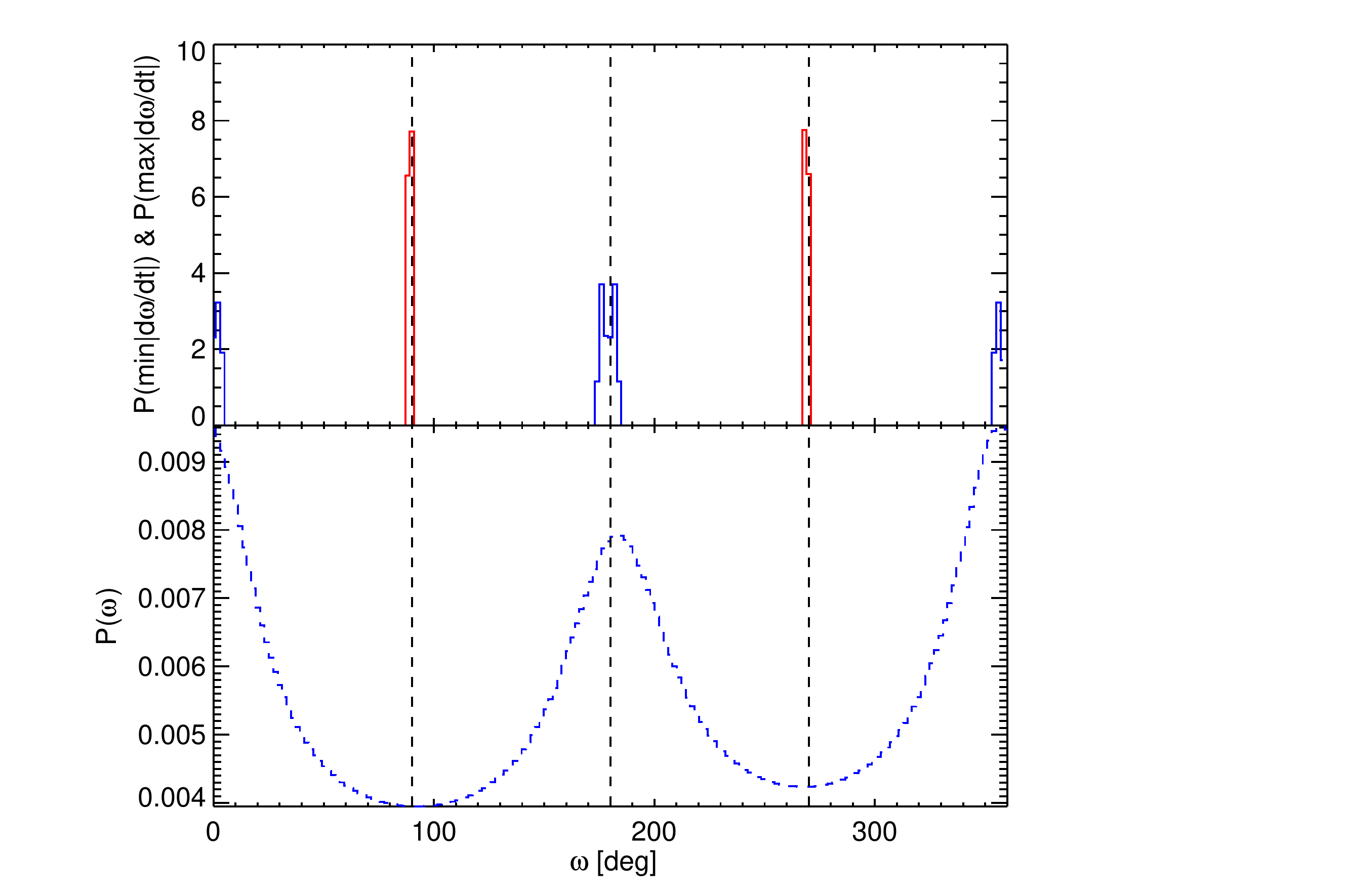}
\caption{\label{fig_Pw_recon_2} Top panel: (relative) $P\left(\mathrm{min}\left|\frac{\mathrm{d}\omega}{\mathrm{d}t}\right|\right)$ (blue) and $P\left(\mathrm{max}\left|\frac{\mathrm{d}\omega}{\mathrm{d}t}\right|\right)$ (red) for 2 $\mathrm{\mu m}$ particles from Europa. Bottom panel: constructed (relative) $P (\omega)$.}
\end{figure}

For the case without fixed points of $\left|\left\langle\frac{\mathrm{d}\omega}{\mathrm{d}t}\right\rangle\right|$, the distribution of $\omega$ can be constructed as follows
\begin{linenomath*}
\begin{equation}
P(\omega) =\left|\frac{\mathrm{d}\Omega}{\mathrm{d}\omega}\right|P(\Omega) = \left|\frac{\mathrm{d}\Omega}{\mathrm{d}t}\right|\left|\frac{\mathrm{d}\omega}{\mathrm{d}t}\right|^{-1} P(\Omega).
\end{equation}
\end{linenomath*}	
Taking an average over $\Omega$,
\begin{linenomath*}
\begin{equation} \label{equ_w_recon}
\langle P(\omega)\rangle = \int_0^{2\pi}\left|\frac{\mathrm{d}\Omega}{\mathrm{d}t}\right|\left|\frac{\mathrm{d}\omega}{\mathrm{d}t}\right|^{-1} P(\Omega)\mathrm d\Omega .
\end{equation}
\end{linenomath*}
From the distribution of orbital elements of 0.3 $\mathrm{\mu m}$ particles from Europa, we pick values for $a=24 \ R_\mathrm{J}$, $e=0.4$ and $i=30^\circ$ that correspond to the peaks in their respective distributions (Fig.~\ref{fig_aeiw_p3}). Based on Eq.~(\ref{equ_w_recon}), the construction of a distribution of $\omega$ with this method is displayed in Fig.~\ref{fig_Pw_recon_p3}. The histograms in the upper panel of this figure show that the minima of $\left|\left\langle\frac{\mathrm{d}\omega}{\mathrm{d}t}\right\rangle\right|$ gather around $90^\circ$ and $270^\circ$, and the maxima of $\left|\left\langle\frac{\mathrm{d}\omega}{\mathrm{d}t}\right\rangle\right|$ gather around $0^\circ$ and $180^\circ$. $P(\omega)$ (bottom panel of Fig.~\ref{fig_Pw_recon_p3}) has peaks close to 90$^\circ$ and 270$^\circ$, and the locations of minima are close to 0$^\circ$ and 180$^\circ$, which is consistent with Fig.~\ref{fig_PwO_p3}. The amplitude of the constructed distribution differs from that of the full dynamics because we only used the peak values of $a$, $e$ and $i$. But these are themselves distributed quantities, which will result in a broadening of the distribution of $\omega$. In a similar manner, for 2 $\mathrm{\mu m}$ sized particles from Europa we take the peak values of the distribution $a=18 \ R_\mathrm{J}$, $e=0.86$ and $i=4^\circ$ (Fig.~\ref{fig_aeiw_2}). In this case, the locations of peaks of the constructed distribution of $\omega$ are close to $0^\circ$ and $180^\circ$, while the locations of minima are close to $90^\circ$ and $270^\circ$ (Fig.~\ref{fig_Pw_recon_2}), which is consistent with Fig.~\ref{fig_PwO_2}.

For particles $\geqslant$ 5 $\mathrm{\mu m}$, the above analysis is not appropriate for two reasons. First, for sizes 5 $\mathrm{\mu m}$ and 10 $\mathrm{\mu m}$, there exists a non-negligible portion of grains that is still in orbit around Jupiter after 8,000 years. This means that the statistics for these sizes are incomplete. Second, for dust particles $\geqslant$ 30 $\mathrm{\mu m}$, the distribution of $\Omega$ is nonuniform, so the distribution of $\omega$ depends on the solar longitude. This implies that the distribution of $\omega$ is time-dependent, a behavior that is not resolved with the methods of this paper, relying on a steady state assumption.

\section{Conclusions}
In this paper the dynamics of dust particles ejected from the four Galilean moons are investigated. To this end, high accuracy orbital integrations of dust particles are performed. The relevant perturbation forces are considered in the dynamical model, including the Lorentz force, solar radiation pressure, Poynting-Robertson drag, solar gravity, the satellites' gravity, plasma drag, and the effect of non-spherical Jovian gravity. The average lifetime of dust particles is analyzed. The dynamics of the 0.05 $\mathrm{\mu m}$ (and smaller) particles is studied in an analytical way. Two sharp jumps in the average lifetime with the increase of the grain size are found for dust from all Galilean moons, and the physical origin for the jumps is explained. The dust configurations are found to be asymmetric with respect to the solar direction. For small particle sizes, the dust configuration is displaced towards the Sun, while for larger sizes it is displaced away from the Sun. Transport of the dust between the four Galilean moons and to Jupiter is evaluated, and it is discovered that a small fraction of the orbits can become retrograde due to the effect of solar radiation pressure and the Lorentz force. Finally, the distribution of orbital elements in the Jovicentric equatorial inertial frame is studied using a semi-analytical method.

%%% End of body of article:

%%%%%%%%%%%%%%%%%%%%%%%%%%%%%%%%
%% Optional Appendix goes here
%
% \appendix resets counters and redefines section heads
% but doesn't print anything.
% After typing \appendix
%
%\section{Here Is Appendix Title}
% will show
% Appendix A: Here Is Appendix Title
%
%%%%%%%%%%%%%%%%%%%%%%%%%%%%%%%%%%%%%%%%%%%%%%%%%%%%%%%%%%%%%%%%
%
% Optional Glossary or Notation section, goes here
%
%%%%%%%%%%%%%%
% Glossary is only allowed in Reviews of Geophysics
% \section*{Glossary}
% \paragraph{Term}
% Term Definition here
%
%%%%%%%%%%%%%%
% Notation -- End each entry with a period.
% \begin{notation}
% Term & definition.\\
% Second term & second definition.\\
% \end{notation}
%%%%%%%%%%%%%%%%%%%%%%%%%%%%%%%%%%%%%%%%%%%%%%%%%%%%%%%%%%%%%%%%
%
%  ACKNOWLEDGMENTS

\begin{acknowledgments}
This work was supported by the European Space Agency in the project Jovian Micrometeoroid Model (JMEM) (contract number: 4000107249/12/NL/AF) at the University of Oulu. M.~Sachse and F.~Spahn were also supported by Deutsches Zentrum f{\"u}r Luft- und Raumfahrt (50 OH 1401). We acknowledge the Finnish CSC -- IT Center for Science Ltd.~for providing computing time on their Taito cluster. The data used are listed in the references, tables, and are available from the authors upon request (xiaodong.liu@oulu.fi). We acknowledge the comments and suggestions from two anonymous reviewers that helped us improve the paper.
\end{acknowledgments}

\end{article}

%
%
%% Enter Figures and Tables here:
%
% DO NOT USE \psfrag or \subfigure commands.
%
% Figure captions go below the figure.
% Table titles go above tables; all other caption information
%  should be placed in footnotes below the table.
%
%----------------
% EXAMPLE FIGURE
%
 %\begin{figure}
 %\noindent\includegraphics[width=20pc]{samplefigure.eps}
 %\caption{Caption text here}
 %\label{figure_label}
 %\end{figure}
 
%  \begin{figure}
%  \noindent\includegraphics[width=20pc]{QprMieWarren.eps}
%  \caption{Caption text here}
%  \label{figure_label}
%  \end{figure} 

%
% ---------------
% EXAMPLE TABLE
%
\clearpage

\begin{table}
\caption{Physical and orbital properties of the Galilean moons\tablenotemark{a}}
\centering
\begin{tabular}{lrrrr}
\hline
satellites  & $R_\mathrm{m}$ [km] & $a_\mathrm{m}$ [$R_\mathrm{J}$] & $r_\mathrm{h}$ [$r_\mathrm{m}$] & $v_\mathrm{esc}$ [m/s] \\
\hline
Io	&  1829.4   &  5.90 	&  5.77		&  2552.59 \\
Europa		&  1562.6   &  9.39 	&  8.74		&  2024.66 \\
Ganymede     	&  2631.2   &  14.98	&  12.06	&  2741.50 \\
Callisto    	&  2410.3   &  26.35	&  20.82	&  2440.73 \\
\hline
\end{tabular}
\tablenotetext{a}{Here $v_\mathrm{esc}$ is the two-body escape velocity of the source moon. The parameters $R_\mathrm{m}$, $a_\mathrm{m}$ and $r_\mathrm{h}$ are radius, semi-major axis and Hill radius of the source moon, respectively. For Jupiter, the reference radius $R_\mathrm{J}=$71492 km, $J_2=$1.469562477069651E-2, $J_4=-$5.913138887463315E-4, and $J_6=$2.077510523748891E-5 \citep{jup310}. The values of $R_\mathrm{m}$ and $a_\mathrm{m}$ are taken from kernels provided by the NASA's Jet Propulsion Laboratory for the NAIF SPICE toolkit. The values of $r_\mathrm{h}$ are calculated based on Eq.~\ref{eq:hill_radius}.}
\label{tab:galilean_moons}
\end{table}

\begin{table}
\caption{Transport of dust from Io after 8000 years (the largest probabilities for each grain size are underlined)}
\centering
\begin{tabular}{lccccccc}
\hline
$r_\mathrm{g}$ ($\mu \mathrm m$)  & still in orbit  & hit Jupiter  & hit Io & hit Europa & hit Ganymede & hit Callisto  & escape  \\
\hline
 0.05	&0		&0		&0		&0		&0		&0		&\underline{1.00E+0} \\ 
 0.1	&0		&0		&0		&0		&0		&0		&\underline{1.00E+0} \\ 
 0.3	&0		&1.94E--1	&1.80E--3	&4.88E--4	&9.54E--3	&2.88E--3	&\underline{7.91E--1} \\ 
 0.6	&0		&2.80E--1	&5.56E--2	&3.93E--2	&6.76E--2	&3.22E--3	&\underline{5.54E--1} \\
 1	&0		&\underline{4.64E--1}	&1.03E--1	&9.87E--2	&7.29E--2	&6.97E--3	&2.55E--1 \\ 
 2	&7.49E--4	&\underline{5.88E--1}	&2.53E--1	&6.95E--2	&6.06E--2	&2.71E--3	&2.63E--2 \\ 
 5	&4.78E--2	&1.95E--1	&\underline{4.86E--1}	&1.45E--1	&9.22E--2	&2.42E--2	&1.01E--2 \\ 
 10	&3.84E--2	&3.88E--2	&\underline{5.00E--1}	&1.82E--1	&1.94E--1	&3.24E--2	&1.42E--2 \\ 
 30	&1.21E--2	&1.15E--2	&\underline{6.99E--1}	&1.59E--1	&9.00E--2	&1.88E--2	&9.80E--3 \\ 
 100	&3.05E--3	&2.13E--3	&\underline{8.23E--1}	&1.07E--1	&5.94E--2	&2.26E--3	&3.44E--3 \\ 
 300	&6.23E--3	&2.38E--3	&\underline{8.71E--1}	&8.13E--2	&3.28E--2	&3.91E--3	&2.88E--3 \\ 
 1000	&8.37E--5	&2.20E--3 	&\underline{8.67E--1}	&9.32E--2	&2.84E--2 	&5.41E--3	&3.44E--3 \\ 
 10000	&1.36E--3	&3.88E--3	&\underline{8.61E--1}	&8.72E--2	&4.14E--2	&2.27E--3	&2.83E--3 \\ 
 \hline
\end{tabular}
\label{tab_Io_sink}
\end{table}

\begin{table}
\caption{Transport of dust from Europa after 8000 years (the largest probabilities for each grain size are underlined)}
\centering
\begin{tabular}{lccccccc}
\hline
$r_\mathrm{g}$ ($\mu \mathrm m$)  & still in orbit  & hit Jupiter  & hit Io & hit Europa & hit Ganymede & hit Callisto  & escape  \\
\hline
 0.05	&0		&0		&0		&0		&0		&0		&\underline{1.00E+0}  \\
 0.1	&0		&0		&0		&0		&0		&0		&\underline{1.00E+0}  \\
 0.3	&0		&1.90E--1	&1.72E--3	&8.48E--4	&2.01E--2	&2.28E--3	&\underline{7.85E--1}  \\ 
 0.6	&0		&2.88E--1	&1.50E--2	&7.04E--2	&9.40E--2	&1.04E--2	&\underline{5.23E--1}  \\
 1	&0		&\underline{5.03E--1}	&3.47E--2	&1.02E--1	&9.49E--2	&5.90E--3	&2.60E--1  \\ 
 2	&0		&\underline{6.78E--1}	&9.06E--2	&1.33E--1	&5.77E--2	&8.03E--3	&3.29E--2  \\ 
 5	&9.52E--2	&1.69E--1	&7.61E--2	&2.46E--1	&\underline{2.98E--1}	&7.89E--2	&3.70E--2  \\ 
 10	&1.12E--1	&3.38E--2	&7.13E--2	&\underline{3.99E--1}	&3.03E--1	&7.22E--2	&9.44E--3  \\ 
 30	&1.52E--2	&4.69E--3	&1.30E--1	&\underline{5.25E--1}	&2.58E--1	&4.71E--2	&2.06E--2  \\ 
 100	&8.22E--3	&4.31E--3	&1.80E--1	&\underline{5.22E--1}	&2.50E--1	&3.13E--2	&4.82E--3  \\
 300	&1.84E--2	&4.99E--3	&1.65E--1	&\underline{5.09E--1} 	&2.58E--1	&3.70E--2	&8.29E--3  \\
 1000	&7.21E--3	&3.95E--3	&2.15E--1 	&\underline{5.10E--1}	&2.18E--1	&4.03E--2	&5.45E--3  \\ 
 10000	&1.81E--2	&4.84E--3	&2.14E--1	&\underline{4.56E--1}	&2.42E--1	&5.90E--2	&7.13E--3  \\
 \hline
\end{tabular}
\label{tab_Europa_sink}
\end{table}

\begin{table}
\caption{Transport of dust from Ganymede after 8000 years (the largest probabilities for each grain size are underlined)}
\centering
\begin{tabular}{lccccccc}
\hline
$r_\mathrm{g}$ ($\mu \mathrm m$)  & still in orbit  & hit Jupiter  & hit Io & hit Europa & hit Ganymede & hit Callisto  & escape  \\
\hline
 0.05	&0		&0		&0		&0		&0		&0		&\underline{1.00E+0}  \\
 0.1	&0		&0		&0		&0		&0		&0		&\underline{1.00E+0}  \\  
 0.3	&0		&1.75E--1	&0.00E+0	&0.00E+0	&1.07E--2	&6.01E--3	&\underline{8.08E--1}  \\  
 0.6	&0		&3.22E--1	&2.87E--3	&1.24E--2	&7.75E--2	&1.01E--2	&\underline{5.75E--1}  \\  
 1	&0		&\underline{5.40E--1}	&2.64E--2	&1.94E--2	&6.10E--2	&1.30E--3	&3.52E--1  \\ 
 2	&6.73E--3	&\underline{8.17E--1}	&1.94E--2	&6.36E--3	&7.03E--2	&2.21E--3	&7.80E--2  \\  
 5	&2.06E--1	&1.28E--1	&2.35E--2	&5.77E--2	&\underline{4.46E--1}	&9.02E--2	&4.77E--2  \\  
 10	&1.85E--1	&7.26E--2	&3.51E--2	&7.54E--2	&\underline{5.02E--1}	&9.50E--2	&3.49E--2  \\  
 30	&1.01E--1	&2.89E--2	&5.06E--2	&7.39E--2	&\underline{5.87E--1}	&1.11E--1	&4.80E--2  \\   
 100	&4.10E--2	&7.51E--3	&7.69E--2	&1.17E--1	&\underline{5.86E--1}	&1.24E--1	&4.75E--2  \\ 
 300	&4.76E--2	&4.19E--3	&6.35E--2	&1.66E--1	&\underline{5.81E--1}	&8.86E--2	&4.86E--2  \\ 
 1000	&4.45E--2	&9.49E--3	&8.05E--2	&1.41E--1	&\underline{5.69E--1}	&1.02E--1 	&5.33E--2  \\ 
 10000	&6.74E--2	&1.77E--2	&8.27E--2	&1.37E--1	&\underline{5.63E--1}	&9.43E--2	&3.82E--2  \\
 \hline
\end{tabular}
\label{tab_Ganymede_sink}
\end{table}

\begin{table}
\caption{Transport of dust from Callisto after 8000 years (the largest probabilities for each grain size are underlined)}
\centering
\begin{tabular}{lccccccc}
\hline
$r_\mathrm{g}$ ($\mu \mathrm m$)  & still in orbit  & hit Jupiter  & hit Io & hit Europa & hit Ganymede & hit Callisto  & escape  \\
\hline
 0.05	&0		&0		&0		&0		&0		&0		&\underline{1.00E+0}  \\
 0.1	&0		&0		&0		&0		&0		&0		&\underline{1.00E+0}  \\ 
 0.3	&0		&1.57E--1	&1.07E--4	&0.00E+0 	&1.25E--2	&1.21E--3 	&\underline{8.29E--1}  \\    
 0.6	&0		&4.01E--1	&6.29E--3	&3.48E--3	&7.77E--4	&1.37E--2	&\underline{5.74E--1}  \\   
 1	&0		&\underline{6.46E--1}	&5.50E--3	&1.45E--3	&2.75E--2	&6.39E--3	&3.13E--1  \\   
 2	&1.12E--1	&\underline{5.39E--1}	&4.72E--3 	&1.54E--2	&7.11E--2	&3.30E--2	&2.25E--1  \\   
 5	&\underline{4.22E--1}	&1.28E--1	&1.08E--2	&1.45E--3	&1.18E--1	&2.35E--1	&8.47E--2  \\    
 10	&\underline{3.98E--1}	&5.55E--2	&7.03E--3	&6.16E--2	&1.42E--1	&2.44E--1	&9.17E--2  \\   
 30	&2.32E--1	&2.17E--2	&3.28E--2	&2.94E--2	&2.18E--1	&\underline{3.54E--1}	&1.12E--1  \\
 100	&1.42E--1	&6.25E--3	&3.36E--2	&4.86E--2	&2.35E--1	&\underline{4.35E--1}	&9.94E--2  \\   
 300	&2.11E--1	&1.09E--2	&2.26E--2	&7.55E--2	&2.24E--1	&\underline{3.73E--1}	&8.38E--2  \\ 
 10000	&1.99E--1	&1.64E--2	&4.19E--2	&4.78E--2	&2.03E--1	&\underline{3.99E--1}	&9.21E--2  \\ 
 10000	&1.67E--1	&1.36E--2	&3.65E--2	&4.08E--2	&3.03E--1	&\underline{3.46E--1}	&9.37E--2  \\ 
 \hline
\end{tabular}
\label{tab_Callisto_sink}
\end{table}

\begin{table}
\caption{Fraction of retrograde particles per source moon}
\centering
\begin{tabular}{lcccc}
\hline 
$r_\mathrm{g}$ ($\mu \mathrm m$)  & Io & Europa & Ganymede & Callisto  \\
\hline
 0.05  & 0		& 0		& 0		& 5.45E--5  \\
 0.1   & 0		& 0		& 0		& 7.45E--4  \\
 0.3   & 1.79E--2	& 1.05E--2	& 1.04E--2 	& 1.91E--2  \\
 0.6   & 2.43E--2	& 2.43E--2	& 2.85E--2	& 4.67E--2  \\
 1     & 3.09E--2	& 2.18E--2	& 3.64E--2	& 8.38E--2  \\
 2     & 4.38E--2	& 2.95E--2	& 1.20E--2	& 9.70E--4  \\
 5     & 3.72E--4	& 5.30E--4	& 6.95E--4	& 4.98E--4  \\
 10    & 3.15E--6	& 3.84E--6	& 2.93E--5	& 9.28E--5  \\
 30    & 6.17E--7	& 2.49E--5	& 8.93E--6	& 1.51E--4  \\
 100   & 1.29E--5	& 4.80E--6	& 6.31E--6	& 1.14E--4  \\
 300   & 0		& 2.41E--6	& 2.83E--6	& 2.50E--4  \\
 1000  & 5.88E--7	& 9.38E--6	& 2.79E--6	& 1.19E--4  \\
 10000 & 4.64E--9	& 5.80E--6	& 2.87E--6	& 8.63E--5  \\
\hline
\end{tabular}
\label{tab_retro_frac}
\end{table}


\end{document}